\documentclass[12pt]{iopart}
\newcommand{\phol}{P_{L}^{\gamma}}
\newcommand{\phoc}{P_{c}^{\gamma}}
\newcommand{\tarx}{P_{x}^{T}}
\newcommand{\tary}{P_{y}^{T}}
\newcommand{\tarz}{P_{z}^{T}}
\newcommand{\recx}{P_{x'}^{R}}
\newcommand{\recy}{P_{y'}^{R}}
\newcommand{\recz}{P_{z'}^{R}}
\newcommand{\fpf}[2]{F_{#1}^{*}F_{#2}}
\newcommand{\upa}{\textit{unp}}

\newcommand{\ket}[1]{|{#1}\rangle}
\newcommand{\bra}[1]{\langle{#1}|}

\newcommand{\sla}[1]{{\not\! #1}}

\usepackage{iopams} 
\usepackage{setstack}
\usepackage{graphicx}
\usepackage{color}
\usepackage{lscape}
\usepackage{color}
\usepackage{ulem}

\begin{document}

\title[Determining pseudoscalar meson photo-production amplitudes ...]
{Determining pseudoscalar meson photo-production amplitudes from complete experiments}

\author{A. M. Sandorfi$^1$, S. Hoblit$^{2,3}$, H. Kamano$^1$ and T.-S. H. Lee$^{4,1}$}

\address{$^1$ Thomas Jefferson National Accelerator Facility, Newport News, VA 23606, USA}
\address{$^2$ Department of Physics, University of Virginia, Charlottesville, VA 22901, USA}
\address{$^3$ National Nuclear Data Center, Brookhaven National Laboratory, Upton, NY 11973, USA}
\address{$^4$ Physics Division, Argonne National Laboratory, IL 60439, USA}
\begin{abstract}
A new generation of complete experiments is focused on
a high precision extraction of pseudoscalar meson photo-production amplitudes. 
Here, we review the development of the most general analytic form of the cross section,
dependent upon the three polarization vectors of the beam, target and recoil baryon,
including all single, double and triple-polarization terms involving 16 spin-dependent observables.
We examine the different conventions that have been used by different authors, and we present  
expressions that allow the direct numerical calculation of any pseudoscalar meson photo-production
observables  with arbitrary 
spin projections from the Chew-Goldberger-Low-Nambu (CGLN) amplitudes. 
We use this numerical tool to clarify apparent sign differences that
exist in the literature, in particular with the definitions of six double-polarization observables.
We also present analytic expressions that determine the recoil baryon polarization, 
together with examples of their potential use with quasi-$4\pi$ detectors to deduce observables.
As an illustration of the use of the consistent machinery presented in this review, 
we carry out a multipole analysis of 
the $\gamma p\to K^+\Lambda$ reaction and examine the impact of recently published 
polarization measurements. 
When combining data from different experiments, we utilize the Fierz identities to fit 
a consistent set of scales. 
In fitting multipoles, we use a combined Monte Carlo sampling of the amplitude space, 
with gradient minimization, and find a shallow $\chi^2$ valley pitted with a very large 
number of local minima. 
This results in broad bands of multipole solutions that are experimentally indistinguishable. 
While these bands have been noticeably narrowed by the inclusion of new polarization 
measurements, many of the multipoles remain very poorly determined, even in sign, 
despite the inclusion of data on 8 different observables. 
We have compared multipoles from recent PWA codes with our model-independent solution 
bands, and found that such comparisons provide useful consistency tests which
clarify model interpretations.
The potential accuracy of amplitudes that could be extracted from measurements of all 16
polarization observables has been studied with mock data using the statistical variations
that are expected from ongoing experiments.
We conclude that, while a mathematical solution to the problem of determining an amplitude 
free of ambiguities may require 8 observables, as has been pointed out in the literature, 
experiments with realistically achievable uncertainties will require a significantly larger number.
\end{abstract}

\maketitle

\section{\label{sec:introduction}Introduction}

As a consequence of dynamic chiral symmetry breaking, the Goldstone bosons $(\pi, \eta, K)$
dress the nucleon and alter its spectrum. Not surprisingly, pseudoscalar meson
production has been a powerful tool in studying the spectrum of excited nucleon
states. However, such states are short lived and broad so that above the energy
of the first resonance, the $P_{33}$~$\Delta$(1232), the excitation spectrum is a complicated
overlap of many resonances. Isolating any one and separating it from backgrounds
has been a long-standing problem in the literature.

The spin degrees of freedom in meson photoproduction provide signatures of
interfering partial wave strength that are often dramatic and have been useful
for differentiating between models of meson production amplitudes. 
Models that must account for interfering resonance amplitudes and non-resonant 
contributions are often severely challenged by new polarization data. 
Ideally, one would like to partition the problem by first determining 
the amplitudes from experiment, at least to within a phase, and then relying 
upon a model to separate resonances from non-resonant processes. 
Single-pseudoscalar photoproduction is described by 4~complex amplitudes 
(two for the spin states of the photon, two for the nucleon target and two 
for the baryon recoil, which parity considerations reduce to a total of 4). 
They are most commonly expressed in terms of
the Chew-Goldberger-Low-Nambu (CGLN)~\cite{chew} amplitudes.
To avoid ambiguities, it has been shown~\cite{chiang} that 
 angular distribution measurements of
at least 8 carefully chosen observables at each energy for both proton and
neutron targets must be performed.
While such experimental information has not yet been available, even after 50 years 
of photoproduction experiments, a sequence of \textit{complete} experiments 
are now underway at Jefferson Lab~\cite{Frost,HDice}, as well as complementary experiments
from the GRAAL backscattering source in Grenoble~\cite{gr07,gr09} and
the electron facilities in Bonn and Mainz, with
the goal of obtaining a direct determination of the amplitude to within a
phase, for at least a few production channels, notably $K\Lambda$ and possibly $\pi N$.

The four CGLN amplitudes can be expressed in Cartesian ($F_i$), Spherical or Helicity ($H_i$), or 
Transversity ($b_i$) representations. 
While the latter two choices afford some theoretical simplifications when predicting asymmetries 
from models~\cite{barker}, when working in the reverse direction, fitting asymmetries to extract 
amplitudes, such simplifications are largely moot. 
The four amplitudes in each of these representations are angle dependent. 
Extracting them directly from experiment would require separate fits at each angle, which greatly 
limits the data that can be used and requires some model-dependent scheme to constrain an arbitrary 
phase that could be angle-dependent. 
The solution to this intractable situation is a Wigner-Eckhart style factorization into reduced matrix 
elements, multipoles, and simple angle-dependent coefficients from angular momentum algebra. 
One can then fit the multipoles directly, which both facilitates the search for resonance behavior 
and allows the use of full angular distribution data at a fixed energy to constrain angle-independent 
quantities. 
The price is a significant increase in the number of fitting parameters, but since the excited 
states of the nucleon are associated with discrete values for their angular momentum, this expansion 
of variables is inevitable. 
Here we restrict our considerations to the CGLN $F_i$ representation, which has the simplest 
decomposition into multipoles~\cite{chew}, equations (\ref{eq:mp-f1})-(\ref{eq:mp-f4}) below.

In single-pseudoscalar meson photoproduction there are 16 possible observables, the
unpolarized differential cross section ($d\sigma_0$), 
three asymmetries which to leading order enter the general 
cross section scaled by a single polarization of either beam, target or recoil 
($\Sigma$, $T$, $P$), and three sets of four asymmetries whose leading polarization dependence
in the general cross section involves two polarizations of either 
beam-target (BT), beam-recoil (BR), or target-recoil (TR), as in \cite{barker}. 
Expressions for at least some of these observables 
in terms of the CGLN $F_{i}$ appear already 
in earlier papers \cite{donn66,donn72,fasano,drech92,knoch}.
In all cases we have found in the literature, the magnitudes of the expressions relating the
CGLN $F_{i}$ to experimental observables are identical, but the signs of some appear to differ.
This is only now becoming a significant issue since the
sign differences occur in double-polarization observables
for which little data have been available until very recently. 
There is also a set of \textit{Fierz} identities interrelating the 16 polarization 
observables, the most complete list being given in \cite{chiang}. 
We have found many of the signs in the expressions of this list appear to be incompatible 
with several of these papers.
As we will see below, much of this confusion has its origin in the same symbol, or observable name,
being used by different authors to represent different experimental quantities.

Our purpose here is two fold. 
First we assemble a complete set of relations, defining observables in terms of 
specific pairs of measurable quantities and 
providing the most general form of the cross sections in terms of all observables. 
We then give a consistent set of relations between these experimental observables and 
the CGLN amplitudes and electromagnetic multipoles~\cite{chew}.
Next, as an illustration of the use of these relations along the path to determining 
an amplitude,
we use recently published results
on 8 different observables to carry out a multipole analysis of 
the $\gamma p\to K^+ \Lambda$ reaction, free of model assumptions, and examine
the uniqueness of the resulting solutions.
Finally, we use mock data to study the potential uniqueness
of amplitudes that could be extracted from complete sets of all 16 observables.

There are several coordinate systems in use in the literature and in section~\ref{sec:kine} we 
define ours, which is the same as used in the seminal paper by Barker, Donnachie and 
Storrow (BDS)~\cite{barker}. 
In section~\ref{sec:theory} we present explicit and complete formulae that allow the direct calculation 
of matrix elements with arbitrary spin projections from CGLN amplitudes or multipoles. 
In section~\ref{sec:gcs} we present the most general analytic form of the cross section, 
dependent on the three polarization vectors of the beam, the target and the recoil baryon. 
The derivation of this cross section expression is summarized in \ref{apx_eqs}, 
and the experimental definitions of the observables in terms of cross sections with explicit 
polarization orientations is tabulated in \ref{apx_tab}. 
Using these definitions, we give in section~\ref{sec:review} 
a summary of the variations in similar formulae that appear in literature.
While the beam and target polarizations can be controlled in an experiment, the recoil polarization 
is on a very different footing, in that it arises as a consequence of
the angular momentum of the entrance channel and the reaction physics,
neither of which is under experimental control.
Expressions that determine the recoil baryon polarization are developed in section~\ref{sec:recoil}. 
To evaluate the analytic relations between observables and amplitudes we next use numerical 
calculations of the expressions in section~\ref{sec:theory} to fix signs and present the complete set 
of equations in section~\ref{sec:cgln} that determine the 16 observables from the CGLN amplitudes. 
The 37 Fierz identities that interrelate the observables are 
discussed in section~\ref{sec:fierz} and presented with consistent
signs in \ref{apx_fier}.
In section~\ref{sec:anal} we utilize the machinery we have assembled to carry out a multipole 
analysis of the $\gamma p\to K^+\Lambda$ reaction. 
(Born terms for this process are summarized in \ref{apx_born}.)
In so doing we test the nature of the $\chi^2$ valley, discuss the role of 
the arbitrary phase and examine the impact of recently published polarization data and 
the uniqueness of the multipole solutions from resent data. 
The accuracy of the data needed for a precise model independent extraction
of amplitudes is then investigated in section~\ref{sec:mock}
from a study with mock data on all possible 16 observables 
with varying levels of statistical precision.
Section~\ref{sec:summary} concludes with a brief summary.

\section{\label{sec:kine}Kinematics and coordinate definitions}

\begin{figure}[t]
\centering
\includegraphics[clip,width=\textwidth]{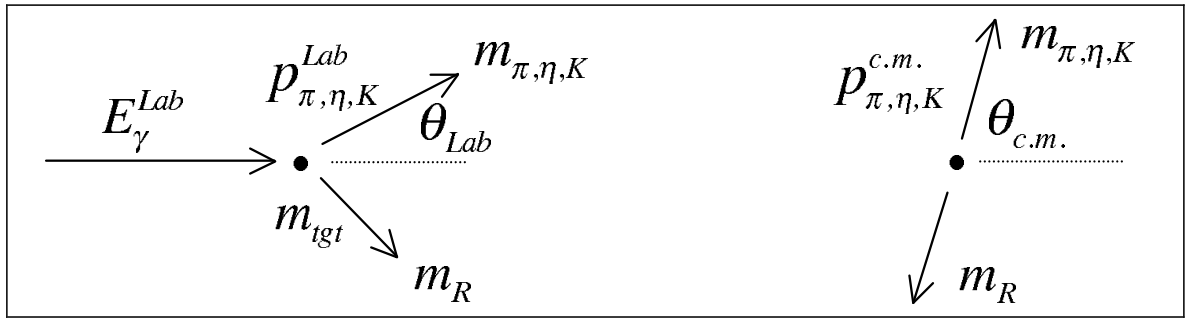}
\caption{\label{fig:kinem} Kinematic variables in meson photoproduction in Lab and c.m. frames.}
\end{figure}

The kinematic variables of meson photoproduction 
used in our derivations are specified in figure~\ref{fig:kinem}. 
Some useful relations are :
\begin{itemize}
\item{The total center of mass (c.m.) energy:
\begin{equation}
W = \sqrt{s}  = \sqrt{ m_{{\rm tgt}}(m_{{\rm tgt}} + 2E_{\gamma}^{{\rm Lab}}) }.
\end{equation} }
\item{The laboratory (Lab) energy needed to excite the hadronic system with total c.m. energy $W$:
\begin{equation}
E_{\gamma}^{{\rm Lab}} = \frac{W^{2} - m_{{\rm tgt}}^{2}}{2m_{{\rm tgt}}}.
\end{equation} }
\item{The energy of the photon in the c.m. frame:
\begin{equation}
E_{\gamma}^{{\rm c.m.}}  = \frac{W^{2} - m_{{\rm tgt}}^{2}}{2W} = q.
\end{equation} }
\item{The magnitude of the 3-momentum of the meson in the c.m. frame:
\begin{equation}
\fl
\left| p_{\pi ,\eta ,K}^{{\rm c.m.}} \right| = \frac{W}{2}\left\{ {\left[ {1 -
\left( {\frac{{m_{\pi ,\eta ,K}  + m_R }}{W}} \right)^2 }
\right]\left[ {1 - \left( {\frac{{m_{\pi ,\eta ,K}  - m_R }}{W}} \right)^2 } \right]} \right\}^{1/2}
= k.
\end{equation} }
\item{The density of state factor:
\begin{equation}
\rho_0  = \left| p_{\pi ,\eta ,K}^{{\rm c.m.}} \right| / E_{\gamma}^{{\rm c.m.}} = k/q.
\end{equation} }
\end{itemize}

\begin{figure}[t]
\centering
\includegraphics[clip,width=\textwidth]{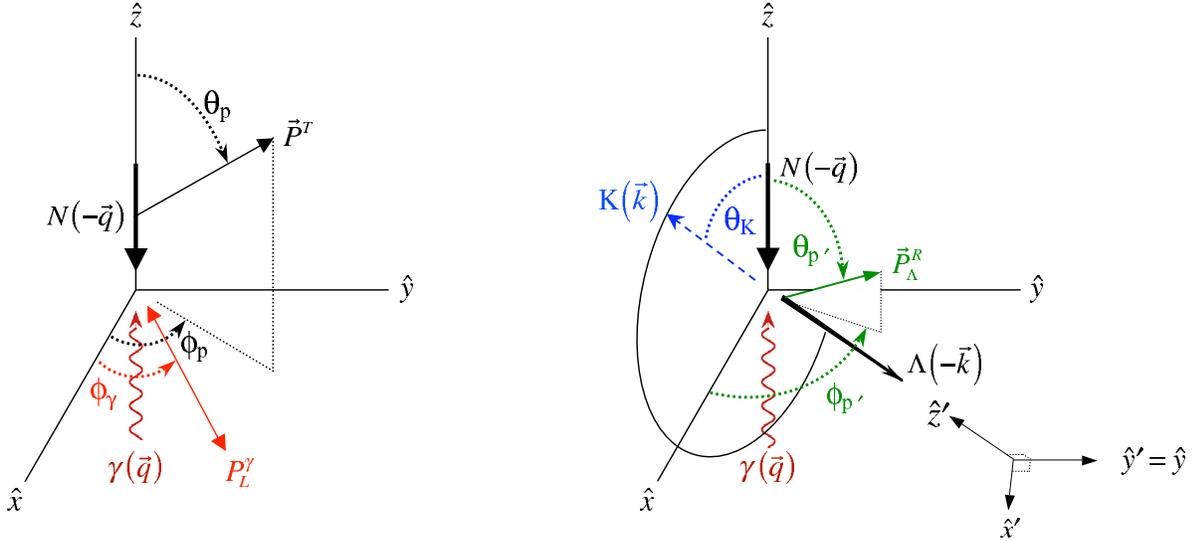}
\caption{\label{fig:coord}(Color online) 
The c.m. coordinate system and angles used to specify polarizations
in the reaction, $\gamma(\vec{q},\vec{P}^{\gamma}) + N(-\vec{q},\vec{P}^{T}) \rightarrow
K(\vec{k}) + \Lambda(-\vec{k},\vec{P}^R_{\Lambda})$. The left (right) side
is for the initial $\gamma N$ (final $K\Lambda$) system;
$\hat z$ is along the photon beam direction;
$\hat y$ is perpendicular to the $\langle \hat x - \hat z\rangle$ reaction plane
and $\hat x = \hat y \times \hat z$; 
$\hat z'$ is along the meson momentum and $\hat x'$ is in the 
$\langle \hat x - \hat z\rangle$ plane, rotated down from $\hat z$ by $\theta_K + \pi/2$.
}
\end{figure}

The definitions of polarization angles used in our derivation
are shown in figure~\ref{fig:coord}, using the
case of K $\Lambda$ production as an example. 
The $\langle \hat{x} - \hat{z} \rangle$ plane is
the reaction plane in the center of mass. 
The figure illustrates the case of linear $\gamma$ polarization, with 
the alignment direction $P_{L}^{\gamma}$ (parallel to the oscillating electric 
field of the photon) in the $\langle \hat{x} - \hat{y} \rangle$ plane 
at an angle $\phi_{\gamma}$, rotating from $\hat{x}$ towards $\hat{y}$. 
The target nucleon polarization $\vec{P}^{T}$ is specified by polar
angle $\theta_{p}$ measured from $\hat{z}$, and azimuthal angle $\phi_{p}$ in
the $\langle \hat{x} - \hat{y} \rangle$ plane, rotating from $\hat{x}$ towards $\hat{y}$. 
The recoil $\Lambda$ baryon is in the $\langle \hat{x} - \hat{z} \rangle$ plane; 
its polarization $\vec{P}_{\Lambda}^{R}$ is at polar $\theta_{p'}$, 
measured from $\hat{z}$, and azimuthal $\phi_{p'}$ in the 
$\langle \hat{x} - \hat{y} \rangle$ plane, rotating from $\hat{x}$ to $\hat{y}$.
Following BDS~\cite{barker}, observables involving recoil polarization are specified in 
the rotated coordinate system with $\hat z'=+\hat k$, along the meson c.m. momentum and 
opposite to the recoil momentum, $\hat y'=\hat y$, and $\hat x'=\hat y'\times \hat z'$ 
in the scattering plane at a polar angle of $\theta_K + (\pi/2)$ relative to $\hat z$.

The case of circular photon polarization can potentially lead to some confusion.
Most particle physics literature designates circular states as $r$, for right circular
(or $l$, for left circular), referring to the fact that with $r$ polarization the
electric vector of the photon appears to rotate clockwise \textit{when the photon is
traveling away from the observer}. However, when the same photon is viewed by
an observer facing the incoming photon the electric vector appears to rotate
counter-clockwise. For this reason optics literature traditionally designates
this same state as $l$ circularly polarized. Nonetheless, both conventions agree
on the value of the photon helicity~\cite{jackson} $h = \vec{S}\cdot\vec{P} / |\vec{P}| = \pm 1$
and so we use only the helicity designations here, $\vec{P}_{c}^{\gamma} = +1 (-1)$
when 100\% of the photon spins are
parallel (anti-parallel) to the photon momentum vector.

\section{\label{sec:theory}Calculation of polarization observables}
As discussed in section~\ref{sec:introduction}, all publications give similar formulae for polarization
observables, but conflicting signs occur in some terms with very lengthy expressions.
It is very difficult, if not impossible, to resolve this problem by repeating the same
algebraic procedures used in previous works.
To resolve these sign problems, it is necessary to develop completely different and
yet simple formulae which can be used to calculate numerically all spin observables
of pseudoscalar meson photoproduction.
This numerical tool will then allow us to check unambiguously the analytic expressions
for spin observables in all previous publications.
In this section, we present the derivation of such formulae using the case of $K\Lambda$
photoproduction as an example.

Let us first consider the case when all beam, target, and 
recoil polarizations are 100\% polarized in certain directions.
With variables specified as in figure~\ref{fig:coord}, 
the differential cross section for
$\gamma(\vec{q},\hat P^\gamma)+N(-\vec{q},m_{s_N}) \rightarrow 
K(\vec{k})+\Lambda (-\vec{k},m_{s_\Lambda})$
in the center of mass frame can be written as
\begin{equation}
\frac{d\sigma}{d\Omega}(\hat P^\gamma,m_{s_N},m_{s_\Lambda}) 
=\frac{1}{(4\pi)^2}\frac{k}{q}
\frac{m_Nm_\Lambda}{W^2}
|\bar{u}_\Lambda(-\vec{k},m_{s_\Lambda})I^\mu
\epsilon_\mu
u_N(-\vec{q},m_{s_N})|^2,
\label{eq:dsdo}
\end{equation}
where $W=q+E_N(q)=E_K(k)+E_\Lambda(k)$;
$\epsilon_\mu = (0,\hat P^\gamma)$ with $|\hat P^\gamma|=1$
is the photon polarization vector;
$m_{s_\Lambda}$ and $m_{s_N}$ are the spin substate quantum numbers
of the $\Lambda$ and the nucleon along the $z$-direction, respectively;
$\bar{u}_\Lambda I^\mu\epsilon_\mu u_N$
is normalized to the usual invariant amplitude
calculated from a Lagrangian 
in the convention of Bjorken and Drell~\cite{bjdr}.
For example, for a simplified Lagrangian density 
$L(x)=-(f_{K\Lambda N}/m_K)\bar{\psi}_\Lambda(x)\gamma_5\gamma_\mu\psi_N(x)\partial^\mu\phi_{K}(x)
+e_N\bar{\psi}_N(x)\gamma^\mu\psi_N(x)A_\mu(x)$, the $s$-channel 
$\gamma(q) + N(p) \rightarrow N(p'+k) \rightarrow K^+(k) + \Lambda(p')$
contribution to $I^\mu$ is 
$\rmi e_N(f_{K\Lambda N}/m_K)\sla{k}\gamma_5[(\sla{k}+\sla{p'})-m_N]^{-1} \gamma^\mu$.
By averaging over all initial state polarizations and summing over final state polarizations
in (\ref{eq:dsdo}), we can obtain the unpolarized cross section:
\begin{equation}
d\sigma_0 \equiv \frac{1}{4}
\sum_{m_{s_N}=\pm 1/2} \sum_{m_{s_\Lambda}=\pm 1/2} \sum_{\gamma{\rm -spins}}
\frac{d\sigma}{d\Omega}(\hat P^\gamma,m_{s_N},m_{s_\Lambda}) ,
\label{eq:unp-dcs}
\end{equation}
where the symbol $\sum_{\gamma{\rm -spins}}$ implies taking summation over two photon polarization
states, with polarization vectors perpendicular to each other for linearly polarized 
photons and with helicity $\pm 1$ states for circularly polarized photons.

The CGLN amplitude~\cite{chew} is defined by
\begin{eqnarray}
\bar{u}_\Lambda(-\vec{k},m_{s_\Lambda})I^\mu \epsilon_\mu
 u_N(-\vec{q},m_{s_N}) =-\frac{4\pi W}{\sqrt{m_Nm_\Lambda}}
\bra{m_{s_\Lambda}}F_{{\rm CGLN}}\ket{m_{s_N}},
\label{eq:cgln-def}
\end{eqnarray}
where $\ket{m_s}$ is the usual eigenstate of the Pauli operator $\sigma_z$,
and
\begin{eqnarray}
F_{{\rm CGLN}} = \sum_{i=1,4} O_i F_i(\theta_K, E),
\label{eq:cgln}
\end{eqnarray}
with
\begin{eqnarray}
O_1 &=& -\rmi \vec{\sigma}\cdot \hat P^\gamma, 
\label{eq:o1}\\
O_2 &=&
-[\vec{\sigma}\cdot\hat{k}] [\vec{\sigma}\cdot (\hat{q}\times \hat P^\gamma)], 
\label{eq:o2}\\
O_3 &=& 
-\rmi [\vec{\sigma}\cdot\hat{q}][ \hat{k}\cdot\hat P^\gamma],
\label{eq:o3}\\
O_4 &=&
-\rmi [\vec{\sigma}\cdot\hat{k}][\hat{k}\cdot\hat P^\gamma].
\label{eq:o4}
\end{eqnarray}
Here we have defined $\hat{k}=\vec k/|\vec k|$ and $\hat{q}=\vec q/|\vec q|$.
We then obtain
\begin{eqnarray}
\frac{d\sigma}{d\Omega}(\hat P^\gamma,m_{s_N},m_{s_\Lambda})
=\frac{k}{q} |\bra{m_{s_\Lambda}}F_{{\rm CGLN}}\ket{m_{s_N}}|^2.
\label{eq:dsdo-cgln}
\end{eqnarray}

The formulae for calculating CGLN amplitudes from multipoles
are well known~\cite{chew} and are given below:
\begin{eqnarray}
\fl
F_1 = \sum_{l=0}[P_{l+1}'(x) E_{l+} + P_{l-1}'(x) E_{l-} + lP_{l+1}'(x)M_{l+} + (l+1)P_{l-1}'(x) M_{l-}], 
\label{eq:mp-f1}\\
\fl
F_2 = \sum_{l=0}[(l+1)P_l'(x)M_{l+} + lP_l'(x) M_{l-}],
\label{eq:mp-f2}\\
\fl
F_3 = \sum_{l=0}[P_{l+1}''(x) E_{l+} + P_{l-1}''(x) E_{l-} - P_{l+1}''(x) M_{l+} + P_{l-1}''(x) M_{l-}],
\label{eq:mp-f3}\\
\fl
F_4 = \sum_{l=0}[- P_{l}''(x) E_{l+} - P_{l}''(x) E_{l-} + P_{l}''(x) M_{l+} - P_{l}''(x) M_{l-}].
\label{eq:mp-f4}
\end{eqnarray}
where $x=\hat{k}\cdot\hat{q} = \cos\theta_K$, $l$ is the
orbital angular momentum of the $K\Lambda$ system, and $P_{l}'(x)=dP_{l}(x)/dx$ and
$P_{l}''(x)=d^2P_l(x)/dx^2$ are the derivatives of the Legendre function
$P_l(x)$, with the understanding that $P_{-1}'= P_{-1}''=0$.
In practice, the sum runs to a limiting value of $l_{max}$ which depends on the energy.

In order to calculate the 16 polarization observables in an arbitrary experimental geometry,
we develop a form for the cross section with arbitrary spin projections for initial and 
final baryon states, 
$\gamma(\vec{q}, \hat P^\gamma) + N(-\vec{q},\hat P^T) 
\to K (\vec{k}) + \Lambda(-\vec{k},\hat P^R)$, 
as specified in figure~\ref{fig:coord}, where $\hat P^T$ ($\hat P^R$) is the unit 
vector specifying the direction of the target (recoil) spin polarization.
Here linear photon polarization must be in the $\langle\hat x-\hat y\rangle$ plane
and circular photon polarization must be aligned with $\hat z$, while
$\hat P^T$ and $\hat P^R$ can be in any directions.
The corresponding cross section is obtained by simply replacing 
$|\bra{m_{s_\Lambda}}F_{{\rm CGLN}}\ket{m_{s_N}}|^2$
in (\ref{eq:dsdo-cgln}) with $|\bra{\hat P^R} F_{{\rm CGLN}} \ket{\hat P^T}|^2$:
\begin{equation}
d\sigma^{{\rm B,T,R}}(\hat P^\gamma,\hat P^T,\hat P^R)\equiv
\frac{d\sigma}{d\Omega}(\hat P^\gamma,\hat P^T,\hat P^R)
=\frac{k}{q}
|\bra{\hat P^R} F_{{\rm CGLN}} \ket{\hat P^T}|^2,
\label{eq:spin-obs}
\end{equation}
where $\ket{\hat P^T}$ ($\bra{\hat P^R}$) is a state of the initial (final) 
spin-$1/2$ baryon with the spin pointing in the $\hat P^T$ ($\hat P^R$) direction.
We note that if $\hat P^T$ ($\hat P^R$) is in the direction of the momentum of 
the initial (final) baryon, then $\ket{\hat P^T}$ ($\bra{\hat P^R}$) is 
the usual helicity state as defined, for example, by Jacob and Wick~\cite{jw59}. 
We need to consider more general spin orientations for all possible experimental geometries.
The spin state $\ket{\hat s}$ quantized in the direction
of an arbitrary vector $\hat s =(1, \theta,\phi)$ is defined by
\begin{equation}
\vec S\cdot \hat s  \ket{\hat s} = +\frac{1}{2}\ket{\hat s},
\label{eq:hel}
\end{equation}
where $\vec S$ is the spin operator. 
For the considered spin-1/2 baryons,
$\vec S$ is expressed with the Pauli matrix: $\vec S = \vec \sigma/2$.

We next derive explicit formulae for  calculating the matrix element 
$\bra{\hat P^R} F_{{\rm CGLN}} \ket{\hat P^T}$
in terms of the CGLN amplitudes $F_i$ in (\ref{eq:mp-f1})-(\ref{eq:mp-f4}).
We note that the spin state $\ket{\hat s}$ is related to the usual eigenstate
of $z$-axis quantization by rotations:
\begin{eqnarray}
\ket{\hat s} = \sum_{m=\pm 1/2} D^{(1/2)}_{m,+1/2}
(\phi,\theta,-\phi) \ket{m},
\label{eq:hel-v}
\end{eqnarray}
where $\ket{m}$ is defined as $S_z \ket{\pm {1}/{2}}
=(\pm 1/2)\ket{\pm {1}/{2}}$, and
\begin{eqnarray}
D^{(1/2)}_{m,\lambda}(\phi,\theta,-\phi) =
\exp[-\rmi(m-\lambda)\phi]
d^{1/2}_{m,\lambda}(\theta).
\label{eq:big-d}
\end{eqnarray}
We use the phase convention of Brink and Satchler~\cite{bs68} where,
\begin{equation}
\eqalign{
d^{1/2}_{+1/2,+1/2}(\theta)=d^{1/2}_{-1/2,-1/2}(\theta)=\cos\frac{\theta}{2}\cr
d^{1/2}_{-1/2,+1/2}(\theta)=-d^{1/2}_{+1/2,-1/2}(\theta)=\sin\frac{\theta}{2}.
}
\label{eq:wig-d}
\end{equation}
Equation~(\ref{eq:hel-v}) can be easily verified by explicit calculations
using the definition (\ref{eq:hel})
and the properties~(\ref{eq:big-d}) and~(\ref{eq:wig-d})
for the special cases where $\hat{s} =\hat{x}$, $\hat{y}$, $\hat{z}$, 
together with the usual definition of the Pauli matrices,
$(\sigma_i)_{mm'}$ [$i=x,y,z$ and $m$ (row), $m'$ (column) $=\pm 1/2,\pm1/2$],
\begin{equation}
\sigma_x= \left(\begin{array}{cc}0&1\\ 1&0 \end{array}\right),\qquad
\sigma_y= \left(\begin{array}{cc}0&-\rmi\\ \rmi&0 \end{array}\right),\qquad
\sigma_z= \left(\begin{array}{cc}1&0\\ 0&-1 \end{array}\right).
\label{eq:pauli}
\end{equation}

From figure~\ref{fig:coord}, the momenta and linear photon polarization are expressed as
\begin{eqnarray}
\vec{q}&=& q(0,0,1), \label{eq:q} \\
\vec{k}&=& k(\sin\theta_K, 0, \cos\theta_K), \label{eq:k}\\
\hat P^\gamma_L&=&(\cos\phi_\gamma, \sin\phi_\gamma, 0). \label{eq:esilon}
\end{eqnarray}
Circular photon polarizations of helicity $\lambda_\gamma$ are expressed as
\begin{equation}
(\hat P_c^\gamma)_{\lambda_\gamma=\pm 1} =
\mp\frac{1}{\sqrt{2}}(\hat x \pm \rmi\hat y) .
\label{eq:circular}
\end{equation}
For the initial and final baryon polarizations, we use the spherical variables,
as in figure~\ref{fig:coord}:
\begin{eqnarray}
\hat P^T&=&(1,\theta_p,\phi_p) \label{eq:pt},\\
\hat P^R&=&(1,\theta_{p'},\phi_{p'}) \label{eq:pr}.
\end{eqnarray}

By using (\ref{eq:q})-(\ref{eq:esilon}), we can rewrite
$O_i$ in (\ref{eq:o1})-(\ref{eq:o4}) as
\begin{eqnarray}
O_i =\sum_{n=0,3}C_{i,n}(\theta_K,\phi_\gamma) \sigma_n,
\label{eq:cgln-o}
\end{eqnarray}
where $\sigma_0 = \mathbf{1}$, $\sigma_1=\sigma_x$, $\sigma_2=\sigma_y$,
$\sigma_3=\sigma_z$. 
The explicit form of $C_{i,n}$ is given in table~\ref{tab:cin}.

\begin{table}[tb]
\caption{$C_{i,n}(\theta_K,\phi_\gamma)$ of (\ref{eq:cgln-o}) and~(\ref{eq:spin-mx-1}).}
\label{tab:cin}
\begin{indented} 
\item[] \begin{tabular}{@{}lllll}
\br
   &$n=0$&$n=1$&$n=2$&$n=3$\\
\mr
$i=1$&0&$-\rmi\cos\phi_\gamma$&$-\rmi\sin\phi_\gamma$& 0 \\
$i=2$&$ \sin\theta_K \sin\phi_\gamma$& $\rmi \cos\theta_K \cos\phi_\gamma$& $i\cos\theta_K \sin\phi_\gamma$& $-\rmi \sin\theta_K \cos\phi_\gamma$ \\
$i=3$&0&0&0&$-\rmi \sin\theta_K \cos\phi_\gamma$ \\
$i=4$&0&$-\rmi\sin^2\theta_K \cos\phi_\gamma$& 0&$-\rmi\sin\theta_K \cos\theta_K \cos\phi_\gamma$\\
\br
\end{tabular}
\end{indented}
\end{table}

By using (\ref{eq:hel-v}) and (\ref{eq:cgln}) and~(\ref{eq:cgln-o}),
the photoproduction matrix element can then be calculated as
\begin{equation}
\bra{\hat P^R }F_{{\rm CGLN}}\ket{\hat P^T }
=
\sum_{n=0,3} G_n(\theta_K,\phi_\gamma)\bra{\hat P^R}\sigma_n\ket{\hat P^T},
\label{eq:spin-mx}
\end{equation}
with 
\begin{equation}
G_n(\theta_K,\phi_\gamma) = 
\sum_{i=1,4} F_i(\theta_K,E) C_{i,n}(\theta_K,\phi_\gamma),
\label{eq:spin-mx-1}
\end{equation}
and
\begin{eqnarray}
\fl \bra{\hat P^R} \sigma_n \ket{\hat P^T} =
\sum_{m_{s_\Lambda},m_{s_N}=\pm 1/2}
\! \! \! \!
D^{(1/2)*}_{m_{s_\Lambda},+1/2}(\phi_{p'},\theta_{p'},-\phi_{p'}) D^{(1/2)}_{m_{s_N},+1/2}(\phi_p,\theta_{p},-\phi_{p})
\bra{m_{s_\Lambda}} \sigma_n \ket{m_{s_N}},
\nonumber\\
\label{eq:spin-mx-2}
\end{eqnarray}
where $\bra{m_{s_\Lambda}} \sigma_n \ket{m_{s_N}}=(\sigma_n)_{m_{s_\Lambda},m_{s_N}}$ 
are the elements of the Pauli matrices of (\ref{eq:pauli}).

We may now start with any set of multipoles and use 
(\ref{eq:mp-f1})-(\ref{eq:mp-f4}) to calculate the CGLN amplitudes, 
which are then used to calculate the matrix element 
$\bra{\hat P^R} F_{{\rm CGLN}} \ket{\hat P^T}$ with the help of
(\ref{eq:spin-mx})-(\ref{eq:spin-mx-2}).
Equation~(\ref{eq:spin-obs}) then allows us 
to calculate all possible  polarization observables,
for the case of unit polarization vectors with arbitrary orientation.

With non-unit polarization vectors, the general cross section 
can be expressed in terms of (\ref{eq:spin-obs}) as,
(see also \ref{apx_eqs}),
\begin{eqnarray}
\fl
d\sigma^{{\rm B,T,R}}(\vec P^\gamma,\vec P^T,\vec P^R)
=
\sum_{\hat P=\hat P^\gamma_1,\hat P^\gamma_2} \sum_{\hat Q=\pm \hat P^T} \sum_{\hat R=\pm \hat P^R} 
\mathfrak{p}^\gamma_{\hat P} \mathfrak{p}^T_{\hat Q} \mathfrak{p}^R_{\hat R}~
d\sigma^{{\rm B,T,R}}(\hat P,\hat Q,\hat R).
\label{eq:general-dcs}
\end{eqnarray}
Here the vector $\vec P^X$ specifies the degree and direction of the polarization 
of particle $X=\gamma,T,R$. 
For the target (T) and recoil (R) baryons, this is just 
$\vec P^X = (\mathfrak{p}^X_{+\hat P^X}-\mathfrak{p}^X_{-\hat P^X})\hat P^X$, where
$\mathfrak{p}^X_{\pm\hat P^X}$ ($X=T,R$) is the probability of observing $X$ with its polarization
vector pointing in the $\pm \hat P^X$ direction.
For the photons ($\gamma$), however, the non-unit polarization vector can be expressed as
$\vec P^\gamma = (\mathfrak{p}^\gamma_{\hat P^\gamma_1}-\mathfrak{p}^\gamma_{\hat P^\gamma_2})\hat P^\gamma$. 
Here, $\hat P^\gamma_1$ ($\equiv\hat P^\gamma$) and $\hat P^\gamma_2$ are orthogonal 
polarization directions, $90^\circ$ apart for linear polarization, and opposite 
helicity states for circular polarization.
Then $\mathfrak{p}^\gamma_{\hat P^\gamma_1}$ ($\mathfrak{p}^\gamma_{\hat P^\gamma_2}$)
is a probability observing photons with its polarization vector pointing in the
$\hat P^\gamma_1$ ($\hat P^\gamma_2$) direction.
To clarify (\ref{eq:general-dcs}), consider the case that all beam, target, and recoil 
particles are unpolarized as an example.
In this case the probabilities of finding spin projection in each of two possible directions
are equal and hence 
$\mathfrak{p}^{T,R}_{\pm \hat P^{T,R}}=\mathfrak{p}^\gamma_{\hat P^\gamma_1,\hat P^\gamma_2}=1/2$,
which leads to $\vec P^{\gamma,T,R} = \vec 0$.
Then we have
\begin{eqnarray}
d\sigma^{{\rm B,T,R}}(\vec 0,\vec 0,\vec 0) &=&
\frac{1}{8}
\left[
 d\sigma^{{\rm B,T,R}}(\hat P^\gamma_1,+\hat P^T,+\hat P^R)
+d\sigma^{{\rm B,T,R}}(\hat P^\gamma_1,+\hat P^T,-\hat P^R)
\right.
\nonumber\\
&&
+d\sigma^{{\rm B,T,R}}(\hat P^\gamma_1,-\hat P^T,+\hat P^R)
+d\sigma^{{\rm B,T,R}}(\hat P^\gamma_1,-\hat P^T,-\hat P^R)
\nonumber\\
&&
+d\sigma^{{\rm B,T,R}}(\hat P^\gamma_2,+\hat P^T,+\hat P^R)
+d\sigma^{{\rm B,T,R}}(\hat P^\gamma_2,+\hat P^T,-\hat P^R)
\nonumber\\
&&
\left.
+d\sigma^{{\rm B,T,R}}(\hat P^\gamma_2,-\hat P^T,+\hat P^R)
+d\sigma^{{\rm B,T,R}}(\hat P^\gamma_2,-\hat P^T,-\hat P^R)
\right]
\nonumber\\
&=& \frac{1}{2}d\sigma_0,
\label{eq:unp-dcs2}
\end{eqnarray}
where $d\sigma_0$ is the unpolarized cross section defined in (\ref{eq:unp-dcs}).
The factor $(1/2)$ in the last equation appears because the polarization of 
the final recoil particles is also averaged in (\ref{eq:unp-dcs2}).

\section{\label{sec:gcs}General cross section}
While the formulae presented in the previous section can be used numerically to
calculate any observable of pseudoscalar meson photoproduction, 
it is more convenient to analyze the data using
an analytic expression for the general cross section of equation~(\ref{eq:general-dcs}).
In terms of the polarization vectors of figure~\ref{fig:coord}, 
and with signs verified numerically using (\ref{eq:general-dcs}) of section~\ref{sec:theory},
the most general form of the cross section can be written as,
\begin{eqnarray}
d\sigma^{{\rm B,T,R}}(\vec P^\gamma,\vec P^T,\vec P^R) & = & \frac{1}{2} 
\left\{
      d\sigma_{0}  \left[ 1 - \phol\tary\recy\cos(2\phi_\gamma) \right] \right. \nonumber \\
&&~~~ +  \hat{\Sigma} \left[-\phol\cos(2\phi_{\gamma}) + \tary\recy \right] \nonumber \\
&&~~~ +  \hat{T}      \left[ \tary - \phol\recy\cos(2\phi_\gamma) \right] \nonumber \\
&&~~~ +  \hat{P}      \left[ \recy - \phol\tary\cos(2\phi_\gamma) \right] \nonumber \\
&&~~~ +  \hat{E}      \left[-\phoc\tarz + \phol\tarx\recy\sin(2\phi_\gamma) \right] \nonumber \\
&&~~~ +  \hat{G}      \left[ \phol\tarz\sin(2\phi_{\gamma}) + \phoc\tarx\recy \right] \nonumber\\
&&~~~ +  \hat{F}      \left[ \phoc\tarx + \phol\tarz\recy\sin(2\phi_\gamma) \right] \nonumber \\ 
&&~~~ +  \hat{H}      \left[ \phol\tarx\sin(2\phi_{\gamma}) - \phoc\tarz\recy \right] \nonumber \\
&&~~~ +  \hat{C}_{x'} \left[ \phoc\recx -\phol\tary\recz\sin(2\phi_\gamma) \right] \nonumber \\
&&~~~ +  \hat{C}_{z'} \left[ \phoc\recz +\phol\tary\recx\sin(2\phi_\gamma) \right] \nonumber \\
&&~~~ +  \hat{O}_{x'} \left[ \phol\recx\sin(2\phi_{\gamma}) +\phoc\tary\recz \right] \nonumber \\
&&~~~ +  \hat{O}_{z'} \left[ \phol\recz\sin(2\phi_{\gamma}) -\phoc\tary\recx \right] \nonumber \\
&&~~~ +  \hat{L}_{x'} \left[ \tarz\recx +\phol\tarx\recz\cos(2\phi_\gamma) \right] \nonumber \\
&&~~~ +  \hat{L}_{z'} \left[ \tarz\recz -\phol\tarx\recx\cos(2\phi_\gamma) \right] \nonumber \\
&&~~~ +  \hat{T}_{x'} \left[ \tarx\recx -\phol\tarz\recz\cos(2\phi_\gamma) \right] \nonumber \\
&&~~~ +  \left. \hat{T}_{z'} \left[ \tarx\recz +\phol\tarz\recx\cos(2\phi_\gamma) \right]
\right\}.
\label{eq:gcs}
\end{eqnarray}
The derivation of this analytic expression is summarized
in \ref{apx_eqs}, where we follow the formalism of Fasano, Tabakin and Saghai 
(FTS)~\cite{fasano}, expanding their treatment to include the complete set of 
triple polarization cases. 
In this expression we have designated the product of an asymmetry and $d\sigma_0$
with a caret, so that $\hat A = A d\sigma_0$. 
These products are referred to as \textit{profile functions} in \cite{chiang,fasano}. 
One can of course pull a common factor of $d\sigma_0$ out in front of the above expression, 
in which case all the profile functions are replaced by their corresponding asymmetries. 
However, we keep the above form since it is the profile functions that are most simply determined 
by the CGLN amplitudes. 
(The definition of each of these profile functions in terms of measurable quantities
is given by \ref{apx_tab}.)
The second, third and fourth terms ($\hat\Sigma$, $\hat T$, $\hat P$) are commonly referred to 
as single-polarization observables, since their leading coefficients contain only a single 
polarization vector. 
The subsequent 12 terms are grouped into 3 sets, each containing four terms, referred to as \{BT, BR, TR\}
according to the combination of polarization vectors appearing in their leading coefficients.
Two of the leading terms have negative coefficients.
The first arises because we have taken
for the numerator of the beam asymmetry ($\Sigma$) the somewhat more common definition
of ($\sigma_{\perp} - \sigma_{\parallel}$), rather than its negative.
[Here $\perp (\parallel)$ corresponds to $\vec P^{\gamma}_{L} = \hat{y}$ ($\vec P^\gamma_L = \hat{x}$) in the
left panel of figure~\ref{fig:coord}.]
In the second leading term with a negative coefficient,
we have taken the numerator of the $E$ asymmetry as the difference of
cross sections with anti-parallel and parallel photon and target spin alignments
($\sigma_{\rm A}-\sigma_{\rm P}$).
This follows a convention first introduced by Worden~\cite{worden} and 
propagated through many (though not all) subsequent papers, and has been used in
recent experimental evaluations of the GDH sum rules~\cite{legs09}. 
The specific measurements needed to construct each of these observables are tabulated 
in \ref{apx_tab}.

Recoil observables are generally specified in the rotated coordinate system with 
$\hat{z}' = +\hat{k}$. 
Occasionally, a particular recoil observable will have a more transparent interpretation 
in the unprimed coordinate system of figure~\ref{fig:coord}~\cite{g1c07}. 
Since a baryon polarization transforms
as a standard three vector, the \textit{unprimed} and \textit{primed}
observables are  simply related:
\begin{equation}
\eqalign{
A_{x}  =  +A_{x'}\cos\theta_{K}+ A_{z'}\sin\theta_{K} \cr
A_{z}  =  -A_{x'}\sin\theta_{K} + A_{z'}\cos\theta_{K} ,
}\label{eq:br-1}
\end{equation}
and
\begin{equation}
\eqalign{
A_{x'}  =  +A_x\cos\theta_{K} - A_z\sin\theta_{K} \cr
A_{z'}  =  +A_x\sin\theta_{K} + A_z\cos\theta_{K} ,
}\label{eq:br-inv}
\end{equation}
where $A$ represents any one of the BR or TR observables.

It is convenient to arrange the observables entering the general cross section in tabular form, 
as in table~\ref{tab:pol-obs}.
The four rows correspond to different states of beam polarization, either ignoring the incident 
polarization entirely (labeled unpolarized in table~\ref{tab:pol-obs}), or in one of three standard 
Stokes vector components that characterize an ensemble of photons with polarization $P^\gamma$, linear 
at $\pm 45^\circ$ to the reaction plane (which enters the cross section with a $\sin(2\phi)$ dependence), 
linear either in or perpendicular to the reaction plane (which enters the cross section with a 
$\cos(2\phi)$ dependence), or circular.
The columns of the table give the polarization of the target, recoil, or target + recoil 
combination. 
One can readily construct from this table the terms that enter the general cross section for 
any given combination of polarization conditions. 
These consist of the terms involving all applicable polarization vectors, as well as those 
that survive when initial states are averaged and/or final states are summed. 
We consider two examples as an illustration. 
First, for a circularly polarized beam on an unpolarized target with an analysis of the three 
components of recoil polarization, the general cross section contains terms from the average of 
initial states (first row of table~\ref{tab:pol-obs}) and from the polarized initial state (forth row). 
Contributing terms come from only those columns that do not require knowing the target polarization state. 
Thus the cross section for this condition becomes 
$(1/2)[(d\sigma_0 + P^R_{y'} \hat P) + P^\gamma_c (P^R_{x'} \hat C_{x'} + P^R_{z'} \hat C_{z'})]$. 
Alternatively, with linear beam polarization in or perpendicular to the reaction plane, 
a longitudinally polarized target (along $\hat z$) and an analysis of recoil polarization along 
the meson (kaon) momentum ($\hat z'$), the general cross section is given by the terms in the 
first (\textit{unpolarized}) and third rows that are either independent of target and 
recoil polarization ($d\sigma_0$,$-\Sigma$) or in columns associated with polarization 
along $\hat z$ and/or $\hat z'$, namely 
$(1/2)[(d\sigma_0 + \tarz\recz\hat L_{z'}) + \phol\cos(2\phi_\gamma)(-\hat\Sigma -\tarz\recz\hat T_{x'})]$.

\begin{landscape}
\begin{table}
\caption{
Polarization observables in pseudoscalar meson photoproduction. 
Each observable appears twice in the table. 
The 16 entries in italics indicate the leading polarization dependence of each observable 
in the general cross section. 
The three underlined entries ($\hat P$, $\hat T$, $\hat \Sigma$) are nominal 
\textit{single-polarization} quantities that can be measured with double-polarization. 
Those in bold are the unpolarized cross section and 12 nominal double-polarization quantities 
that can be measured with triple-polarization. (See text.)
}
\label{tab:pol-obs}
\begin{indented}
\item[] \begin{tabular}{@{}lll l lll l lll l lllllllll}
\br
\multicolumn{2}{c}{Beam ($P^\gamma$)}&   &~~& \multicolumn{3}{c}{Target ($P^T$)} &~~&
\multicolumn{3}{c}{Recoil ($P^R$)} &~~&\multicolumn{9}{c}{Target ($P^T$) + Recoil ($P^R$)} \\
\multicolumn{3}{c}{}               &~~& \multicolumn{3}{c}{}               &~~& 
$x'$ & $y'$ & $z'$ &~~& 
$x'$ & $x'$ & $x'$ &
$y'$ & $y'$ & $y'$ & 
$z'$ & $z'$ & $z'$ \\
\multicolumn{2}{c}{}& &~~&
$x $ & $y $ & $z $ &~~&
\multicolumn{3}{c}{} &~~& 
$x $ & $y $ & $z $ & 
$x $ & $y $ & $z $ & 
$x $ & $y $ & $z $ \\
\mr
\multicolumn{2}{c}{unpolarized}& $d\sigma_0$&~~&
&$\hat T$&&~~&
&$\hat P$&&~~&
$\hat T_{x'}$&&$\hat L_{x'}$&
&\underline{$\hat \Sigma$}&&
$\hat T_{z'}$&&$\hat L_{z'}$\\ \\
\multicolumn{2}{c}{$\phol\sin(2\phi_\gamma)$}&&~~&
$\hat H$&&$\hat G$&~~&
$\hat O_{x'}$&&$\hat O_{z'}$&~~&
&$\mathbf{\hat{C}_{z'}}$&&
$\mathbf{\hat{E}}$&&$\mathbf{\hat{F}}$&
&$\mathbf{-\hat{C}_{x'}}$& \\ \\
\multicolumn{2}{c}{$\phol\cos(2\phi_\gamma)$}&$-\hat\Sigma$  &~~&
&\underline{$-\hat{P}$}&&~~&
&\underline{$-\hat{T}$}&&~~&
$\mathbf{-\hat{L}_{z'}}$&&$\mathbf{\hat{T}_{z'}}$&
&$\mathbf{-d\sigma_0}$&&
$\mathbf{\hat{L}_{x'}}$&&$\mathbf{-\hat{T}_{x'}}$ \\ \\
\multicolumn{2}{c}{circular $\phoc$}&&~~&
$\hat F$&&$-\hat E$&~~&
$\hat C_{x'}$&&$\hat C_{z'}$&~~&
&$\mathbf{-\hat{O}_{z'}}$&&
$\mathbf{\hat{G}}$&&$\mathbf{-\hat{H}}$&
&$\mathbf{\hat{O}_{x'}}$&\\
\br
\end{tabular}
\end{indented}
\end{table}
\end{landscape}

\section{\label{sec:review}Variations within the existing literature}

The form of the general cross section expression in equation~(\ref{eq:gcs})
has been derived analytically in~\ref{apx_eqs} and checked numerically with the
tools of section~\ref{sec:theory}.
At this point, it is instructive to summarize the
variations in similar formulae in the literature which 
have already caused some confusions in analyzing
recent data and must be resolved for future development.
The most frequently quoted works that discuss the relation between observables and CGLN
amplitudes are the following four: Barker-Donnachie-Storrow (BDS)~\cite{barker}, 
Adelseck-Saghai (AS)~\cite{adel}, Fasano-Tabakin-Saghai (FTS)~\cite{fasano}, and 
Kn\"{o}chlein-Drechsel-Tiator (KDT)~\cite{knoch}. 
A few of the differences between them are summarized in the
following subsections.

\subsection{\label{sec:bds}BDS}

The coordinate system of the BDS paper is the same as ours in figure~\ref{fig:coord} above. 
The photon beam momentum is along $+\hat z$; 
$\langle \hat z - \hat x\rangle$ is the reaction plane containing the meson momentum $\vec p_m$ 
emerging at a center of mass angle measured from $\hat z$ rotating towards $\hat x$; 
$(\vec p_\gamma \times \vec p_m )/ |\vec p_\gamma \times \vec p_m | = + \hat y$ 
and 
$[(\vec p_\gamma \times \vec p_m) \times \vec p_\gamma]/
|(\vec p_\gamma \times \vec p_m) \times \vec p_\gamma| = +\hat x$.
The recoil baryon polarization is specified in a rotated \textit{primed}-coordinate system, 
with $+\hat z'$ along the meson momentum, $\vec p_m$; 
$\hat y' = \hat y$ and $\hat x'$ lies in the $\langle \hat z - \hat x \rangle$ plane, rotated
down from $\hat x$ by $\theta_{{\rm c.m.}}$. 
It has since become common to indicate the use of this rotated system by including a \textit{prime} in 
the symbol of observables that involve recoil, e.g., $C_{z'}$, $O_{x'}$, etc., although the 
prime is not used in the BDS paper.
The BDS paper is certainly a seminal work on this subject but, in its published form,
it contains an unfortunate piece of typesetting 
that has lead to some confusion. Page 348 of that journal article ends with the sentence, 
``\textit{The precise relation between observables and the experiments we consider is as follows.}'' 
The next page 349 contains table I with several columns, the ``Usual symbol'' for the observables, 
their decomposition into ``Helicity'' and ``Transversity'' amplitudes, and in the fourth column 
the ``Experiment required'' to measure each observable. 
This forth column utilizes a notation that is somewhat condensed, but at least appears clear for linear
polarization at $45^\circ$ to the reaction plane. 
For example the \textit{experiment required} to determine the $H$
asymmetry is listed as $\{L(\pm 1/4);x;-\}$, which would imply the following ratio of cross sections
with polarized beam, target and recoil,
\begin{eqnarray*}
\fl
\qquad
\frac{
 d\sigma^{{\rm B,T,R}} (\phi_\gamma^L=+\pi/4,\vec P^T=+\hat x, {\rm sum~f.s.})
-d\sigma^{{\rm B,T,R}} (\phi_\gamma^L=-\pi/4,\vec P^T=+\hat x, {\rm sum~f.s.})
}{
 d\sigma^{{\rm B,T,R}} (\phi_\gamma^L=+\pi/4,\vec P^T=+\hat x, {\rm sum~f.s.})
+d\sigma^{{\rm B,T,R}} (\phi_\gamma^L=-\pi/4,\vec P^T=+\hat x, {\rm sum~f.s.})
},
\end{eqnarray*}
where unobserved final recoil polarization states are summed. 
However, equation (2) in BDS~\cite{barker} at the top of the following page 350 gives 
the $H$-dependence of the cross section as,
\begin{eqnarray}
d\sigma^{\rm{B,T,R}}&=&d\sigma_0\left\{ 1+P^T_x\left[-P^\gamma_L H\sin(2\phi_\gamma^L)\right] + \cdots\right\},
\end{eqnarray}
and using this to evaluate the above ratio results in $-H$. 
The sense of rotation for the angle $\phi_\gamma^L$ is not defined, but we assume it is measured 
from the $x$ axis rotating toward the $y$ axis. 
[The opposite sense would introduce another negative sign in terms proportional to $\sin(2\phi_\gamma^L)$.] 
We regard the equation for the cross section as the most definitive. 
Thus, one should take the ``Helicity'' and ``Transversity'' expansions of the observables in the second 
and third columns of table I in BDS~\cite{barker} literally, but the required experiment 
in column four 
as schematic only, leaving the sign of the specific combination of measurements to be determined from 
their equations (2)-(4). 
While rather convoluted, we believe this represents the correct reading of the BDS paper.
Finally, we note that equations (3) and (4) in BDS~\cite{barker}, which give their cross sections 
for polarized beam and recoil, and for polarized target and recoil, respectively, are both missing 
a factor of $1/2$. 
This is easily seen by averaging over initial states and summing over final states, which for 
the equations as written results in twice the unpolarized cross section, $2\sigma_0$.

\subsection{\label{sec:as}AS}

While the BDS coordinates were focused on the meson, the coordinate system of the AS paper is
focused on the final state baryon. 
The photon beam momentum is along $-\hat z$. 
In the $\langle \hat z - \hat x\rangle$ reaction plane the recoil baryon emerges at 
a center of mass angle measured from $\hat z$ rotating towards $\hat x$; 
$\hat y$ is still defined with the meson momentum as 
$(\vec p_\gamma\times\vec p_m)/|\vec p_\gamma\times\vec p_m| =+\hat y$, 
but now 
$[(\vec p_\gamma\times \vec p_m)\times(-\vec p_\gamma)]
/|(\vec p_\gamma\times \vec p_m)\times(-\vec p_\gamma)|=+\hat x$.
The sense of rotation for the linear photon polarization angle $\phi_\gamma^L$ is defined from 
the $\hat x$ axis rotating toward $\hat y$. 
Relative to $(\vec p_\gamma \times\vec p_m)\times\vec p_\gamma$, 
a linear polarization orientation of $+\pi/4$ in AS coordinates
corresponds to $-\pi/4$ in BDS and the present work. 
The \textit{primed}-coordinate system is taken with $+\hat z'$ along the baryon momentum;
$\hat y'=\hat y$ and $\hat x'$ lies in the $\langle \hat z - \hat x\rangle$ plane, 
rotated up from $\hat x$ by $\theta_{\rm c. m.}$.
The observables involving components of the recoil polarization refer to the \textit{primed} coordinates,
although \textit{primes} are not included in their notation.
The AS paper includes a general expression for the cross section in terms of beam, target and
recoil polarizations. 
However, as discussed in section~\ref{sec:recoil} below, their expression has at least one
misprint in its last line, with two terms involving $P_z^R$ and $O_z$ but none with $P_z^R$ 
and $O_x$. 
As evident in our equation~(\ref{eq:gcs}), each of these observables appears in the general
cross section with two coefficients, one dependent upon $P_x^R$ and the other upon $P_z^R$.

\subsection{\label{sec:fts}FTS}

The coordinate system of FTS is the same as that of BDS and of the present work. 
Circular polarization states are designated as $r$ and $l$. 
Although the Stokes vector for the photon beam is taken from optics 
(which associates $r$ circular polarization with helicity $-1$), their table~I associates
$r$ beam polarization with helicity $+1$. 
The sense of rotation for the linear photon polarization angle $\phi_\gamma^L$ is defined from 
the $\hat x$ axis rotating toward $\hat y$. 
The observables involving components of the recoil polarization are designated with primed symbols. 
FTS does not give an explicit expression for the cross section in terms of observables and polarizations, 
but the paper does list explicit definitions of observables in terms of measurable quantities. 
The FTS paper also gives explicit equations relating the observables to the CGLN amplitudes.

\subsection{\label{sec:kdt}KDT}

The coordinate system of KDT is the same as that of BDS and of the present work. 
Circular photon polarization ($P_\odot$) is referred to as right handed;
although helicity is not discussed, we have assumed (in Table~\ref{tab:comparison} below)
that their right-handed state corresponds to $h=+1$.
The direction  of rotation for the angle $\phi_\gamma^L$ is not defined;
in the evaluation below we have assumed their azimuthal polarization angle rotates
from $\hat x$ toward $\hat y$.
Cross section equations are given for the cases of \textit{beam}+\textit{target} polarization, 
\textit{beam}+\textit{recoil} polarization and \textit{target}+\textit{recoil} polarization. 
As in BDS, the latter two are missing a factor of $1/2$, as is easily verified 
by averaging over initial states and summing over final states. 
KDT provides explicit equations to relate each observable to the CGLN amplitudes.
\\
\\

To completely define an observable in terms of measurable quantities one needs a specification
of the coordinate system and either the equation for the cross section in terms of polarizations and
observables, or an explicit definition of the observables in terms of measurable cross sections. 
As an example of some of the variations that have resulted from different conventions, consider the
beam+target asymmetries. We can define these as coordinate-independent ratios with directions
specified by only photon ($\vec p_\gamma$) and meson ($\vec p_m$) momenta.
\begin{eqnarray}
R_E&=&
\left[
 d\sigma^{{\rm B,T,R}}_1 (P^\gamma_h=+1,\vec P^T=-\hat p_\gamma, {\rm sum~f.s.})
\right.
\nonumber\\
&&
\left.
-d\sigma^{{\rm B,T,R}}_2 (P^\gamma_h=+1,\vec P^T=+\hat p_\gamma, {\rm sum~f.s.})
\right]
\nonumber\\
&&
/[d\sigma_1 + d\sigma_2],
\label{eq:re}
\end{eqnarray}
\begin{eqnarray}
R_F&=&
\left[
 d\sigma^{{\rm B,T,R}}_1 (P^\gamma_h=+1,\vec P^T=\hat p_1, {\rm sum~f.s.})
\right.
\nonumber\\
&&
\left.
-d\sigma^{{\rm B,T,R}}_2 (P^\gamma_h=-1,\vec P^T=\hat p_1, {\rm sum~f.s.})
\right]
\nonumber\\
&&
/[d\sigma_1 + d\sigma_2],
\label{eq:rf}
\end{eqnarray}
\begin{eqnarray}
R_G&=&
\left[
 d\sigma^{{\rm B,T,R}}_1 (\phi_\gamma^L =+\pi/4~{\rm from}~\hat p_1~{\rm toward}~\hat p_2,
\vec P^T=+\hat p_\gamma, {\rm sum~f.s.})
\right.
\nonumber\\
&&
\left.
-d\sigma^{{\rm B,T,R}}_2 (\phi_\gamma^L =+\pi/4~{\rm from}~\hat p_1~{\rm toward}~\hat p_2,
\vec P^T=-\hat p_\gamma, {\rm sum~f.s.})
\right]
\nonumber\\
&&
/[d\sigma_1 + d\sigma_2],
\label{eq:rg}
\end{eqnarray}
\begin{eqnarray}
R_H&=&
\left[
 d\sigma^{{\rm B,T,R}}_1 (\phi_\gamma^L =+\pi/4~{\rm from}~\hat p_1~{\rm toward}~\hat p_2,
\vec P^T=+\hat p_1, {\rm sum~f.s.})
\right.
\nonumber\\
&&
\left.
-d\sigma^{{\rm B,T,R}}_2 (\phi_\gamma^L =+\pi/4~{\rm from}~\hat p_1~{\rm toward}~\hat p_2,
\vec P^T=-\hat p_1, {\rm sum~f.s.})
\right]
\nonumber\\
&&
/[d\sigma_1 + d\sigma_2],
\label{eq:rh}
\end{eqnarray}
with 
\begin{equation*}
\hat p_1 \equiv
\frac{(\vec p_\gamma\times \vec p_m)\times\vec p_\gamma}{|(\vec p_\gamma\times\vec p_m)\times\vec p_\gamma|},
\qquad
\hat p_2 \equiv
\frac{\vec p_\gamma\times\vec p_m}{|\vec p_\gamma\times\vec p_m|}.
\end{equation*}

The variable \textit{names} of these ratios as used by different authors are listed in table I. 
As evident there, the same symbol has been used in different papers to refer to different quantities, 
with common magnitudes but varying signs. 
This creates the potential for spiraling confusion when a third party combines equations 
from different papers. 

\begin{table}[tb]
\caption{Ratios of cross sections involving beam and target polarizations and 
the names given these quantities by different authors.}
\label{tab:comparison}
\begin{indented} 
\item[] \begin{tabular}{@{}lrrrrr}
\br
    &BDS~\cite{barker}&AS~\cite{adel}&FTS~\cite{fasano}&KDT~\cite{knoch}&Present work\\
\mr
$\vec P_\gamma$ & $+\hat z$ & $-\hat z$ & $+\hat z$ & $+\hat z$ & $+\hat z$ \\
\mr
$R_E$ & $ E$ & $ E$ & $-E$ & $ E$ & $ E$ \\
$R_F$ & $ F$ & $-F$ & $ F$ & $ F$ & $ F$ \\
$R_G$ & $ G$ & $ G$ & $ G$ & $ G$ & $ G$ \\
$R_H$ & $-H$ & $ H$ & $ H$ & $-H$ & $ H$ \\
\br
\end{tabular}
\end{indented}
\end{table}

The present work has avoided the confusions associated with variations in 
formulae from different papers by developing a consistent and self-contained set of expressions that 
(a) define each observable in terms of measurable cross sections (\ref{apx_tab}), 
(b) provide the most general expression for the cross
section in terms of the 16 observables and the beam, target and recoil polarization states, 
both derived analytically (\ref{apx_eqs}) and checked numerically 
(sections~\ref{sec:theory} and~\ref{sec:gcs}),
and (c) provide the defining relations between the 16 spin observables and the CGLN amplitudes
(section~\ref{sec:cgln} below).

\section{\label{sec:recoil} Recoil polarization}

As a first application of the consistent expression of the general cross section presented 
in section \ref{sec:gcs}, we analyze the potential of experiments measuring 
recoil polarization.
The general expression in (\ref{eq:gcs}) displays a level of symmetry in the three polarization vectors, 
$\vec P^\gamma$, $\vec P^T$ and $\vec P^R$. 
However, while the first two are parameters that are under experimental control, 
the recoil polarization is not. 
Rather, $\vec P^R$ is a consequence of the angular momentum brought into the entrance channel through
$\vec P^\gamma$ and $\vec P^T$, and the reaction physics. 
The relations determining $\vec P^R$ are readily derived. 
We start by regrouping terms in the general cross section expression to display the explicit 
dependence on $\vec P^R$ and recast (\ref{eq:gcs}) as,
\begin{equation}
d\sigma^{{\rm B,T,R}}(\vec P^\gamma,\vec P^T,\vec P^R) 
= \frac{1}{2} \left[ A^0 + (P^R_{x'})A^{x'} +(P^R_{y'})A^{y'} +(P^R_{z'})A^{z'} \right],
\label{eq:gcs-2}
\end{equation}
where
\begin{eqnarray*}
A^0 &=& 
  d\sigma_0   - \phol\cos(2\phi_\gamma)\hat\Sigma + \tary\hat T 
\nonumber\\
&&
- \phol\tary\cos(2\phi_\gamma)\hat P 
- \phoc\tarz\hat E
+ \phol\tarz\sin(2\phi_\gamma)\hat G 
\nonumber\\
&&
+ \phoc\tarx\hat F 
+ \phol\tarx\sin(2\phi_\gamma)\hat H,
\end{eqnarray*}
\begin{eqnarray*}
A^{x'} &=& 
  \phoc\hat C_{x'} + \phol\sin(2\phi_\gamma)\hat O_{x'} 
+ \tarz\hat L_{x'} + \tarx\hat T_{x'}
\nonumber\\
&&
+ \phol\tary\sin(2\phi_\gamma)\hat C_{z'} - \phoc\tary\hat O_{z'} 
\nonumber\\
&&
- \phol\tarx\cos(2\phi_\gamma)\hat L_{z'} + \phol\tarz\cos(2\phi_\gamma)\hat T_{z'},
\end{eqnarray*}
\begin{eqnarray*}
A^{y'} &=& 
  \hat P + \tary\hat \Sigma
- \phol\cos(2\phi_\gamma)\hat T 
\nonumber\\
&&
- \phol\tary\cos(2\phi_\gamma)d\sigma_0
+ \phol\tarx\sin(2\phi_\gamma)\hat E + \phoc\tarx\hat G 
\nonumber\\
&&
+ \phol\tarz\sin(2\phi_\gamma)\hat F - \phoc\tarz\hat H,
\end{eqnarray*}
\begin{eqnarray*}
A^{z'} &=& 
  \phoc\hat C_{z'} + \phol\sin(2\phi_\gamma)\hat O_{z'} 
+ \tarz\hat L_{z'} + \tarx\hat T_{z'}
\nonumber\\
&&
- \phol\tary\sin(2\phi_\gamma)\hat C_{x'} + \phoc\tary\hat O_{x'} 
\nonumber\\
&&
+ \phol\tarx\cos(2\phi_\gamma)\hat L_{x'} - \phol\tarz\cos(2\phi_\gamma)\hat T_{x'}.
\end{eqnarray*}
The recoil polarization $\vec P^R$ can be resolved as the vector sum of three component vectors,
$P^R_{x'}\hat x'$, $P^R_{y'}\hat y'$, $P^R_{z'}\hat z'$. 
Considering first $P^R_{x'}\hat x'$, this is the degree of polarization along 
$\hat x'$ and is given by
\begin{equation}
P^R_{x'} = \mathfrak{p}^R_{x',+} - \mathfrak{p}^R_{x',-} ,
\end{equation}
where $\mathfrak{p}^R_{x',\pm}$ is the probability for observing the recoil with spin along 
$\pm \hat x'\equiv(\pm 1,0,0)'$. 
Using (\ref{eq:gcs-2}), we evaluate this as the ratio of cross sections, 
\begin{equation}
P^R_{x'} = 
\frac{
 d\sigma^{{\rm B,T,R}}(\vec P^\gamma,\vec P^T,+1\hat x')
-d\sigma^{{\rm B,T,R}}(\vec P^\gamma,\vec P^T,-1\hat x')}
{d\sigma^{{\rm B,T,R}}(\vec P^\gamma,\vec P^T,+1\hat x')
+d\sigma^{{\rm B,T,R}}(\vec P^\gamma,\vec P^T,-1\hat x')}
=\frac{A^{x'}}{A^0}.
\end{equation}
The $\hat y'$ and $\hat z'$ recoil components are evaluated in a similar manner. 
Thus, the components of the recoil polarization are determined from (\ref{eq:gcs-2}), 
in terms of combinations of the profile functions and initial polarizations, as
\begin{equation}
P^R_{x'} =\frac{A^{x'}}{A^0},\qquad
P^R_{y'} =\frac{A^{y'}}{A^0},\qquad
P^R_{z'} =\frac{A^{z'}}{A^0}.
\label{eq:recoil}
\end{equation}
These recoil components determine the orientation of the recoil vector, $\vec P^R$, 
and its magnitude,
\begin{equation}
|\vec P^R| =\frac{1}{A^0}\sqrt{(A^{x'})^2+(A^{y'})^2+(A^{z'})^2}.
\end{equation}
It is worth clarifying the relationship between (\ref{eq:gcs}) or~(\ref{eq:gcs-2}) and 
(\ref{eq:recoil}). Equations~(\ref{eq:gcs}) and~(\ref{eq:gcs-2}) display
the general dependence of the cross section upon the three polarization vectors, 
each of which is in a superposition of two spin states. 
If any one polarization is not observed, either by not experimentally preparing it 
($\vec P^\gamma$ or $\vec P^T$) or by not detecting it ($\vec P^R$), then the terms proportional 
to that polarization average or sum to zero and drop out of the cross section. 
The action of preparing or detecting a polarization forces the corresponding magnetic 
substate population into a particular distribution, which in the case of the recoil 
polarization is given by (\ref{eq:recoil}). 
A particular consequence of this is that one may not substitute 
(\ref{eq:recoil}) back into (\ref{eq:gcs-2}) to obtain a cross section 
that appears to be independent of recoil polarization.

An expression similar in spirit to (\ref{eq:gcs-2}) but different in form is given 
by Adelseck and Saghai in \cite{adel}. 
However, the coordinate system is very different and
there is at least one obvious misprint, with two terms involving 
$P^R_z$ and $O_z$ but none with $P^R_z$ and $O_x$, as discussed in section~\ref{sec:as}.

In practice, the recoil polarization is measured either following a secondary scattering 
or, in the case of hyperon channels, through the angular distribution of their weak decays. 
$K\Lambda\to K\pi^- p$ production provides a particularly efficient channel for recoil 
measurements. 
In the rest frame of the decaying $\Lambda$, the angular distribution of the decay proton 
follows $(1/2)[1+\alpha|\vec P^\Lambda|\cos(\Theta_p)]$, where $\Theta_p$ is the angle between
the proton momentum and the lambda polarization direction~\cite{ly57}. 
Since the analyzing power in this decay is quite high, $\alpha = 0.642 \pm 0.013$~\cite{pdg10}, 
recoil measurements in modern quasi-$4\pi$ detectors can be carried out without 
significant penalty in statistics. 
As a result, such measurements provide information on combinations of observables through 
(\ref{eq:recoil}). 
It is instructive to consider a few examples.
\begin{enumerate}
\item 
Unpolarized beam and target, $P^\gamma_{L,c}= P^T=0$:
In this case, $A^0=d\sigma_0$, $A^{x'}=0$, $A^{y'}=\hat P$ and $A^{z'}=0$, so that
\begin{equation}
\vec P^R = (0, P=\hat P/d\sigma_0, 0).
\label{eq:rec-ex1}
\end{equation}
Thus, even when the initial state is completely unpolarized, 
a measured recoil polarization will be perpendicular to the reaction plane.
\item
Unpolarized beam and longitudinally polarized target, $P^\gamma_{L,c} = 0$ and 
$\vec P^T=(0,0,\tarz)$:
In this case, $A^0=d\sigma_0$,
$A^{x'}=\tarz\hat L_{x'}$,
$A^{y'}=\hat P$, and
$A^{z'}=\tarz\hat L_{z'}$, so that
\begin{equation}
\vec P^R = (P^T_z L_{x'},P,P^T_z L_{z'}).
\label{eq:rec-ex2}
\end{equation}
Thus a measurement of the components of the recoil polarization determine the $L_{x'}$, 
$P$ and $L_{z'}$ asymmetries.
\item
Circularly polarized beam ($P^\gamma_c$) and unpolarized target ($P^T =0$):
In this case, $A^0=d\sigma_0$,
$A^{x'}=\phoc\hat C_{x'}$,
$A^{y'}=\hat P$, and
$A^{z'}=\phoc\hat C_{z'}$, so that
\begin{equation}
\vec P^R = (\phoc C_{x'}, P,\phoc C_{z'}).
\label{eq:rec-ex3}
\end{equation}
This is the form assumed in the analysis of the CLAS-g1c data in \cite{g1c07}.
\item
Linearly polarized beam ($\phol$) and unpolarized target ($P^T=0$):
In this case, $A^0=d\sigma_0-\phol\cos(2\phi_\gamma)\hat\Sigma$,
$A^{x'}=\phol\sin(2\phi_\gamma)\hat O_{x'}$,
$A^{y'}=\hat P -\phol\cos(2\phi_\gamma)\hat T$, and
$A^{z'}= \phol\sin(2\phi_\gamma)\hat O_{z'}$, so that
\begin{eqnarray}
\vec P^R&=&
\left(
\frac{\phol\sin(2\phi_\gamma)O_{x'}}{1-\phol\cos(2\phi_\gamma)\Sigma},
\frac{P -\phol\cos(2\phi_\gamma) T}{1-\phol\cos(2\phi_\gamma)\Sigma},
\frac{\phol\sin(2\phi_\gamma)O_{z'}}{1-\phol\cos(2\phi_\gamma)\Sigma}
\right),
\nonumber\\
\label{eq:rec-ex4}
\end{eqnarray}
which is the form assumed in the analysis of the GRAAL data in \cite{gr09},
although the coordinate system is different.
\item
Circularly polarized beam ($P^\gamma_c$) and longitudinally polarized target 
[$\vec P^T=(0,0,\tarz)$]:
In this case, $A^0=d\sigma_0-\phoc\tarz\hat E$,
$A^{x'}=\phoc\hat C_{x'}+\tarz\hat L_{x'}$,
$A^{y'}=\hat P -\phoc\tarz\hat H$, and
$A^{z'}= \phoc\hat C_{z'} +\tarz\hat L_{z'} $, so that
\begin{equation}
\vec P^R=
\left(
\frac{\phoc C_{x'}+\tarz L_{x'}}{1-\phoc\tarz E},
\frac{P -\phoc\tarz H}{1-\phoc\tarz E},
\frac{\phoc C_{z'} +\tarz L_{z'}}{1-\phoc\tarz E}
\right).
\label{eq:rec-ex5}
\end{equation}
Here, measurements with complete knowledge of all spins involved provide the greatest flexibility.
An initial beam-target analysis summing over final states (i.e., ignoring the recoil) 
results in the cross section $A^0$, which determines the $E$ asymmetry and hence 
the denominator in (\ref{eq:rec-ex5}). 
In an analysis averaging over initial target polarizations $\pm P^T_z$, 
measurements of the recoil polarization vector then determine the $C_{x'}$, $P$ and $C_{z'}$
asymmetries. 
Another pass through the data, averaging instead over initial beam polarization states,
$\pm P^\gamma_c$, and with an analysis of the $P^R_{x'}$ and $P^R_{z'}$ recoil components, 
gives the $L_{x'}$ and $L_{z'}$ asymmetries.
Finally, by keeping track of both beam and target polarization states, a measurement 
of the $P^R_{y'}$ recoil component gives the $H$ asymmetry. 
Although the uncertainty in this determination of $H$ will include the
propagation of errors from $P$ and $E$, this is expected to be held to a reasonable level 
in the modern set of experiments that are now under way. 
The significance of this determination is that it does not require the use of a transversely 
polarized target, as would otherwise be required by the leading polarization
dependence of $H$ in (\ref{eq:gcs}). 
In general, the latter would require a completely separate experiment with
different systematics.
\item
Linearly polarized beam ($P^\gamma_L$) and longitudinally polarized target 
[$\vec P^T=(0,0,P^T_z)$]:
In this case, $A^0=d\sigma_0-\phol\cos(2\phi_\gamma)\hat\Sigma+\phol\tarz\sin(2\phi_\gamma)\hat G$,
$A^{x'}= \phol\sin(2\phi_\gamma)\hat O_{x'}+\tarz\hat L_{x'}+\phol\tarz\cos(2\phi_\gamma)\hat T_{z'}$,
$A^{y'}= \hat P -\phol\cos(2\phi_\gamma)\hat T +\phol\tarz\sin(2\phi_\gamma)\hat F$,
and
$A^{z'}=\phol\sin(2\phi_\gamma)\hat O_{z'}+\tarz\hat L_{z'}-\phol\tarz\cos(2\phi_\gamma)\hat T_{x'}$, 
so that
\begin{eqnarray}
\vec P^R &=&
\left(
\frac{\phol\sin(2\phi_\gamma)O_{x'}+\tarz L_{x'}+\phol\tarz\cos(2\phi_\gamma)T_{z'}}
{1-\phol\cos(2\phi_\gamma)\Sigma+\phol\tarz\sin(2\phi_\gamma)G},
\right.
\nonumber\\
&&
~~\,
\frac{P -\phol\cos(2\phi_\gamma)T+\phol\tarz\sin(2\phi_\gamma)F}
{1-\phol\cos(2\phi_\gamma)\Sigma+\phol\tarz\sin(2\phi_\gamma)G},
\nonumber\\
&&
\left.
~~\,
\frac{\phol\sin(2\phi_\gamma)O_{z'}+\tarz L_{z'}-\phol\tarz\cos(2\phi_\gamma)T_{x'}}
{1-\phol\cos(2\phi_\gamma)\Sigma+\phol\tarz\sin(2\phi_\gamma)G}
\right). 
\label{eq:rec-ex6}
\end{eqnarray}
With such data a beam-target analysis summing over final states (i.e., ignoring the recoil) 
determines the cross section $A^0$, and hence the $\Sigma$ and $G$ asymmetries from 
a Fourier analysis of the $\phi_\gamma$ dependence.
This fixes the denominators in (\ref{eq:rec-ex6}). 
With another analysis pass, averaging over initial target polarizations, 
measurements of the recoil polarization vector provide a determination of 
the $O_{x'}$, $P$ and $T$, and $O_{z'}$ asymmetries. 
Another pass through the same data, integrating over $\phi_\gamma$, gives 
the $L_{x'}$, $P$ and $L_{z'}$ asymmetries from measurements of the recoil 
polarization vector. 
Finally, a Fourier analysis of beam polarization states, using the difference 
between opposing target orientations, $P^T_z- P^T_{-z}$, together with 
a measurement of recoil polarization allows the separation of $L_{x'}$ and 
$T_{z'}$, $F$ (which would otherwise require a transversely polarized target), 
and $L_{z'}$ and $T_{x'}$.
\end{enumerate}

Thus, by judicious use of recoil polarization and a polarized beam, all 16 observables 
can be determined with a longitudinally polarized target 
(often in several ways) and in doing so, with largely common systematics.

A corresponding set of expressions can be developed for a transversely polarized target, 
although they are inherently more complicated since, for fixed target polarization 
perpendicular to $+\hat z$, any reaction plane will generally involve both transverse target 
components $P^T_x$ and $P^T_y$.

\begin{enumerate}
\setcounter{enumi}{6}
\item
Unpolarized beam ($P^\gamma_{L,c} = 0$) with a transversely polarized target and 
[$\vec P^T=(P^T_x, P^T_y,0)$]:
In this case, $A^0=d\sigma_0+\tary\hat T$,
$A^{x'} = \tarx\hat T_{x'}$,
$A^{y'} = \hat P + \tary\hat \Sigma$, and
$A^{z'} = \tarx\hat T_{z'}$,
so that
\begin{equation}
\vec P^R = 
\left( 
\frac{\tarx T_{x'}}{1+\tary T},
\frac{P + \tary \Sigma}{1+\tary T}, 
\frac{\tarx T_{z'}}{1+\tary T}
\right).
\label{eq:rec-ex7}
\end{equation}
Here an analysis summing over final states (i.e., ignoring the recoil) results in the cross 
section $A^0$, and a fit varying $P^T_y$ as the reaction plane tilts relative to 
the direction of the target polarization determines the $T$ asymmetry. 
A subsequent analysis of the recoil polarization components then
determines $T_{x'}$, $P$, $\Sigma$, and $T_{z'}$.
\item
Circularly polarized beam ($P^\gamma_c$) and transverse target polarization 
[$\vec P^T=(P^T_x,P^T_y,0)$]:
In this case, $A^0=d\sigma_0+\tary\hat T+\phoc\tarx \hat F$,
$A^{x'} = \phoc\hat C_{x'} + \tarx\hat T_{x'} -\phoc\tary\hat O_{z'}$,
$A^{y'} = \hat P + \tary\hat \Sigma +\phoc\tarx \hat G$,
and
$A^{z'} = \phoc\hat C_{z'} + \tarx\hat T_{z'} +\phoc\tary\hat O_{x'}$, 
so that
\begin{eqnarray}
\vec P^R &=& 
\left( 
\frac{\phoc C_{x'} + \tarx T_{x'} -\phoc\tary O_{z'}}{1+\tary T+\phoc\tarx F}, 
\frac{ P + \tary \Sigma +\phoc\tarx G}{1+\tary T+\phoc\tarx F}, 
\right.
\nonumber\\
&&
\left.
\frac{\phoc C_{z'} + \tarx T_{z'} +\phoc\tary O_{x'}}{1+\tary T+\phoc\tarx F}
\right).
\label{eq:rec-ex8}
\end{eqnarray}
In this case, a beam-target analysis summing over final states (i.e., ignoring the recoil) 
results in the cross section $A^0$ containing the terms in the $T$ and $F$ asymmetries, 
and these can be separated by first averaging over initial photon states, which removes $F$. 
A subsequent analysis, reconstructing the recoil polarization while averaging over initial 
circular photon states allows one to deduce $T_{x'}$ and $T_{z'}$ from
$P^R_{x'}$ and $P^R_{z'}$. 
Alternatively, with fixed beam polarization and recoil analysis, a fit varying
$P^T_x$ and $P^T_y$ as the reaction plane tilts in azimuth relative to the direction 
of the transversely polarized target determines all of the asymmetries in the numerators 
of (\ref{eq:rec-ex8}).
\end{enumerate}

We leave it to the reader to write out the final combination of linearly polarized beam and 
transverse target polarization. 
There the recoil polarization components involve ratios of 4 to 5 terms each. 
It remains to be seen if sequential analyses of such data are of practical use, 
given limitations on statistics.

\section{\label{sec:cgln}Relating observables to CGLN amplitudes}

To extract nucleon resonances, one needs to extract amplitudes from observables.
Because of the apparent variations in the  available literature, as summarized
in section \ref{sec:review}, there exists sign differences in formula relating
observables to CGLN amplitudes. With the formula presented in sections
\ref{sec:theory} and \ref{sec:gcs},
we are now in a position to clarify this issue. This is done by using 
any set of multipole amplitudes
to calculate the four CGLN amplitudes from (\ref{eq:mp-f1})-(\ref{eq:mp-f4})
and then with these, evaluate (a) the polarization observables by using the formulae
described in section~\ref{sec:theory} and the
spin orientations specified in the tables of \ref{apx_tab}, and
(b) the same observables calculated from the analytic expressions, as 
found in \cite{donn66,donn72,fasano,drech92,knoch}.
As expected, the absolute magnitudes from the two methods
are the same, but some of their signs are different.
In doing so, we are able to fix the signs of the
analytic expressions for the experimental conditions
specified in figure~\ref{fig:coord} and \ref{apx_tab}. Our results are:

\numparts
\begin{eqnarray}
\fl d\sigma_{0} = + \Re e \left\{ \fpf{1}{1} + \fpf{2}{2} + \sin^{2}\theta(\fpf{3}{3} + \fpf{4}{4})/2 
\right. \nonumber\\
\left. + \sin^{2}\theta(\fpf{2}{3} + \fpf{1}{4} + \cos\theta\fpf{3}{4}) - 2\cos\theta\fpf{1}{2} \right\} \rho_0 ,
\label{eq:obs-cgln-begin} \\
\fl \hat{\Sigma}  =  -\sin^{2}\theta\Re e\left\{ \left(\fpf{3}{3} +\fpf{4}{4}\right)/2 
                   + \fpf{2}{3} + \fpf{1}{4} + \cos\theta\fpf{3}{4}\right\}\rho_0 ,\\
\fl \hat{T}       =  + \sin\theta\Im m\left\{\fpf{1}{3} - \fpf{2}{4} + \cos\theta(\fpf{1}{4} - \fpf{2}{3})
                   - \sin^{2}\theta\fpf{3}{4}\right\}\rho_0 , \\
\fl \hat{P}       =  -\sin\theta\Im m\left\{ 2\fpf{1}{2} + \fpf{1}{3} - \fpf{2}{4}
                  - \cos\theta(\fpf{2}{3} -\fpf{1}{4}) - \sin^{2}\theta\fpf{3}{4}\right\}\rho_0 ,\\
\fl \hat{E}       =  + \Re e\left\{ \fpf{1}{1} + \fpf{2}{2} - 2\cos\theta\fpf{1}{2}
                   + \sin^{2}\theta(\fpf{2}{3} + \fpf{1}{4}) \right\}\rho_0 ,\\
\fl \hat{G}       =  + \sin^{2}\theta\Im m\left\{\fpf{2}{3} + \fpf{1}{4}\right\}\rho_0 ,\\
\fl \hat{F}       =  + \sin\theta\Re e\left\{\fpf{1}{3} - \fpf{2}{4} - \cos\theta(\fpf{2}{3} - \fpf{1}{4})\right\}\rho_0 ,\\
\fl \hat{H}       =  - \sin\theta\Im m\left\{2\fpf{1}{2} + \fpf{1}{3} - \fpf{2}{4}
                   + \cos\theta(\fpf{1}{4} - \fpf{2}{3})\right\}\rho_0 ,\\
\fl \hat{C}_{x'}  =  - \sin\theta\Re e\left\{\fpf{1}{1} - \fpf{2}{2} - \fpf{2}{3} + \fpf{1}{4} 
                   - \cos\theta(\fpf{2}{4} - \fpf{1}{3})\right\}\rho_0 , \\
\fl \hat{C}_{z'}  =  -\Re e\left\{2\fpf{1}{2} - \cos\theta(\fpf{1}{1} + \fpf{2}{2})
                   + \sin^{2}\theta(\fpf{1}{3} + \fpf{2}{4})\right\}\rho_0 ,\\
\fl \hat{O}_{x'}  =  -\sin\theta\Im m\left\{\fpf{2}{3} - \fpf{1}{4} + \cos\theta(\fpf{2}{4} - \fpf{1}{3})\right\}\rho_0 ,\\
\fl \hat{O}_{z'}  =  + \sin^{2}\theta\Im m\left\{\fpf{1}{3} + \fpf{2}{4}\right\}\rho_0 ,\\
\fl \hat{L}_{x'}  =  + \sin\theta\Re e\left\{\fpf{1}{1} - \fpf{2}{2} - \fpf{2}{3} + \fpf{1}{4}
                   + \sin^{2}\theta(\fpf{4}{4} - \fpf{3}{3})/2 \right. \nonumber \\
                 \mbox{} \left. + \cos\theta(\fpf{1}{3} - \fpf{2}{4})\right\}\rho_0 ,\\
\fl \hat{L}_{z'}  =  + \Re e\left\{2\fpf{1}{2} - \cos\theta(\fpf{1}{1} + \fpf{2}{2})
                   + \sin^{2}\theta(\fpf{1}{3} + \fpf{2}{4} + \fpf{3}{4}) \right. \nonumber \\
                 \mbox{} \left. + \cos\theta\sin^{2}\theta(\fpf{3}{3} + \fpf{4}{4})/2 \right\}\rho_0 ,\\
\fl \hat{T}_{x'}  =  -\sin^{2}\theta\Re e\left\{\fpf{1}{3} + \fpf{2}{4} + \fpf{3}{4}
                   + \cos\theta(\fpf{3}{3} + \fpf{4}{4})/2 \right\}\rho_0 ,\\
\fl \hat{T}_{z'}  =  + \sin\theta \Re e \left\{\fpf{1}{4} - \fpf{2}{3}
                   + \cos\theta(\fpf{1}{3} - \fpf{2}{4})
                 + \sin^{2}\theta(\fpf{4}{4} - \fpf{3}{3})/2 \right\}\rho_0 .
\label{eq:obs-cgln-end}
\end{eqnarray}
\endnumparts

A comparable set of expressions are given by Fasano, Tabakin and Saghai 
(FTS) in \cite{fasano}. 
With the conventions discussed in section~\ref{sec:fts},
and allowing for their different
definition of the $E$ beam-target asymmetry (as in table~\ref{tab:comparison}), the above expressions 
are consistent with those of \cite{fasano}.

Comparing the above relations to those given by Kn\"{o}chlein, Drechsel 
and Tiator (KDT) (Appendix B and C of \cite{knoch}), 
six of these equations have different signs,
the BT observable $H$, the TR observable
$L_{x'}$ and all four of the BR observables 
$C_{x'}$, $C_{z'}$, $O_{x'}$ and $O_{z'}$. 
The KDT paper~\cite{knoch} is listed in the MAID on-line meson production 
analysis~\cite{MAID,drech07,mart} as the defining reference 
for their connection between CGLN amplitudes and polarization observables. 
To check if these differences persist in the MAID code we have downloaded 
MAID multipoles, used the relations in (\ref{eq:mp-f1})-(\ref{eq:mp-f4})
to construct from these the four CGLN $F_{i}$ amplitudes, 
and then used our equations~(\ref{eq:obs-cgln-begin})-(\ref{eq:obs-cgln-end})
above to construct observables. 
Comparing the results to direct predictions of
observables from the MAID code, we find the same six sign differences.
However, KDT give a form of the general
cross section with leading polarization terms in \cite{knoch}
and there, the equations for these six observables appear with a negative coefficient, as opposed to
our form of the cross section in (\ref{eq:gcs}).
This is equivalent to interchanging the $\sigma_1$ and $\sigma_2$ measurements
of \ref{apx_tab} that are needed to construct these six quantities.
(Such differences were already discussed in section~\ref{sec:review} above,
with the $H$ asymmetry as an example.)
Thus, KDT use the same six observable names as the present work to refer measurable quantities
of the same magnitude but opposite sign.

We have conducted a similar test with the GWU/VPI SAID on-line analysis 
code~\cite{SAID,arndt}, downloading SAID multipoles, using the relations 
in (\ref{eq:mp-f1})-(\ref{eq:mp-f4}) to construct from these 
the four CGLN $F_{i}$ amplitudes, 
and then using our equations~(\ref{eq:obs-cgln-begin})-(\ref{eq:obs-cgln-end}) above to construct
observables. When the results are compared to direct predictions of observables
from the SAID code, again the same 6 observables 
$\left\{ H,C_{x'},C_{z'},O_{x'},O_{z'},L_{x'}\right\}$ differ in sign.
For the definition of observables, SAID refers to the Barker, Donnachie
and Storrow paper~\cite{barker}.
As discussed in section~\ref{sec:review}, the BDS definitions of asymmetries should be
deduced from their equations (2)-(4) and these have signs consistent with KDT.
Thus SAID also uses the same six observable names as the present work to refer
to quantities of the same magnitude but opposite sign.

We have repeated this same test with the Bonn-Gatchina (BoGa) on-line 
PWA~\cite{boga10}, downloading BoGa multipoles, using the relations of 
(\ref{eq:mp-f1})-(\ref{eq:mp-f4}) to construct the four CGLN amplitudes, and 
then using our (\ref{eq:obs-cgln-begin})-(\ref{eq:obs-cgln-end}) to construct observables. 
Comparing these to direct predictions of observables from the BoGa code, 
the results are identical, except for the $E$ asymmetry which is of opposite sign. 
However, for the definition of observables the BoGa on-line site refers 
to FTS of \cite{fasano}, whose definitions are the same as in our 
\ref{apx_tab} except for a sign change in the $E$ asymmetry, as in table~\ref{tab:comparison}.
Thus, we conclude that the relations between observables and amplitudes used 
in the BoGa analysis is completely consistent with the present work.

\begin{figure}[t]
\centering
\includegraphics[clip,width=0.5\textwidth]{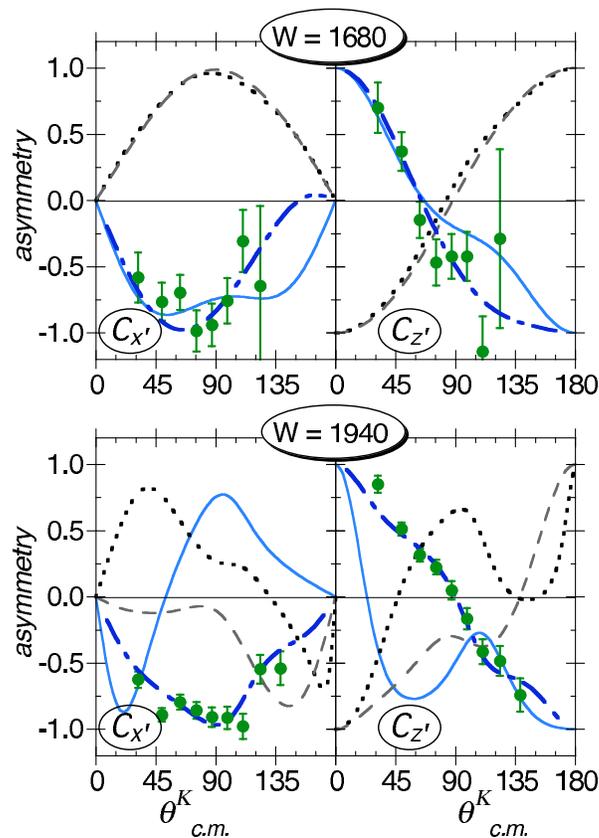}
\caption{\label{fig:cfits}(Color online) $C_{x'}$ (left) and $C_{z'}$ (right) for the
$\gamma p \rightarrow K^{+}\Lambda$ reaction at $W=1680$ MeV
(top) and $W=1940$ MeV (bottom). Kaon-MAID predictions are dashed (black)~\cite{MAID,drech07,mart},
SAID predictions are dotted (black)~\cite{SAID,arndt}, BoGa
predictions are dot-dashed (blue)~\cite{boga10}
and predictions from JSLT~\cite{bruno} are solid (blue). 
The green circles are from \cite{g1c07}.}
\end{figure}

New data are emerging from the current generation of polarization experiments
which make these sign differences an important issue.
In \cite{g1c07}, recent results for the $C_{x'}$ and $C_{z'}$ asymmetries 
have been compared with the direct predictions of the Kaon-MAID code,
with predictions from an earlier version of the BoGa multipoles and
with predictions from Juli\'a-D\'iaz, Saghai, Lee and Tabakin (JLST)~\cite{bruno}.
As an illustration, in figure~\ref{fig:cfits} we have replotted figures~8 and~9 from~\cite{g1c07}
for two energies, transformed to the primed kaon axes using equation~(\ref{eq:br-inv}),
and added the predictions from SAID.
The MAID (black dashed) and SAID (black, dotted) curves for $C_{z'}$ 
approach $-1$ at $\theta_K= 0^\circ$, while the BoGa (blue, dot-dashed)
and JSLT (blue, solid) curves approach $+1$, along with the data (green circles) from~\cite{g1c07}.

The behavior of $C_{z'}$ at $\theta_K= 0^\circ$ is a simple reflection of angular momentum
conservation.
Using the definition from~\ref{apx_tab},
$C_{z'} = \{\sigma_{1}(+1,0,+z') - \sigma_{2}(+1,0,-z')\}/\{\sigma_{1} + \sigma_{2}\}$.
When the incident photon spin is oriented along $+\hat{z}$, only
those target nucleons with anti-parallel spin can contribute to the production
of spin zero mesons at $\theta_K=0$, and the projection of the total angular
momentum along  $\hat{z}$ is $+ 1/2$. 
Thus, the recoil baryon must have its spin oriented
along $+\hat{z} = +\hat{z}'$ at $\theta_K=0$, so that $\sigma_{2}$ must vanish. 
The recent measurements on $K^{+} \Lambda$ production~\cite{g1c07} clearly 
show this asymmetry approaching $+1$ at $\theta_K=0^\circ$. 

The MAID and SAID predictions appear to have the wrong limits for $C_{z'}$ at 0 and 180 degrees.
Also shown are predictions using the multipoles of Juli\'{a}-D\'{i}az, Saghai, Lee and Tabakin 
(JSLT) from \cite{bruno}, used with our
expressions to construct observables (solid blue curves). The MAID and SAID sign differences
are also evident in $C_{x'}$, particularly at low energies where only a
few partial waves are contributing - top panels of figure~\ref{fig:cfits}. There it is
clear that the predictions of the different partial wave solutions are essentially
very similar, differing only in sign.
The comparisons of \cite{g1c07}, repeated here in figure~\ref{fig:cfits}, illustrate the
confusion that arises from the use of the same symbol to mean different
experimental quantities by different authors (section~\ref{sec:review}).

\section{\label{sec:fierz}Relations between observables}
   Since photo-production is characterized by 4 complex amplitudes, equation~(\ref{eq:cgln}), 
the 16 observables of equations~(\ref{eq:obs-cgln-begin})-(\ref{eq:obs-cgln-end}) are not independent. 
There are in fact many relations between them. 
The profile functions of (\ref{eq:obs-cgln-begin})-(\ref{eq:obs-cgln-end}) are bilinear products of the CGLN 
amplitudes, and one of the more extensive sets of equalities interrelating them has 
been derived by Chiang and Tabakin from the Fierz identities that relate bilinear 
products of currents~\cite{chiang}. 
Such relations are particularly useful, since they allow the comparison of data on 
one observable with an evaluation in terms of products of other observables. 
Any determination of the amplitude will invariably require combining data on different 
polarization observables which in general come from different experiments, each having 
different systematic scale uncertainties. 
The Fierz identities provide a means of enforcing consistency provided, of course, that they are
consistent with the expressions of general cross sections given in section~\ref{sec:gcs}.

The Fierz identities as derived by Chiang and Tabakin (CT) are  given in terms of 16
quantities, $\check{\Omega}^i$ in~\cite{chiang}, and the first column of table~I in that paper
gives the relation between these quantities and the conventional single, BT, BR and TR observable names.
CT quote FTS for the definition of these observables.
We have numerically checked the 37 Fierz identities of Appendix D in~\cite{chiang}.
Assuming the definitions of observables as given in our~\ref{apx_tab}, or in FTS,
a large number (more than half of them) require revisions in signs. 
If the signs of $\left\{ H,C_{x'},C_{z'},O_{x'},O_{z'},L_{x'}\right\}$ 
are reversed, as in BDS and KDT,
still many of the equations of~\cite{chiang} require revision.
A set of identities that are consistent with our definitions of observables in terms of
measurable quantities, \ref{apx_tab}, and with the form of our general cross section in
equation~(\ref{eq:gcs}), is given below in the first three sections of~\ref{apx_fier}.
As a practical example, in the next section we use two of the identities in a multipole
analysis to fix the scales of different data sets in a fit weighted by their systematic errors.

Another set of relations has been given by Artru, Richard and Soffer (ARS)~\cite{art07,art09}. 
These are different in form but can be derived from our Fierz identities, 
although with some differences in signs. 
A consistent set is listed in \ref{apx_fier4}.

In addition to identities, there are a number of inequalities, such as 
$(P)^2 +(C_{x'})^2 +(C_{z'})^2\leq 1$, which are often referred to as positivity 
constraints~\cite{art07}. 
These involve the sums of the squares of asymmetries, and as such are immune to sign issues. 
They can be particularly useful when extracting sets of asymmetries from fits to experimental 
data~\cite{ire10}, as in the examples discussed in section~\ref{sec:recoil}. 
But since our focus here is amplitude reduction from cross sections and asymmetries, 
we refer the reader to a recent review of such inequalities~\cite{art09}.

\section{\label{sec:anal} Multipole analyses}

The ultimate goal of the new generation of experiments now under way is a complete 
experimental determination of the multipole decomposition of the full amplitude 
in pseudoscalar meson production. 
In this section, we apply the formula presented in 
previous sections to develop this analysis process as model independently as possible.
While the data published to date are still insufficient to satisfy 
the Chiang and Tabakin requirements for removing discrete ambiguities~\cite{chiang}, 
it is instructive to examine the impact of recently published
polarization measurements. 
We focus here on the $\gamma p\to K^+\Lambda$ channel, which so far has provided the
largest number of different observables, as summarized in table~\ref{tab:td1}.

To avoid bias, the first stage in any multipole decomposition 
is a single-energy analysis, one beam/$W$ energy at a time without any assumptions 
on energy-dependent behavior. 
The range of recent published $K^+\Lambda$ measurements is summarized in table~\ref{tab:td1}. 
[Cross section data from the SAPHIR detector at Bonn~\cite{saph04} have an appreciable (20\%) 
angle- and energy-dependent difference from the CLAS experiments. 
This level of incompatibility makes it impossible to include them in the present analyses.]
While some of the data sets span the full nucleon resonance region in extremely fine steps, 
single-energy analyses are limited by the observables with the coarsest granularity, 
which in this case are the $C_{x'}$, $C_{z'}$ measurements (data group 3~\cite{g1c07}).
The only published $O_{x'}$, $O_{z'}$ and $T$ data are from GRAAL 
(data groups 5 and 8~\cite{gr09}). 
The combination of these data sets allows us to combine groups 1-8 at 5 different 
beam energies, with roughly 100 MeV steps in beam energy, for 
which 8 different observables are now available.

\begin{table}[tb]
\caption{\label{tab:td1} Summary of recent published results on $K^+\Lambda$ photoproduction. 
(Systematic uncertainties on the CLAS data are taken from the indicated references.
The systematic errors on the GRAAL measurements reflect their reported uncertainty 
in beam polarization, in the assumed weak-$\Lambda$-decay parameter and in the 
resulting error propagation through the extraction of $O_{x'}$, $O_{z'}$ and $T$.)}
\begin{tabular}{llllll}
\br
Data & Experiment & Observables & $E_\gamma$ range (MeV) & $\Delta E_\gamma$/$\Delta W$& Systematic\\
group&&&$W$ range (MeV) &binning& scale error \\
\mr
1&CLAS-g11a~\cite{g11a10}& $d\sigma_0$       & 938-3814 &     & $\pm 8$\% \\
&&&1625-2835 & 10 & ($E_\gamma$ dep.)\\
2&CLAS-g11a~\cite{g11a10}& $P$               & 938-3814 &     &$\pm 0.05$\\
&&&1625-2835 & 10 &\\
3&CLAS-g1c~\cite{g1c07}  & $C_{x'}$, $C_{z'}$ & 1032-2741& 101 &$\pm 0.03$\\
&&&1679-2454 &&\\
4&CLAS-g1c~\cite{g1c06}  & $d\sigma_0$       & 944-2950 & 25  &$\pm 8$\% \\
&&&1628-2533 && ($E_\gamma$ dep.)\\
5&GRAAL~\cite{gr09}      & $O_{x'}$, $O_{z'}$ & 980-1466 & 50  &$\pm 4$\%\\
&&&1649-1906 &&\\
6&GRAAL~\cite{gr07}      & $P$               & 980-1466 & 50  &$\pm 3$\%\\
&&&1649-1906 &&\\
7&GRAAL~\cite{gr07}      & $\Sigma$          & 980-1466 & 50  &$\pm 2$\%\\
&&&1649-1906 &&\\
8&GRAAL~\cite{gr09}      & $T$               & 980-1466 & 50  &$\pm 5$\%\\
&&&1649-1906 &&\\
9&LEPS~\cite{leps03}     & $\Sigma$          & 1550-2350& 100 &$\pm 3$\%\\
&&&1947-2300 &&\\
\br
\end{tabular}
\end{table}

\subsection{Coordinate transformations}
\label{sec:anala}
There are several different choices for coordinate systems in use and before data from 
the different experiments can be combined in a common analysis we transformed them to 
the system defined in figure~\ref{fig:coord}. 
The beam-recoil data of group 3~\cite{g1c07} were reported in unprimed 
c.m. coordinates relative to the beam direction.
These are related to the primed system of figure~\ref{fig:coord} by 
the relations in equation~(\ref{eq:br-inv}).
The GRAAL papers use the coordinates of Adelseck and Saghai~\cite{adel}. 
Relative to $\hat y'=\hat y$, their $\hat x$ and $\hat z$ axes are reversed from those of 
figure~\ref{fig:coord}, so that, although $\Sigma$, $T$ and $P$ are unchanged in transferring to our
coordinates, $O_{x',z'}$ become the negative of what GRAAL refer to as $O_{x,z}$.
Thus,
\begin{equation}
O_{x',z'} = - O^{{\rm GRAAL}}_{x,z}. \label{eq:o-graal}
\end{equation}

\subsection{Constraining systematic scale uncertainties}
\label{sec:analb}
Each experiment has reported systematic errors that reflect an uncertainty 
in the scale of the entire data set. 
We use a procedure of imposing self-consistence within a collection
of data sets by including their measurement scales as parameters in a fit minimizing
$\chi^2$~\cite{legs01}. 
To fix first the scales of the polarization observables, data groups (2,3,5,6,7,8) of 
table~\ref{tab:td1}, we use the Fierz identities (\ref{eq:L.BR}) and (\ref{eq:S.br}) of 
\ref{apx_fier} to construct the quantities,
\begin{eqnarray}
F_{{\rm L.BR}}&=& \Sigma P - C_{x'}O_{z'} + C_{z'}O_{x'}-T,\nonumber\\
F_{{\rm S.br}}&=& O_{x'}^{2} + O_{z'}^{2} + C_{x'}^{2} + C_{z'}^{2} + \Sigma^{2} - T^{2} + P^{2} - 1,
\label{eq:lbr-sbr}
\end{eqnarray}
both of which have the expectation value of 0 at each angle and energy. Our fitting procedure then
minimizes the $\chi^2$ function,
\begin{eqnarray}
\chi^2 &=& \sum_{E_\gamma}\sum_{\theta_K} 
\left\{
\left[
\frac{F_{{\rm L.BR}}(f_ix^{{\rm exp}}_{i\theta})}{\delta F_{{\rm L.BR}}(f_i\sigma_{x_{i\theta}})}
\right]^2_{i=2,3,5,6,7,8}
+
\left[
\frac{F_{{\rm S.br}}(f_ix^{{\rm exp}}_{i\theta})}{\delta F_{{\rm S.br}}(f_i\sigma_{x_{i\theta}})}
\right]^2_{i=2,3,5,6,7,8}
\right\}
\nonumber\\
&&
+\sum_i\left[\frac{f_i-1}{\sigma_{f_i}}\right]^2,
\label{eq:chi2}
\end{eqnarray}
where the index $i\equiv (2,3,5,6,7,8)$ runs through each of the data groups of asymmetries 
($x^{{\rm exp.}}_{i\theta}$) needed to construct the Fierz relations of~(\ref{eq:lbr-sbr}). 
All data from a set $i$ having a systematic scale error ($\sigma_{f_i}$) are multiplied by a common 
factor ($f_i$) while adding $(f_i-1)^2/\sigma^2_{f_i}$ to the $\chi^2$. 
This last term weights the penalty for choosing a normalization scale different from unity 
by the reported systematic uncertainty of the experiment.

\begin{figure}
\centering
\includegraphics[clip,width=0.67\textwidth]{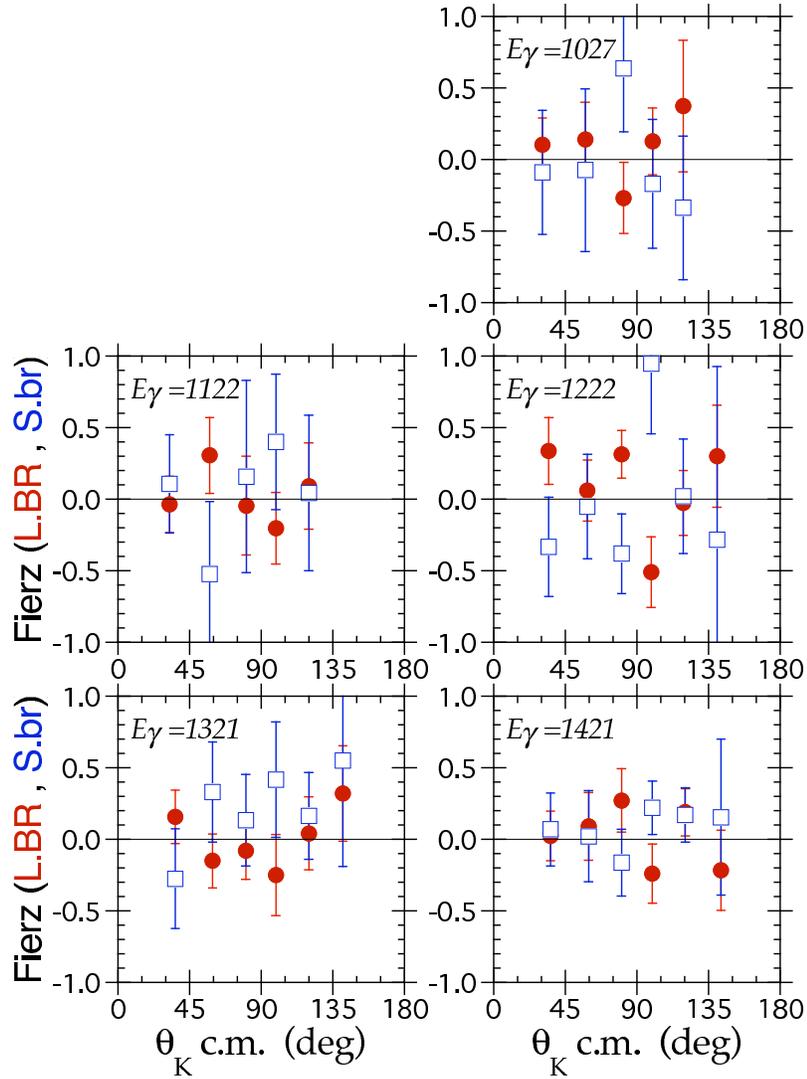}
\caption{\label{fig:lbr-sbr} 
(Color online) Evaluations of the two Fierz relations (\ref{eq:L.BR}) (solid red circles) 
and (\ref{eq:S.br}) (open blue squares) of (\ref{eq:lbr-sbr}),
using the data of table~\ref{tab:td1} and the fitted scales of table~\ref{tab:td2}.
}
\end{figure}

In this procedure polynomial fits are used, where needed, to interpolate the data of 
table~\ref{tab:td1} to a common angle and energy. 
There are two measurements of the recoil polarization asymmetry ($P$), from groups 2 and 6 
in table~\ref{tab:td1}, and a weighted mean of these data, including their scale factors, 
is used in evaluating (\ref{eq:chi2}). 
The scale factors resulting from this fit are listed in table~\ref{tab:td2}. 
All are close to unity.
The resulting evaluations of the Fierz relation, using the scaled data, 
are shown in figure~\ref{fig:lbr-sbr}.

\begin{table}[tb]
\caption{\label{tab:td2} Fitted scales for the data sets of table~\ref{tab:td1}
that are used to construct the relations in (\ref{eq:chi2}).}
\begin{indented}
\item[] \begin{tabular}{@{}lllll}
\br
Data group & Experiment  & Observables & Fitted scale ($f_i$) & Scale error ($\sigma_{f_i}$) \\
\mr 
2&CLAS-g11a & $P$               & 1.000 & 0.049 \\
3&CLAS-g1c  & $C_{x'}$, $C_{z'}$ & 0.984 & 0.025 \\
5&GRAAL     & $O_{x'}$, $O_{z'}$ & 0.997 & 0.035 \\
6&GRAAL     & $P$               & 1.001 & 0.030 \\
7&GRAAL     & $\Sigma$          & 1.001 & 0.020 \\
8&GRAAL     & $T$               & 0.992 & 0.040 \\
\br
\end{tabular}
\end{indented}
\end{table}

While the results in figure~\ref{fig:lbr-sbr} scatter around zero as expected, 
the fluctuations are sometimes appreciable. 
These cannot readily be removed with an energy- and angle-independent scale factor. 
It is likely this results from combining data from different detectors. 
While global uncertainties such as flux normalization and target thickness can be 
readily estimated and easily fitted away in this type of procedure, angle-dependent 
variations in detector efficiencies tend to be the most problematic to control and quantify.

\subsection{Multipole fitting procedure}
\label{sec:analc}
The observables of table~\ref{tab:td1} are determined by the CGLN amplitudes through 
(\ref{eq:obs-cgln-begin})-(\ref{eq:obs-cgln-end}), and these are in turn determined by the multipoles through 
(\ref{eq:mp-f1})-(\ref{eq:mp-f4}). 
Since the multipoles are reduced matrix elements and independent of angle, 
fitting them directly allows the use of complete angular distributions for each observable. 
We fix the scales ($f_i$) of the polarization observables 
$\left\{\Sigma, T, P, C_{x'}, C_{z'}, O_{x'}, O_{z'}\right\}$ to their fitted values 
in table~\ref{tab:td2}, and now vary the multipoles, as well as the scales $f_1$ and $f_4$ 
for the unpolarized cross section ($d\sigma_0$) measurements 
(groups 1 and 4 in table~\ref{tab:td1}) to minimize the $\chi^2$ function,
\begin{equation}
\chi^2 = \sum^{N_s}_{i=1}\left\{\sum^{N_i}_{j=1}
\left[
\frac{f_ix^{{\rm exp}}_{ij}-x^{{\rm fit}}_{ij}(\vec\zeta)}{f_i\sigma_{x_{ij}}}
\right]^2\right\}
+\sum_{i=1,4}\left[\frac{f_i-1}{\sigma_{f_i}}\right]^2,
\label{eq:chi2-2}
\end{equation}
where $N_s$ is the number of independent data sets, each having 
$N_i$ points. 
$x^{{\rm exp}}_{ij}$ and $\sigma_{x_{ij}}$ are the $j$-th experimental datum from 
the $i$-th data set and its associated measurement error, respectively, 
$x^{{\rm fit}}_{ij}(\vec\zeta)$ is the value predicted from the $\vec\zeta$ multipole set being fit, 
and $f_i$ is the global scale parameter associated with the $i$-th data set. 
As before, the last term weights the penalty for choosing a cross section scale different
from unity by the reported systematic uncertainties for data groups 1 and 4~\cite{legs01}.

Thus our fitting procedure is a two-step process, first minimizing (\ref{eq:chi2}) 
by varying the scale factors of the polarization data, and then minimizing 
(\ref{eq:chi2-2}) in a second step by varying the multipoles and the cross section scales. 
These two cannot be combined into a single step in which Fierz relations 
such as (\ref{eq:lbr-sbr}) are minimized by varying multipoles, since all properly 
constructed multipoles will automatically satisfy the Fierz identities.

\begin{figure}
\centering
\includegraphics[clip,width=0.67\textwidth]{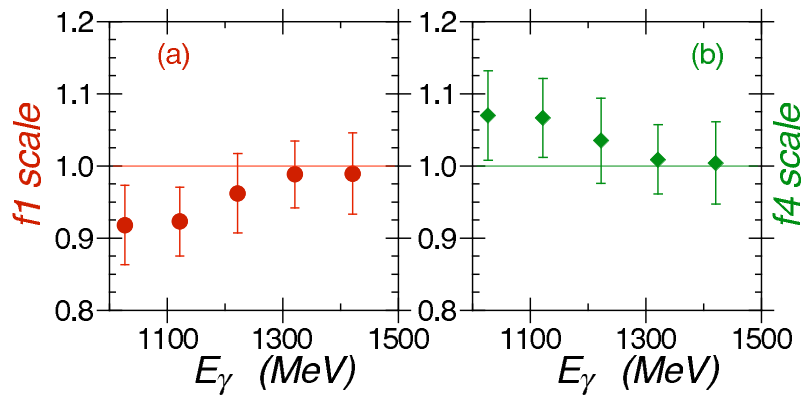}
\caption{\label{fig:f1-f4} (Color online) Fitted scales for 
the cross section ($d\sigma_0$) measurements of \cite{g11a10},
$f_1$ as red circles, and \cite{g1c06}, $f_4$ as green diamonds.
}
\end{figure}

While the cross section experiments report the global systematic uncertainties listed
in table~\ref{tab:td1}, comparisons given in \cite{g11a10} show a clear energy dependence 
to the scale difference between them, which is most pronounced at low energies. 
Accordingly, we have fitted separate cross section scales at each energy and the results 
are plotted in figure~\ref{fig:f1-f4}.

Cross sections for any reaction generally fall with increasing angular momentum, which guarantees
the ultimate convergence of a multipole expansion. 
However, in practice such expansions must be truncated to limit the maximum angular momentum 
to a value that is essentially determined by the statistical precision and breadth of kinematic 
coverage of the data sets. 
The ultimate goal of such work will be the identification of the excited states of the nucleon, 
and this will require, as a minimum, accurate multipole information up to at least $L=2$ 
to be useful. 
As has been shown by Bowcock and Burkhardt~\cite{bb75}, the highest multipole fitted in 
any analysis always tends to accumulate the systematic errors stemming from truncation and 
is essentially guaranteed to be the most uncertain. 
Thus, when focusing on multipoles up to $L=2$ we must vary up to $L=3$ and fix 
the multipoles for $4 \leq L \leq 8$ to their (real) Born values. 
(Details of the Born amplitudes are given in \ref{apx_born}.)

\begin{figure}
\centering
\includegraphics[clip,height=0.9\textheight]{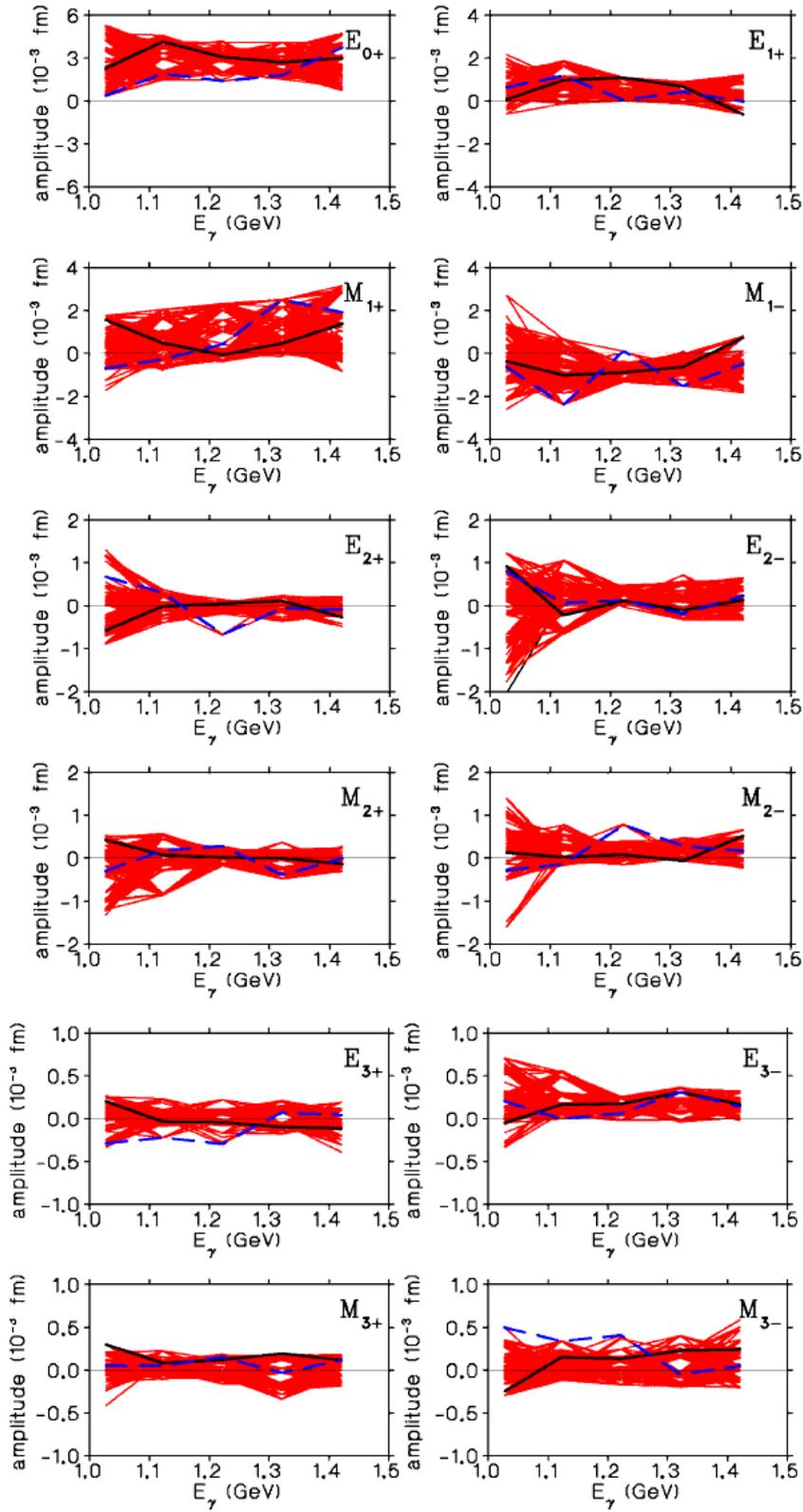}
\caption{\label{fig:mul-re} (Color online) 
Real parts of multipoles for $L=0$ to $3$, fitted to the data of table~\ref{tab:td1} with the
phase of the $E_{0+}$ fixed to 0. The bands show variations in the $\chi^2$/point of 
less than 0.2, as in table~\ref{tab:td3}.
Solutions with the best and largest $\chi^2$, corresponding to the columns of table~\ref{tab:td3},
are shown as solid (black) and long-dashed (blue) curves, respectively.
}
\end{figure}

\begin{figure}
\centering
\includegraphics[clip,height=0.9\textheight]{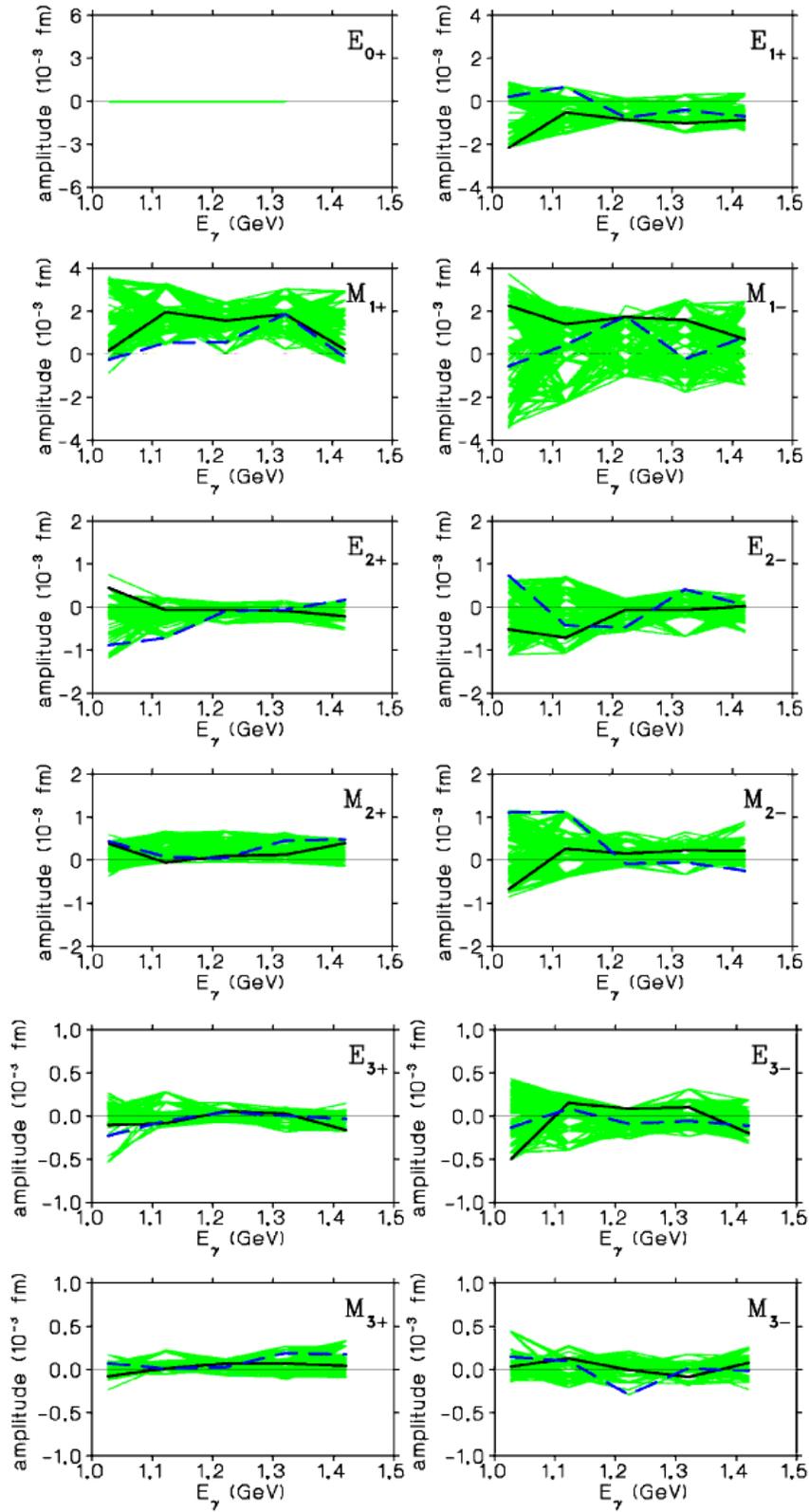}
\caption{\label{fig:mul-im} (Color online) 
Imaginary parts of multipoles for $L=0$ to $3$, fitted to the data of table~\ref{tab:td1} with the
phase of the $E_{0+}$ fixed to 0. The bands show variations in the $\chi^2$/point of 
less than 0.2, as in table~\ref{tab:td3}.
Curves are as in figure~\ref{fig:mul-re}.
}
\end{figure}

To search for a global minimum while allowing for the presence of local minima, 
we use a Monte Carlo sampling of the multipole parameter space. 
Values for the real and imaginary parts of the $0\leq L\leq 3$ multipoles are
chosen randomly and their $\chi^2$ comparison to the data of table~\ref{tab:td1}, 
scaled by the fitted constants in table~\ref{tab:td2} and figure~\ref{fig:f1-f4}, are calculated. 
Whenever the resulting $\chi^2$ is within $10^4$ times the current best value, 
a gradient minimization is carried out. 
We have repeated this procedure for a wide range of Monte Carlo samples, up to $10^7$ per energy, 
and have found a band of solutions with tightly clustered $\chi^2$ that cannot be distinguished 
by the existing data. 
In figures~\ref{fig:mul-re} and~\ref{fig:mul-im} we plot the real and imaginary parts of 
300 multipole solutions for which the gradient search has converged to a minimum. 
The $\chi^2$/point of each solution within these bands is always within 0.2 of the best, 
and is even more tightly clustered at low energies.

The best and largest values of the $\chi^2$/point for these bands are listed in 
table~\ref{tab:td3}. 
(The corresponding multipole solutions are shown as the solid black and blue dashed curves
in figures~\ref{fig:mul-re} and~\ref{fig:mul-im}, respectively.)
The fact that most of the $\chi^2$/point values are substantially less than one 
is a sign that fitting multipoles up to 
$L=3$ provides more freedom than the present collection of data warrant, 
even though the desired physics demands it.

\begin{figure}
\centering
\includegraphics[clip,width=0.4\textwidth]{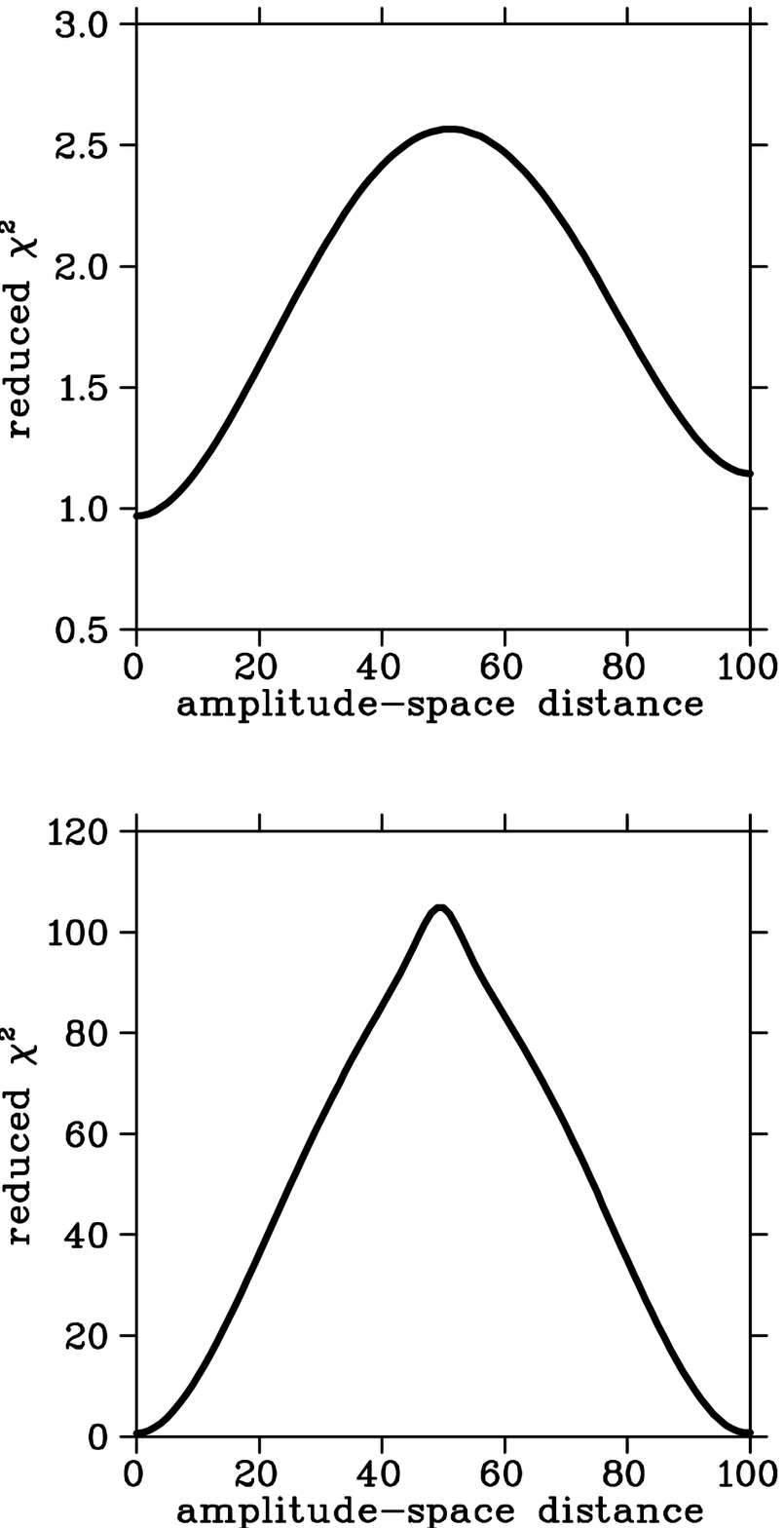}
\caption{\label{fig:chi2-dev}
The $\chi^2$/point calculated by comparing the data of table~\ref{tab:td1} to predictions as 
a hybrid amplitude (\ref{eq:ah}) is tracked between the solutions with the best and 
largest $\chi^2$ in table~\ref{tab:td3} (solid black and dashed blue curves in 
figures~\ref{fig:mul-re} and~\ref{fig:mul-im}, respectively). 
Results are shown for $E_\gamma$ ($W$) energies of 1122 (1728) MeV in the bottom panel and 
1421 (1883) MeV in the top.
}
\end{figure}

The bands in figures~\ref{fig:mul-re} and~\ref{fig:mul-im} reflect a relatively 
shallow valley in the $\chi^2$ space. 
To understand if this valley is smooth, indicating a simple broad minimum, 
or is pitted with many local minima, we have tracked solutions across $\chi^2$. 
This can be done by forming a hybrid amplitude $A_h(x)$ from two solutions $A_1$ and $A_2$:
\begin{equation}
A_h(x) = A_1 \times \left(1-\frac{x}{100}\right) + A_2 \times \left(\frac{x}{100}\right),~~x\in [0,100]  .
\label{eq:ah}
\end{equation}
Here $x$ is an effective distance in amplitude-space. 
For $x=0$, $A_h$ is just $A_1$ while for $x=100$, $A_h$ becomes $A_2$. 
At each value of $x$ between 0 and 100 the hybrid set of multipoles is used to predict 
observables and the $\chi^2$ relative to the data is calculated. 
If the valley between $A_1$ and $A_2$ is smooth and featureless the resulting 
$\chi^2$ map will be similarly featureless. 
We have carried out this exercise for many pairs of solutions and always found 
pronounced peaks in $\chi^2$ for any choice of $A_1$ and $A_2$. 
As an example, the $\chi^2$/point that results from forming a hybrid amplitude out of the 
best and largest (worst) solutions of figures~\ref{fig:mul-re} and~\ref{fig:mul-im} is shown 
in figure~\ref{fig:chi2-dev} for two of the energy bins of table~\ref{tab:td3}. 
(Similar results are obtained at other energies.) 
At $E_\gamma=1122$ MeV ($W=1728$ MeV), in the bottom panel of figure~\ref{fig:chi2-dev}, 
the peak in $\chi^2$ between the two is huge. 
At $E_\gamma=1421$ MeV ($W=1883$ MeV) the intermediate peak is still present, 
though not so tall, probably due to the presence of another local minimum that is nearby 
but off the direct trajectory between the two solutions.

Evidently the bands in figures~\ref{fig:mul-re} and~\ref{fig:mul-im} are created by clusters 
of local minima in $\chi^2$ which, for the present collection of data, are completely 
degenerate and experimentally indistinguishable. 
The 8 observables in table~\ref{tab:td1} do not yet satisfy the Chiang and Tabakin (CT) 
criteria as a minimal set that would determine the photoproduction amplitude free of 
ambiguities~\cite{chiang}. 
Nonetheless, from studies with mock data, as will be described in section~\ref{sec:mock},
 we have found that the presence of multiple 
local minima is essentially universal, even when the CT criteria are satisfied. 
But, as more observables are added with increasing statistical 
accuracy the degeneracy is broken and a global minimum emerges. 
The difficulty then becomes finding it among the pitted landscape in $\chi^2$. 

\begin{table}[tb]
\caption{\label{tab:td3} Best and largest values of the $\chi^2$/point for the solutions in the
bands plotted in figures~\ref{fig:mul-re} and~\ref{fig:mul-im}.
}
\begin{indented}
\item[] \begin{tabular}{@{}lll}
\br
$E_\gamma$ / $W$ (MeV) & Best $\chi^2$/point & Largest $\chi^2$/point \\
\mr
1027 / 1676 & 0.49 & 0.54 \\
1122 / 1728 & 0.59 & 0.62 \\
1222 / 1781 & 0.52 & 0.62 \\
1321 / 1833 & 0.74 & 0.92 \\
1421 / 1883 & 0.97 & 1.15 \\
\br
\end{tabular}
\end{indented}
\end{table}

\subsection{Constraining the arbitrary phase}
\label{sec:anald}
\begin{figure}
\centering
\includegraphics[clip,width=0.67\textwidth]{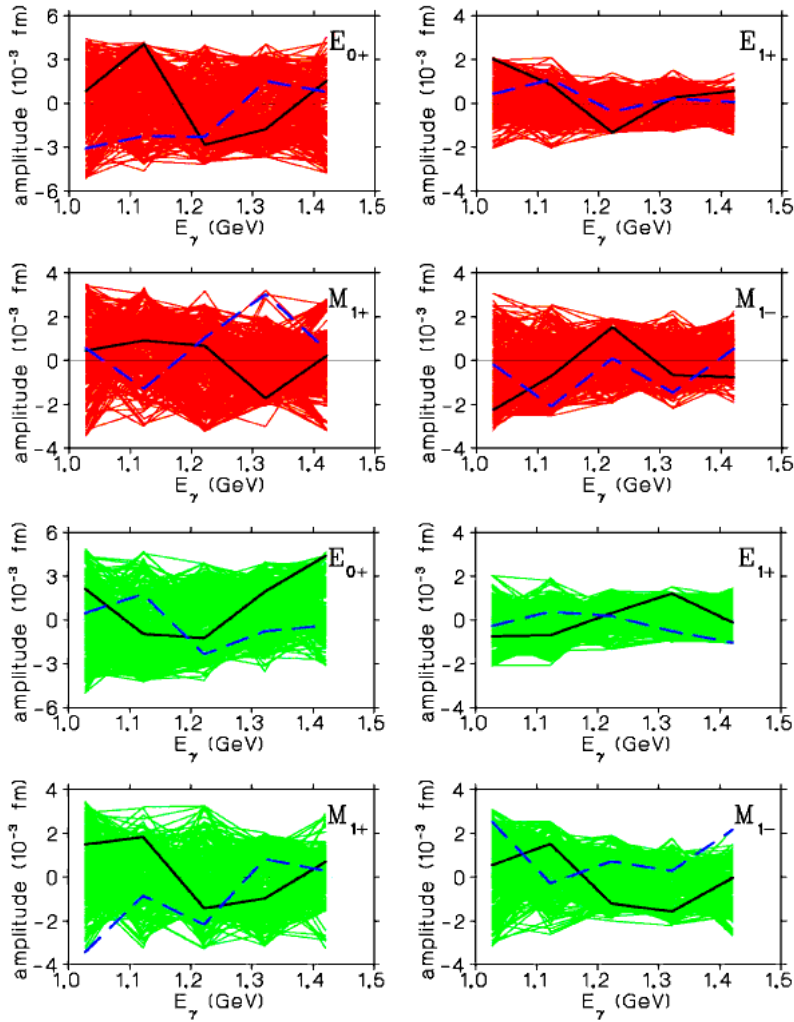}
\caption{\label{fig:mul-v2} (Color online) 
Real parts (top four panels in red) and imaginary parts (bottom four panels in green)
of the $S$ and $P$ wave multipoles, fitted to the data of table~\ref{tab:td1}
without any phase constraints. The bands show variations in the $\chi^2$/point of
less than 0.2.
}
\end{figure}

In determining an amplitude there is one overall phase that can never be constrained, 
and so in fitting the solutions of figures~\ref{fig:mul-re} and~\ref{fig:mul-im} we have chosen 
to fix the phase of the $E_{0+}$ multipole to zero (which sets its imaginary part to zero). 
The consequence of not fixing a phase is illustrated in figure~\ref{fig:mul-v2}, where we plot as
an example the $S$ and $P$ wave multipoles from fits with an unconstrained phase angle.
Again, the solutions within these bands have values for the $\chi^2$/point
that differ by less than 0.2 from that of the best solution.
While these bands appear to be substantially broader, they are in fact just the bands of 
figures~\ref{fig:mul-re} and~\ref{fig:mul-im}, expanded by rotating with a random phase angle. 
The behavior of the $L=2$~($D$) and $L=3$~($F$) waves show a similar broadening.

In practice, the utility of determining a set of multipoles is not diminished by fixing one phase.
Ultimately, such experimentally determined multipoles will be compared to model predictions.
For this, one only has to rotate the model phase to the same reference point, e.g., 
a real $E_{0+}$ in the analysis of figures~\ref{fig:mul-re} and~\ref{fig:mul-im}.
(The result of such an exercise is shown in figures~\ref{fig:sol-re} and~\ref{fig:sol-im}.)

The choice of which multipole phase to fix at zero is somewhat arbitrary. 
From studies with mock data,
we have found that it is sufficient to fix the phase of any one of 
the larger multipoles ($L=0,1$) when the data to be fit have modest statistical accuracy. 
Ultimately, if the data precision is very high, just fixing the higher $L$ multipoles at their 
real Born values is enough to recover the amplitude.

\subsection{Constraints from observables}
\label{sec:anale}
\begin{figure}
\centering
\includegraphics[clip,width=0.67\textwidth]{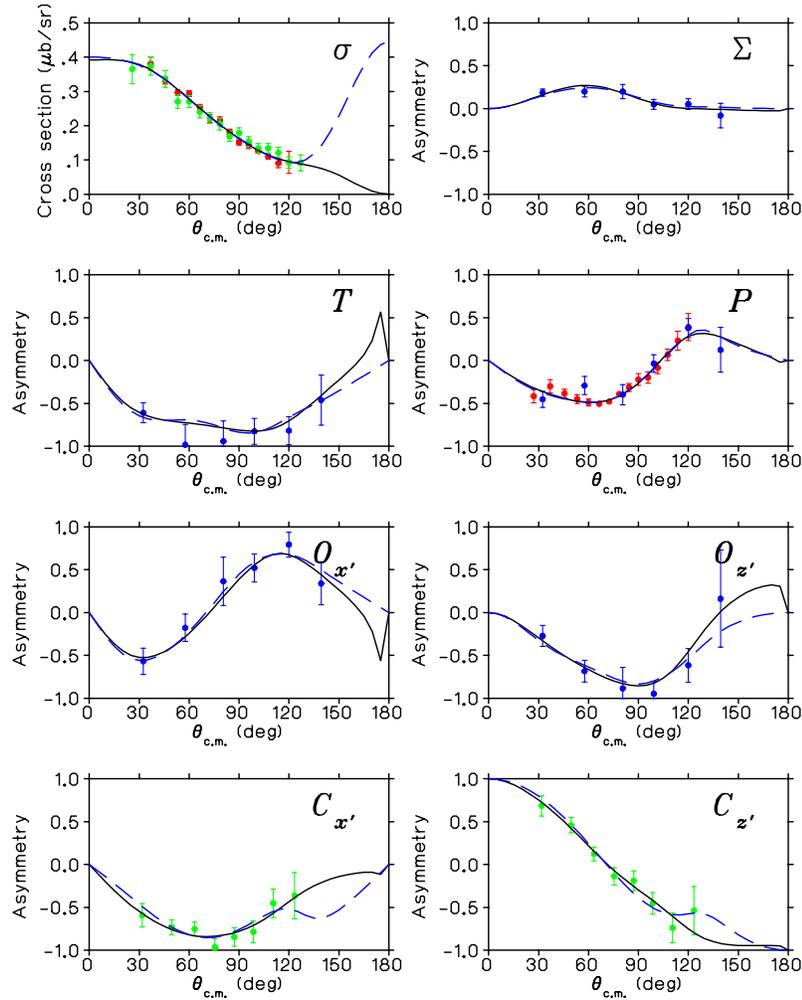}
\caption{\label{fig:pre-v1} (Color online) 
Predictions at $E_\gamma =1122$ MeV ($W=1728$ MeV) compared to the data of table~\ref{tab:td1} for 
the multipole solutions of figures~\ref{fig:mul-re} and~\ref{fig:mul-im}
having the minimum (solid black curves) and largest (long-dashed blue curves) $\chi^2$/point
(table~\ref{tab:td3}).
Data points are from CLAS-g11a~\cite{g11a10} shown in red, CLAS-g1c~\cite{g1c06, g1c07} 
shown in green, and GRAAL~\cite{gr07,gr09} shown in blue.
}
\end{figure}

\begin{figure}
\centering
\includegraphics[clip,width=0.67\textwidth]{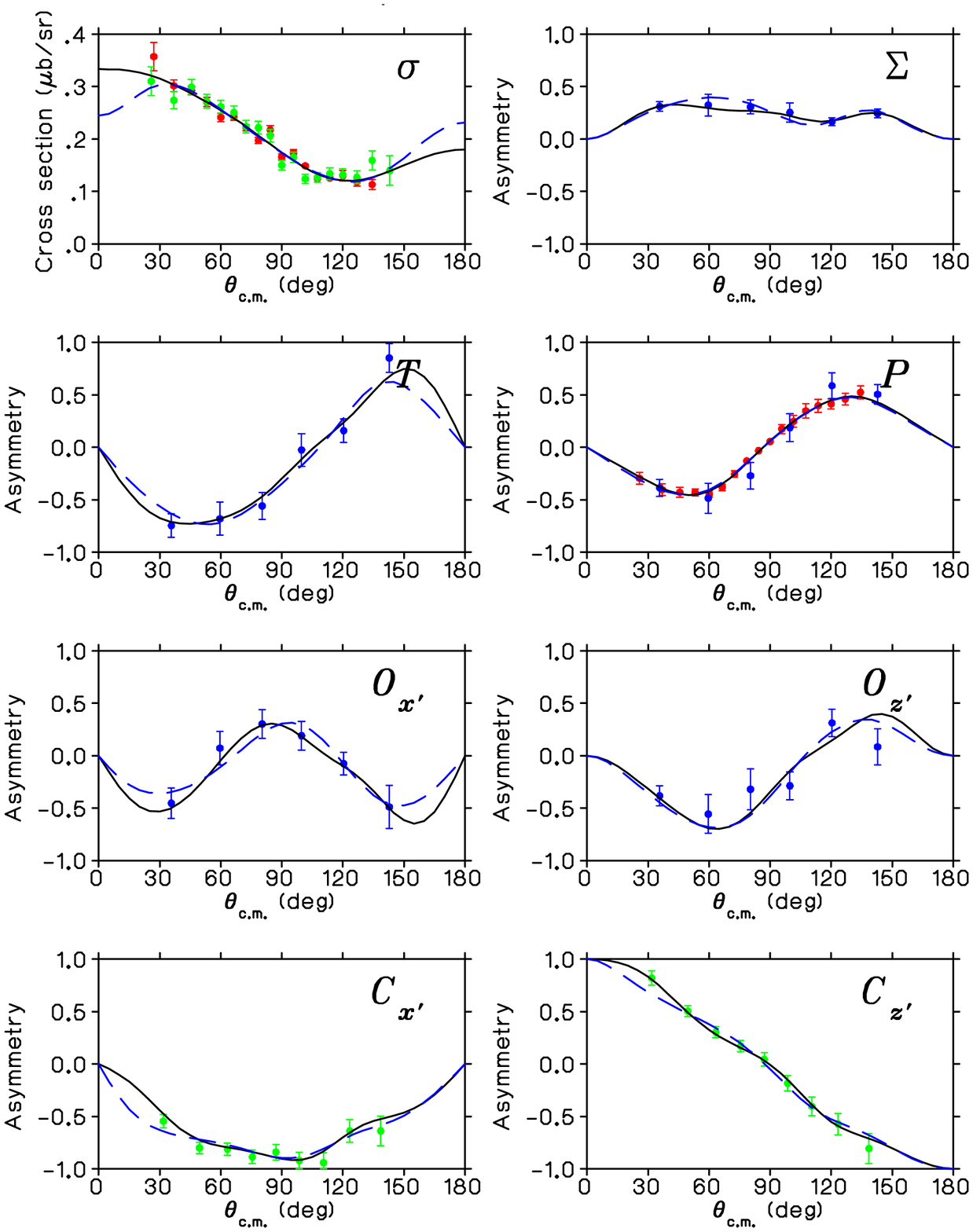}
\caption{\label{fig:pre-v2} (Color online) 
Predictions at $E_\gamma =1421$ MeV ($W=1883$ MeV) compared to the data of table~\ref{tab:td1} for 
the multipole solutions of figures~\ref{fig:mul-re} and~\ref{fig:mul-im}
having the minimum (solid black curves) and largest (long-dashed blue curves) $\chi^2$/point
(table~\ref{tab:td3}).
Data points are plotted as in figure~\ref{fig:pre-v1}.
}
\end{figure}

Predictions of the fitted multipole solutions are compared to the data of table~\ref{tab:td1} 
in figures~\ref{fig:pre-v1} and~\ref{fig:pre-v2} for two beam energies, 1122 and 1421 MeV. 
The best and worst solutions from the bands of figures~\ref{fig:mul-re} and~\ref{fig:mul-im},
in terms of the $\chi^2$/point values of table~\ref{tab:td3}, are shown as 
the solid (black) and dashed (blue) curves, respectively. 
The behavior at other energies is very similar. 
Based on such comparisons with existing published data, the multipole solutions within 
the bands of figures~\ref{fig:mul-re} and~\ref{fig:mul-im} are completely indistinguishable. 
Clearly, despite the presence of 8 polarization observables, 
the multipoles are still very poorly constrained. 
For many of the higher multipoles not even the sign is known.


\begin{figure}
\centering
\includegraphics[clip,width=0.67\textwidth]{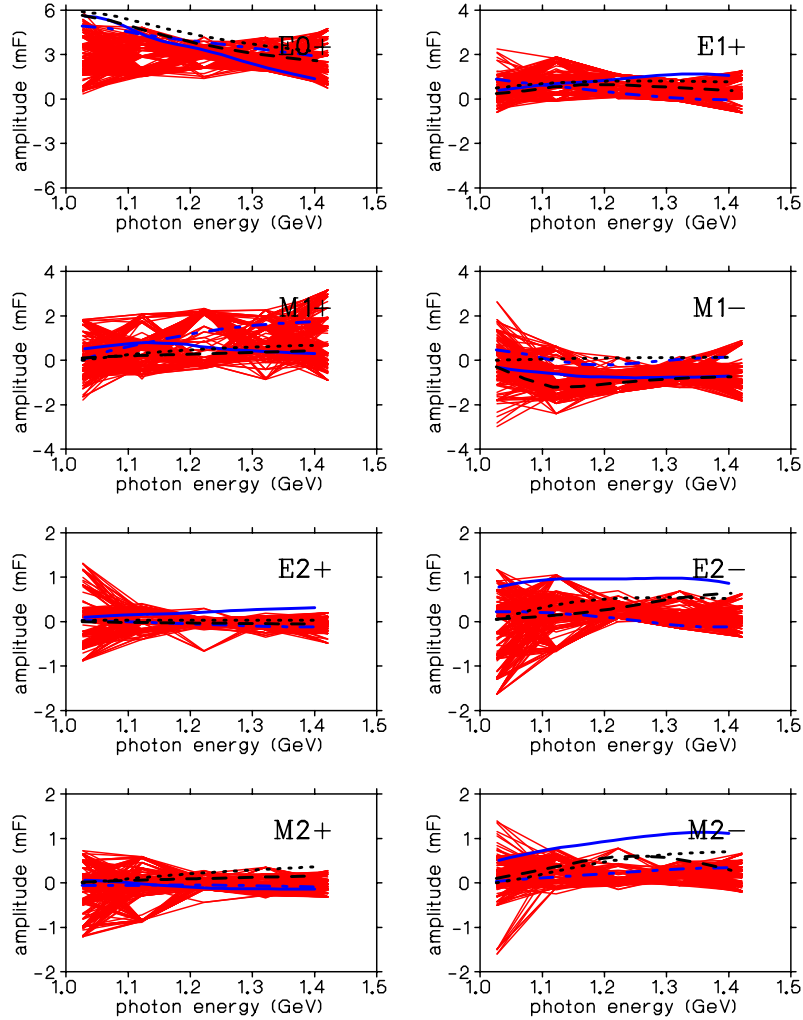}
\caption{\label{fig:sol-re} (Color online) 
The solution bands of figure~\ref{fig:mul-re}, compared to the real parts of PWA 
multipoles of BoGa~\cite{boga10} (blue dashed-dot), Kaon-MAID~\cite{MAID} (black dashed), 
SAID~\cite{SAID} (black dotted) and JSLT~\cite{bruno} (blue solid). 
For this comparison, each PWA has been rotated so that their $E_{0+}$ is real -- see text.
}
\end{figure}

\begin{figure}
\centering
\includegraphics[clip,width=0.67\textwidth]{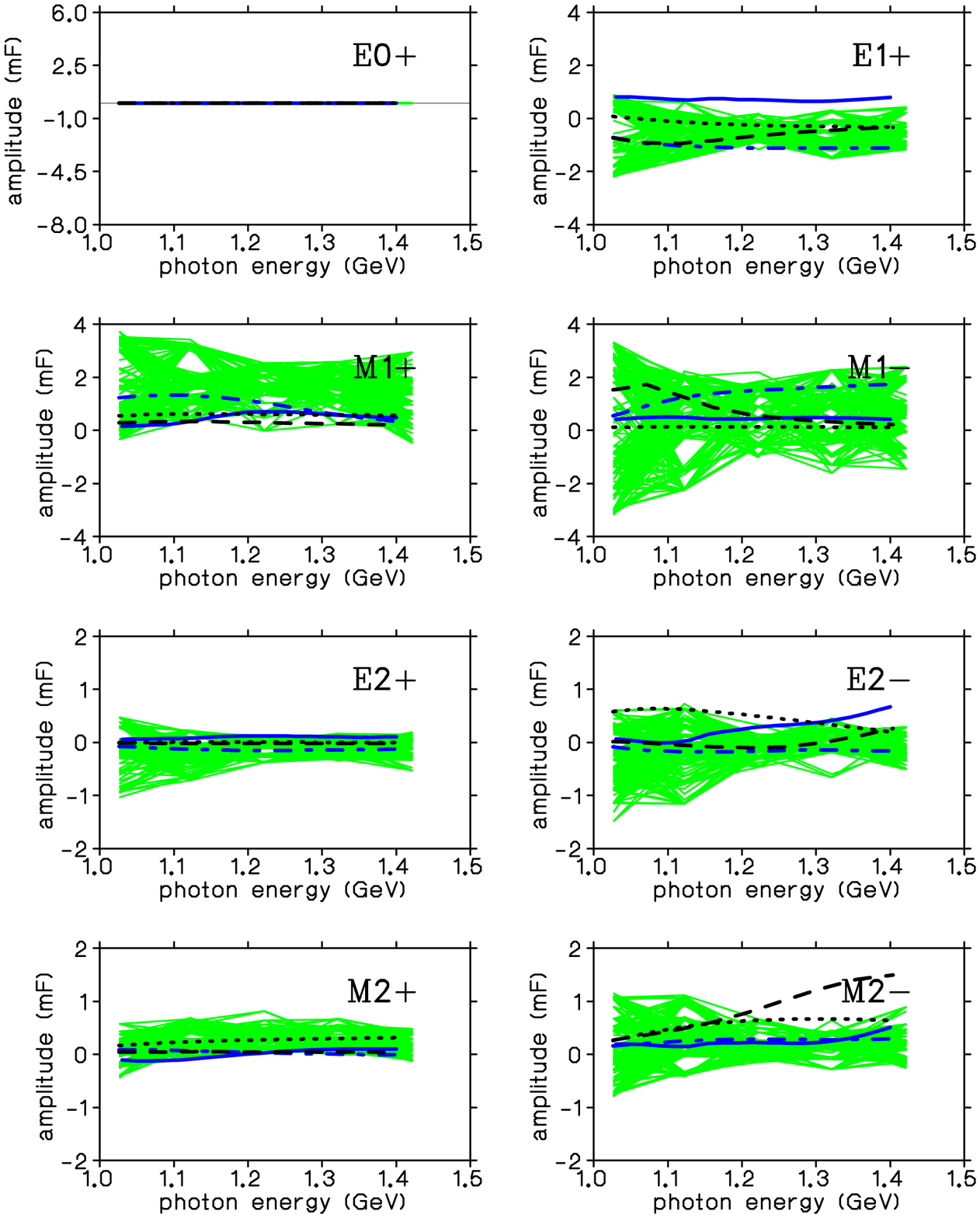}
\caption{\label{fig:sol-im} (Color online) 
The solution bands of figure~\ref{fig:mul-im}, compared to the imaginary parts of PWA 
multipoles of BoGa~\cite{boga10} (blue dashed-dot), Kaon-MAID~\cite{MAID} (black dashed), 
SAID~\cite{SAID} (black dotted) and JSLT~\cite{bruno} (blue solid). 
For this comparison, each PWA has been rotated so that their $E_{0+}$ is real -- see text.
}
\end{figure}

In figures~\ref{fig:sol-re} and~\ref{fig:sol-im} we compare the $S$, $P$ and $D$ wave multipoles 
from existing PWA results 
(BoGa~\cite{boga10}, MAID~\cite{MAID}, SAID~\cite{SAID} and JSLT~\cite{bruno})
with the bands of figures~\ref{fig:mul-re} and~\ref{fig:mul-im}, respectively.
Here we have rotated all multipoles to our common reference point of a real $E_{0+}$. 
(Each set of multipoles has been multiplied by $\exp(-i\delta)$, where $\delta$ is 
the phase of the $E_{0+}$ multipole of the PWA set.) 
For the most part, these PWA lie within our experimental solution bands. 
However, there are a few exceptions at the higher energies, in particular the 
$M_{2-}$ multipole from Kaon-MAID (black dashed curve in figure~\ref{fig:sol-im}) and 
the $E_{2-}$ and $M_{2-}$ multipoles from JSLT (blue solid curves in figure~\ref{fig:sol-re}).
The upper end of our analysis range is near a potentially new $N^\ast(\sim 1900)$. 
The Kaon-MAID~\cite{mart} and JSLT groups~\cite{bruno} have associated an enhancement
in the $K\Lambda$ cross section near 1.9 GeV with
the $D_{13}$ partial wave, which should resonate in either the $E_{2-}$ or $M_{2-}$ multipoles. 
However, our model-independent analysis excludes such conclusions, since their solutions 
lie outside the experimental bands in these partial waves. 
On the other hand, the BoGa analysis~\cite{boga07} has recently modeled the 
$N^\ast(\sim 1900)$ as a $P_{13}$ resonance, which should manifest itself in either 
the $E_{1+}$ or $M_{1+}$ multipoles. 
The BoGa solution is within the experimental solution bands of 
figures~\ref{fig:sol-re} and~\ref{fig:sol-im}. 
(It is also the only PWA analysis that included the CLAS-g1c and GRAAL data sets in fitting
their model parameters.) 
We can conclude that their assignment is consistent with the experimental solution bands, 
but cannot yet confirm it due to the significant width of these bands.


We have investigated a number of possible ways in which additional data may lead to 
narrower multipole bands and improved amplitude determination. 
For the most part, existing data does not extend to extreme angles (near $0^\circ$ and $180^\circ$),
which in general tend to be more sensitive to interfering multipoles of opposite parities. 
In fact, the best and worst solutions at $E_\gamma = 1122$ MeV ($W=1728$ MeV) exhibit 
a dramatic difference in the predicted unpolarized cross section at $180^\circ$ -- compare 
the solid (black) and dashed (blue) curves in figure~\ref{fig:pre-v1}. 
(The extreme angles of the asymmetries are constrained by symmetry to either 0 or $\pm 1$, 
and so contain little additional information.) 
As a test, we have created mock cross section data at $0^\circ$ and $180^\circ$, 
centered on the best solutions of table~\ref{tab:td3} with a statistical error of 
$\pm 0.03 \mu {\rm b/sr}$. 
When the fits are repeated with these mock points added to the CLAS and GRAAL data sets, 
variations such as seen in figure~\ref{fig:pre-v1} disappear, but few of the resulting bands 
of multipole solutions are improved. 
While the $M_{1+}$, $M_{1-}$ and $E_{2-}$ are slightly narrowed at low energies,
generally, there is little improvement over the trends seen in
figures~\ref{fig:mul-re} and~\ref{fig:mul-im}.

The data in table~\ref{tab:td1} span a significant range in statistical precision. 
From preliminary analyses of data from an ongoing generation of new CLAS experiments 
we can anticipate result on the $\Sigma$, $T$, $O_{x'}$ and $O_{z'}$ asymmetries that will have 
at least an order of magnitude improvement over the GRAAL data set. 
To simulate the effect of such an improvement, we have arbitrarily reduced the statistical 
errors on the GRAAL $\Sigma$, $T$, $O_{x'}$ and $O_{z'}$ asymmetries by a factor of 3 and 
repeated the fits. 
Apart from an increase in $\chi^2$, due to undulations in the angular distributions that 
are now artificially beyond the level of statistical fluctuations, there are no significant 
changes in any of the multipole bands of figures~\ref{fig:mul-re} and~\ref{fig:mul-im}.

\begin{figure}
\centering
\includegraphics[clip,width=0.67\textwidth]{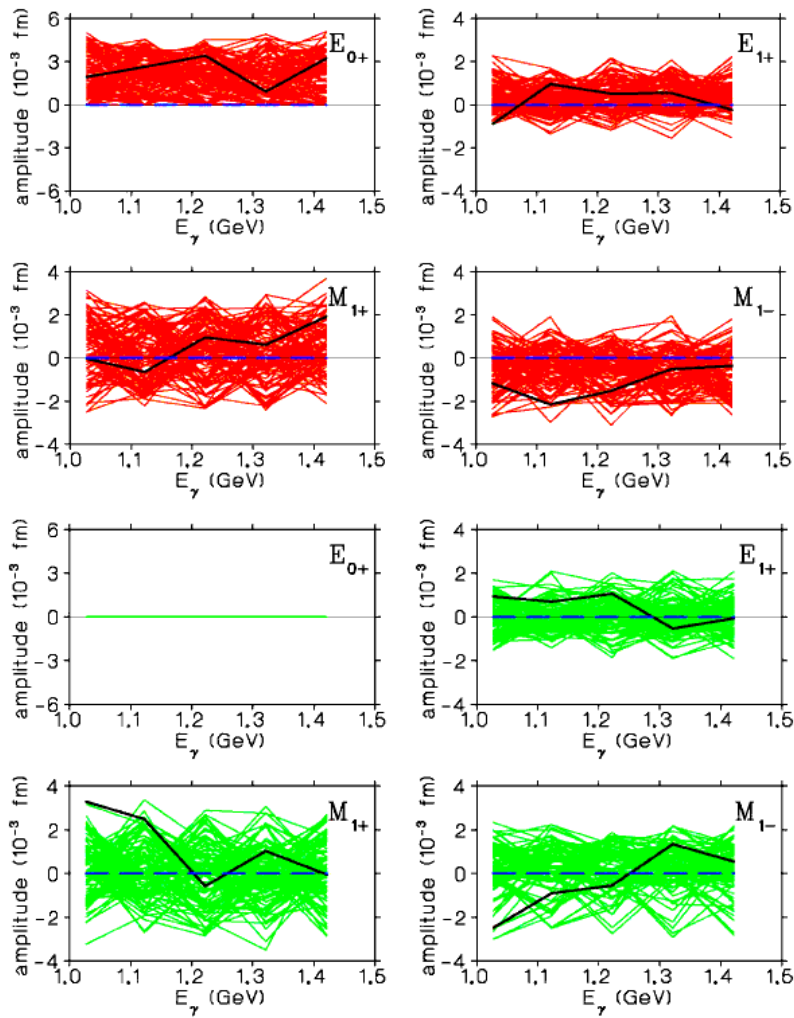}
\caption{\label{fig:mul-v3} (Color online) 
Real parts (top 4 panels in red) and imaginary parts (bottom 4 panels in green) of the $S$
and $P$ wave multipoles, fitted to the CLAS data of table~\ref{tab:td1} 
(excluding the GRAAL measurements).
Solutions with the best (1.07) and largest (1.18) $\chi^2$ are shown 
as solid (black) and long-dashed (blue) curves, respectively.
}
\end{figure}

\begin{figure}
\centering
\includegraphics[clip,width=0.67\textwidth]{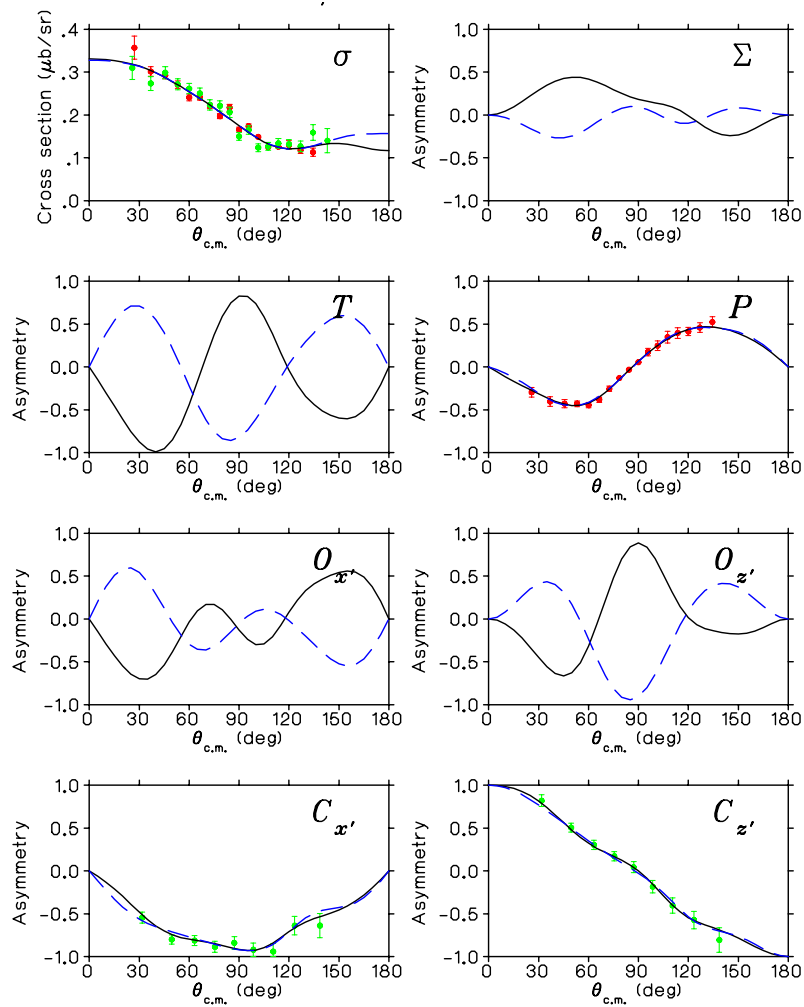}
\caption{\label{fig:pre-v3} (Color online) 
Predictions at $E_\gamma=1421$ MeV ($W=1883$ MeV) from a multipole fit to the CLAS data from
CLAS-g11a~\cite{g11a10} shown in red and CLAS-g1c~\cite{g1c06, g1c07} shown in green, 
excluding the GRAAL results. 
The solid black and long-dashed blue curves show the solutions (figure~\ref{fig:mul-v3}) 
having the minimum (1.07) and largest (1.18) $\chi^2$/point.
}
\end{figure}

Ongoing analyses of new experiments are expected to yield data on all 16 observables. 
The potential impact of such an extensive set is simulated in the next section;
here we can already study the expected trends by
examining the impact 
that the GRAAL measurements of $\left\{\Sigma, T, O_{x'}, O_{z'}\right\}$ have made so far. 
In figure~\ref{fig:mul-v3} we show the $S$ and $P$ wave multipoles obtained if the
GRAAL data are removed from the fitting procedure. 
Comparing these results to figures~\ref{fig:mul-re} and~\ref{fig:mul-im},
it is clear that the $M_{1+}$ band has dramatically 
narrowed with the inclusion of the GRAAL polarization results. 
Lesser but still significant gains occur in the determination of most of the multipoles. 
The range of values for the $\chi^2$/point within these bands are similar to those of 
table~\ref{tab:td3}. 
In figure~\ref{fig:pre-v3} we show the predictions of the band at 1421 MeV beam energy
($W = 1883$ MeV), as represented by the solutions with the minimum 
$\chi^2{\rm /point} = 1.07$ and the maximum $\chi^2{\rm /point} = 1.18$. 
Not surprisingly, predictions for the observables where data have been removed
from the fit are now wildly varied.


There are several conclusions that can be drawn from this analysis, 
along with reasons for genuine hope. 
When the $\chi^2$/point is near or even better than 1, solutions differing in 
the $\chi^2$/point by something like 0.2 are not experimentally distinguishable. 
The existence of bands of multipole solutions, each with small $\chi^2$/point,
indicates a shallow $\chi^2$ surface, pitted with many local minima. 
Certainly the width of the bands evident in figures~\ref{fig:mul-re} and~\ref{fig:mul-im}
precludes using the existing data to \textit{hunt for resonances}. 
However, a comparison of figure~\ref{fig:mul-v3} with figures~\ref{fig:mul-re} and~\ref{fig:mul-im}
indicates the gains resulting from the GRAAL polarization observables are significant,
even though the GRAAL errors are substantially larger than most of the CLAS data.
CLAS data on all 16 photoproduction observables are now under analysis. 
The fact that such data have all been accumulated within a single detector is likely 
to minimize the problems evident in figure~\ref{fig:lbr-sbr}.
Furthermore, with a large number of different observables will come a large number of the Fierz
identities, which can be used to constrain 
and essentially eliminate the effects of systematic scale uncertainties.

\section{\label{sec:mock} The potential of complete experiments -- studies with mock data}
\begin{figure}
\centering
\includegraphics[clip,width=0.67\textwidth]{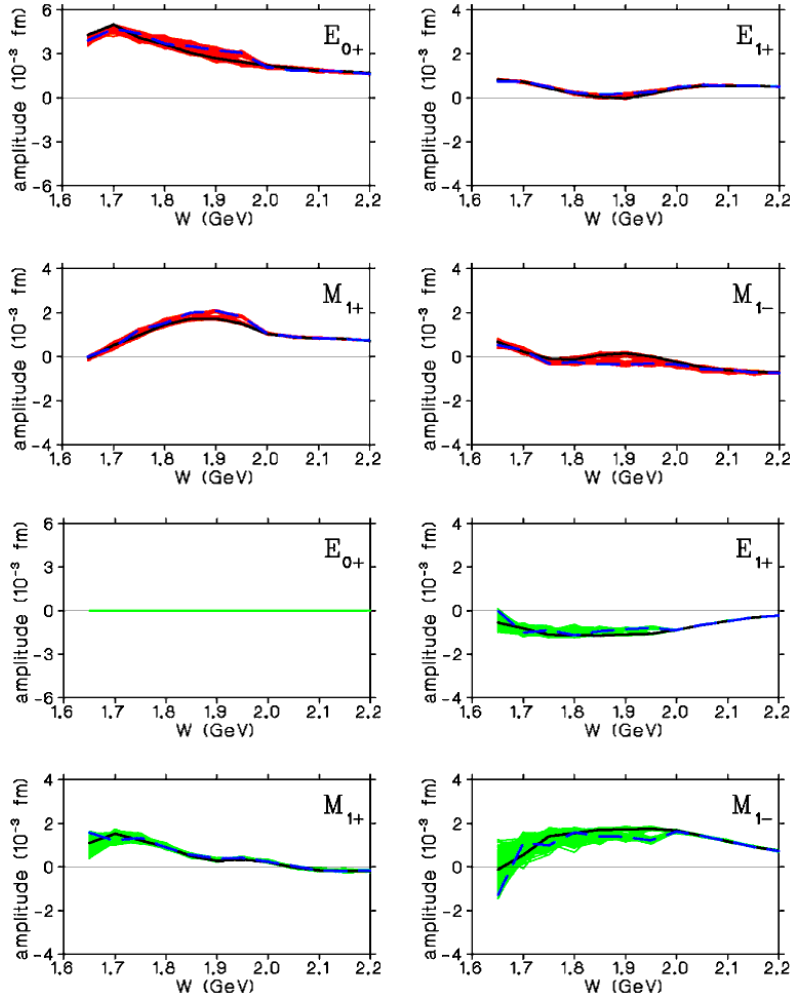}
\caption{\label{fig:mock1} (Color online) 
Real (top four panels in red) and imaginary (bottom four panels in green) parts
of the $S$ and $P$ wave multipoles resulting from fits to mock data with 5\% errors on all 16
observables, with mock data points every $10^\circ$ c.m. Solutions with the best (0.7) and largest (1.3)
$\chi^2/$point are shown as solid (black) and long-dashed (blue) curves, respectively.
}
\end{figure}
\begin{figure}
\centering
\includegraphics[clip,width=0.67\textwidth]{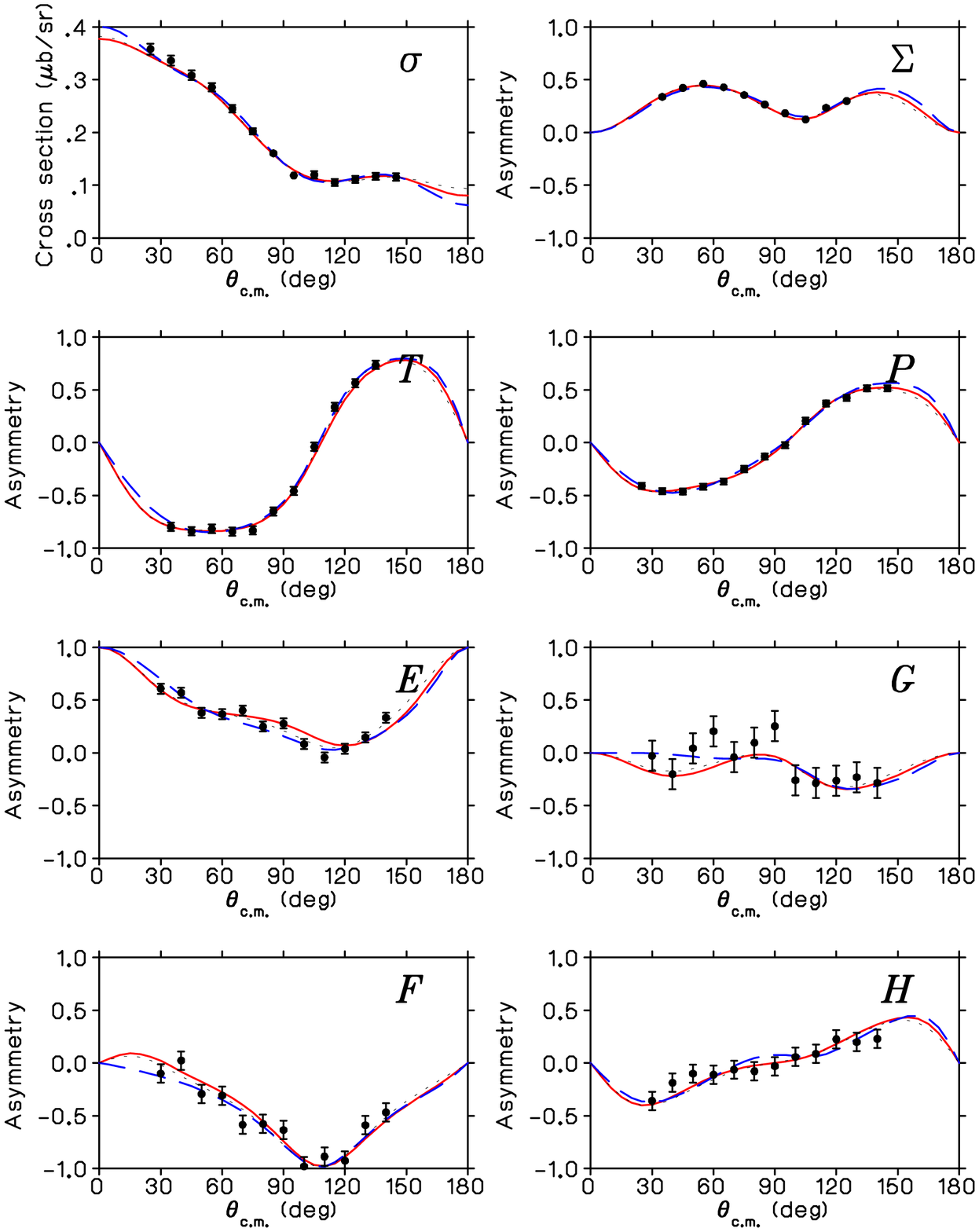}
\caption{\label{fig:mock2} (Color online) 
Mock data on the single- and BT-polarization observables at $W = 1900$ MeV, 
with uncertainties expected from the CLAS set of $K^+\Lambda$ experiment. 
The curves are predictions of multipole solutions fitted to these data with 
the best (0.6) and largest (1.4) $\chi^2$/point, as shown by the solid (red) 
and long-dashed (blue) curves, respectively.
}
\end{figure}
\begin{figure}
\centering
\includegraphics[clip,width=0.67\textwidth]{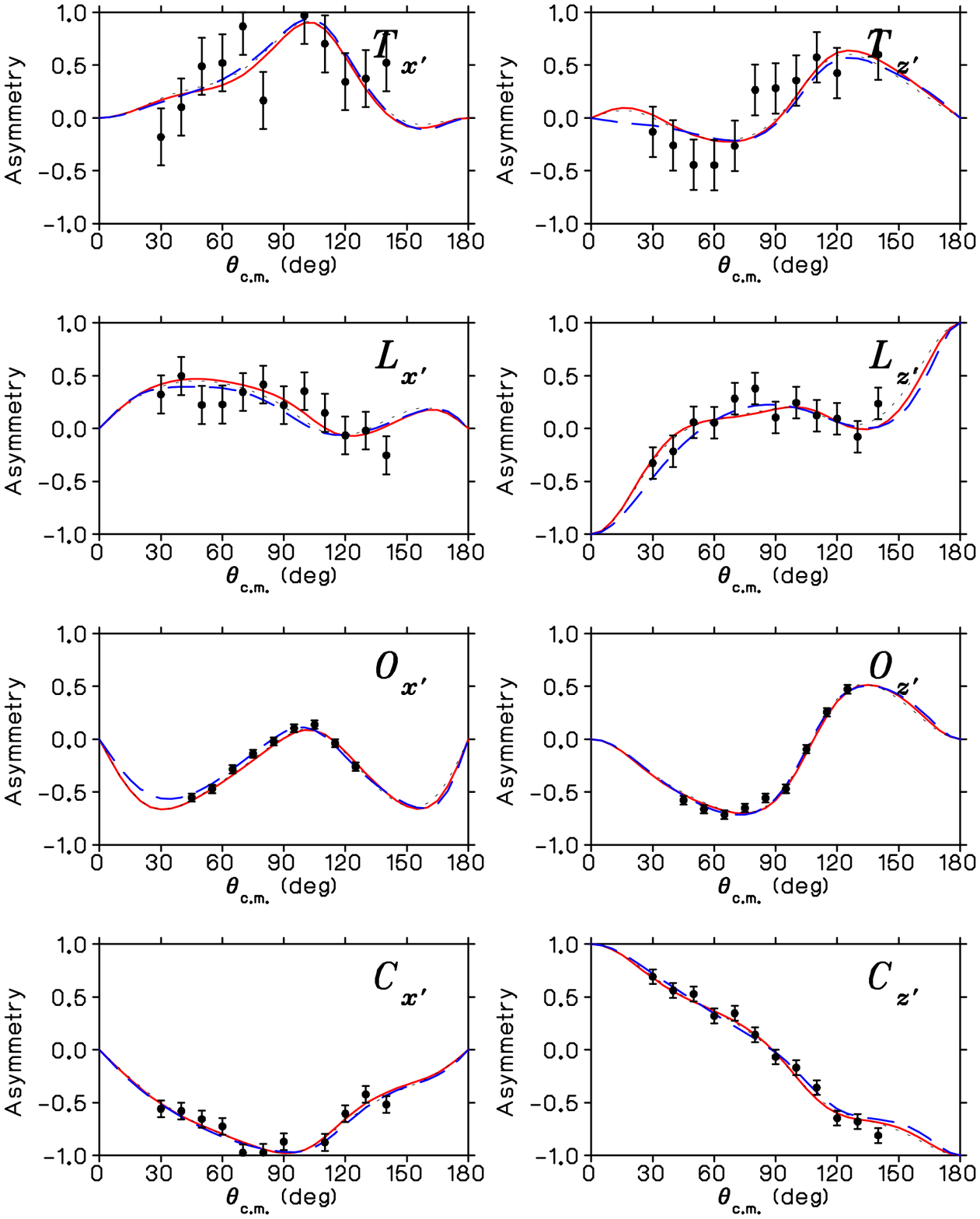}
\caption{\label{fig:mock3} (Color online) 
Mock data on the TR- and BR-polarization observables at $W = 1900$ MeV, 
with uncertainties expected from the CLAS set of $K^+\Lambda$ experiment. 
The curves are predictions of multipole solutions fitted to these data with 
the best (0.6) and largest (1.4) $\chi^2$/point, as shown by the solid (red) 
and long-dashed (blue) curves, respectively.
}
\end{figure}
\begin{figure}
\centering
\includegraphics[clip,width=0.67\textwidth]{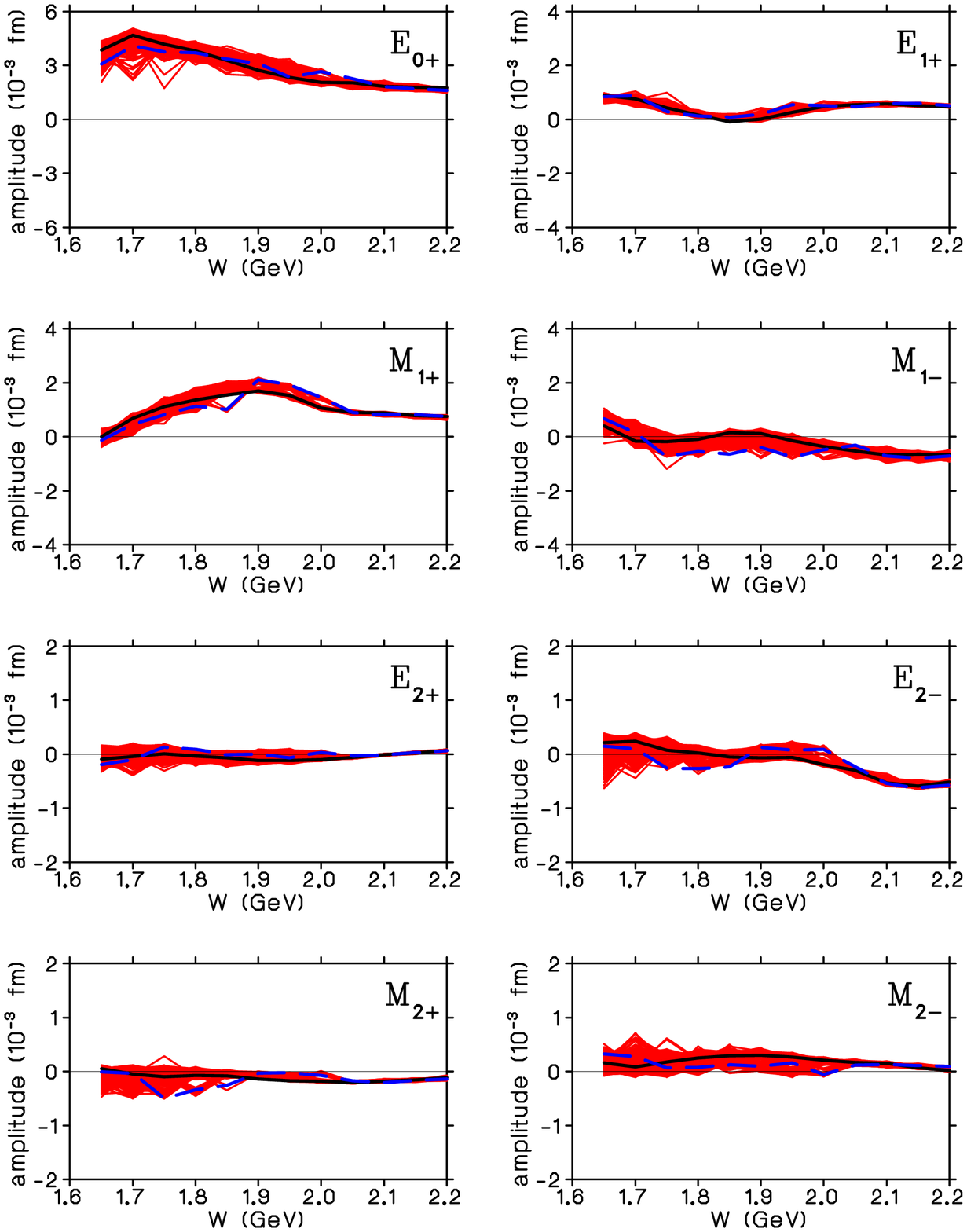}
\caption{\label{fig:mock4} (Color online) 
Real parts of the $S$, $P$ and $D$ wave multipoles resulting from fits to mock
$K^+\Lambda$ data with the precision and kinematic coverage expected from the complete set of CLAS
experiments on all 16 observables. Solutions with the best (typically 0.6) and largest (typically
1.2) $\chi^2$/point are shown as solid (black) and long-dashed (blue) curves, respectively.
}
\end{figure}
\begin{figure}
\centering
\includegraphics[clip,width=0.67\textwidth]{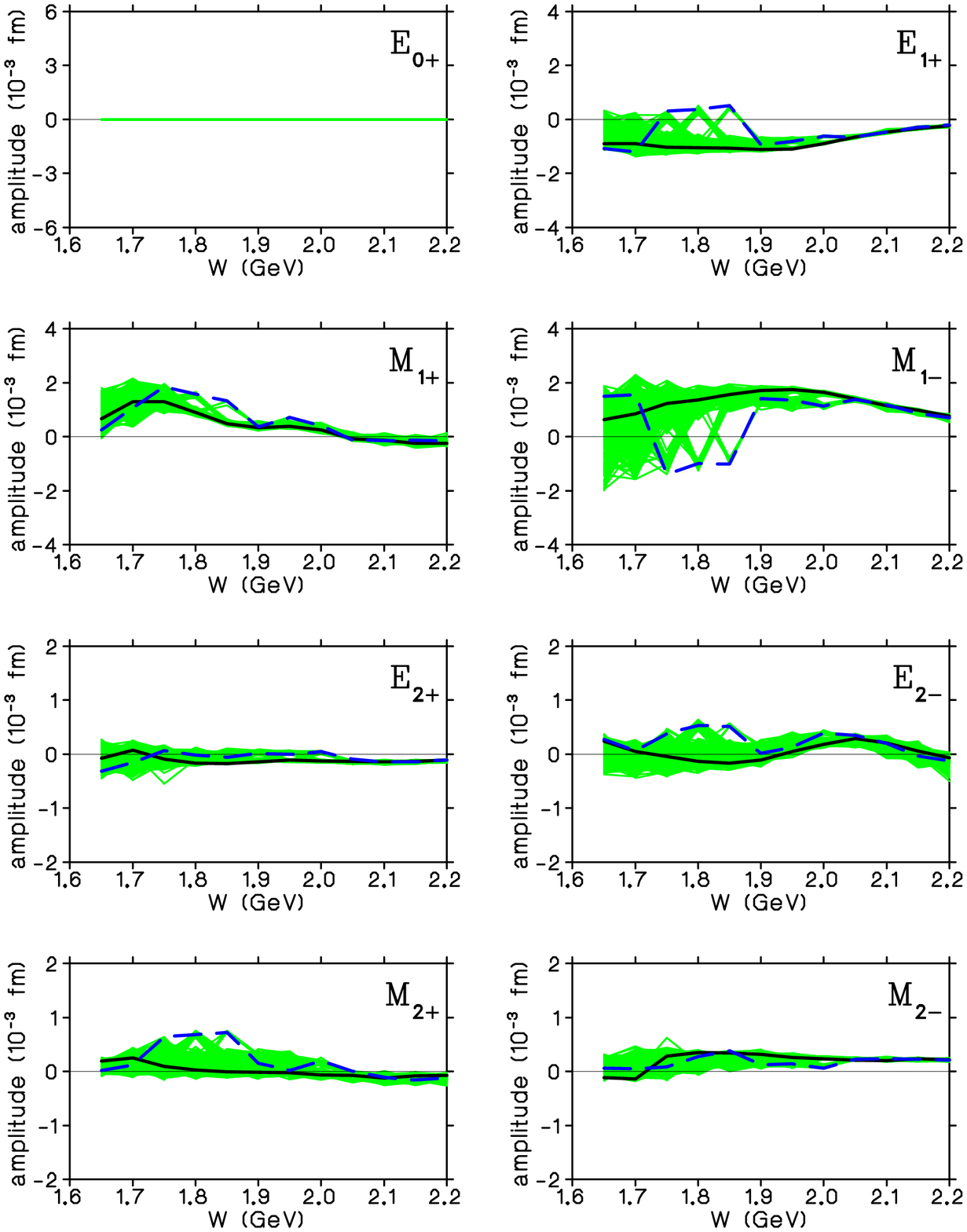}
\caption{\label{fig:mock5} (Color online) 
Imaginary parts of the $S$, $P$ and $D$ wave multipoles resulting from fits to mock
$K^+\Lambda$ data with the precision and kinematic coverage expected from the complete set of CLAS
experiments on all 16 observables. Solutions with the best (typically 0.6) and largest (typically
1.2) $\chi^2$/point are shown as solid (black) and long-dashed (blue) curves, respectively.
}
\end{figure}

To further investigate the potential impact of measuring a complete set of all 16 observables 
on the determination of multipole amplitudes, we have created \textit{mock data} using predictions 
of the BoGa multipoles, Gaussian-smeared to reflect different levels of uncertainty. 
Fitting such mock data with the same procedure described in the previous section, i.e., 
Monte Carlo sampling combined with gradient minimization and a real $E_{0+}$ multipole, 
leads to the following conclusions.
\begin{itemize}
\item 
With data points at every 10 degrees and with 0.1\% errors on each point for every observable, a
level of accuracy that will never be achieved at any facility, two minima are always found, one with
a $\chi^2$/point near 1 and the other substantially larger -- e.g., greater than 50. 
Thus, a unique solution is easily identifiable.
\item 
When the uncertainties on the mock data are increased to 1\% on each point every 10
degrees, a few minima appear. Nonetheless, with the exception of the lowest energies, these are
still widely spaced in $\chi^2$ so that in general the \textit{true solution} can still be identified.
\item 
When the uncertainties on the mock data are increased to 5\%, bands of indistinguishable
solutions from multiple $\chi^2$ minima begin to appear, although the bands are considerably
narrowed from those of figures~\ref{fig:mul-re} and~\ref{fig:mul-im}. 
As an example, the resulting real and imaginary parts of
the $E_{0+}$ to $M_{1-}$ multipoles are shown in figure~\ref{fig:mock1} for 
the c.m. energy range from 1650 to 2200 MeV.
With small errors and bands as narrow as in figure 16, there are typically a few local minima 
for each energy. 
However, the positions of such minima depend on the particular statistical distribution of the 
mock data, due to the complicated structure of the $\chi^2$ space. 
To remove this dependence we have repeated the exercise of creating Gaussian smeared mock 5\% data 
and searching for local minima 300 times, with a different random seed to distribute the mock data each time.
This is the result plotted in figure~\ref{fig:mock1}. 
It should be noted that a real experiment will not have the luxury of being repeated so many times, 
if at all, and so fits to the particular statistical distribution of data that is accumulated will 
have a narrower band width, which will not represent the true uncertainty. 
Nonetheless, the full uncertainty can readily be determined by simulation.
\item 
While actual data sets may attain 5\% uncertainties on some observables, 
others will be considerably larger, notably those involving polarized targets 
which are always significantly shorter than liquid targets. 
To consider a more realistic collection of data, we have created Gaussian-smeared mock data 
with a kinematic coverage typical of the CLAS detector at Jefferson Lab, 
using uncertainties on liquid target measurements taken from the CLAS g1c~\cite{g1c07}, g8~\cite{ire10} 
and g11a~\cite{g11a10} data sets, and with polarized target data errors estimated for 
the g9-FROST running period. 
As an example, the resulting mock data with expected CLAS uncertainties at $W = 1900$ MeV 
are shown in figures~\ref{fig:mock2} and~\ref{fig:mock3}. 
The multipole bands resulting from fits to these mock data are plotted in figures~\ref{fig:mock4} 
and~\ref{fig:mock5}. 
As with the 5\% error study, the Gaussian smearing followed by Monte Carlo and minimization to search 
for local minima has been repeated 300 times to avoid the dependence on the starting 
distribution of the data. 
This had a smaller effect in the resulting multipole bands of figures~\ref{fig:mock4} 
and~\ref{fig:mock5}, since the errors are somewhat larger than the 5\% case of figure~\ref{fig:mock1}. 
Compared to figures~\ref{fig:mul-re} and~\ref{fig:mul-im}, the multipole bands of 
figures~\ref{fig:mock4} and~\ref{fig:mock5} are dramatically narrower. 
Almost all multipoles are well determined. Some, like the imaginary part of
the $M_{1-}$, remain broad at low energies.
But all are well defined above above 1.9 GeV where unobserved $N^\ast$ states
are predicted in various calculations.
From extensive studies we attribute this mainly to the larger number of observables rather than 
to increased statistics on any specific asymmetry. 
These studies give us confidence in the expectation of a well determined amplitude 
from complete experiments, such as those from CLAS. 
This will be a truly significant milestone after over 
fifty years of photo-production experiments.
\end{itemize}

\section{\label{sec:summary} Summary}
It is anticipated that data will soon be available on all 16 pseudoscalar meson 
photoproduction observables from a new generation of ongoing experiments, certainly 
for $K \Lambda$ final states and possibly for $\pi N$ channels as well. 
This will significantly reduce the model dependence in the study of excited baryon 
structure by providing a total amplitude that is experimentally determined to within a phase. 
Such an experimental amplitude can be utilized at two levels, first as a test to validate total
amplitudes associated with different models and second as a starting point that can be 
analytically continued into the complex plane to search for poles. 
Here we have laid the ground work for this by assembling a complete and consistent
set of equations needed for amplitude reduction from experiment and have
demonstrated the first stage of interaction with theoretical models.

In summary, we have used direct numerical evaluations, 
(\ref{eq:spin-mx})-(\ref{eq:general-dcs}), to verify the most general analytic 
form of the cross section, dependent on the three polarization vectors of the beam, target 
and recoil baryon, including all single, double and triple-polarization terms involving 
the 16 possible spin-dependent observables~(\ref{eq:gcs}). 
(Copies of the associated computer code are available upon request~\cite{hkamano}.)
We have explicitly listed the experimental measurements needed to construct each observable 
in pseudoscalar meson photoproduction (\ref{apx_tab}) and 
provided a consistent set of equations relating 
these quantities to the CGLN amplitudes~(\ref{eq:obs-cgln-begin})-(\ref{eq:obs-cgln-end}), 
and from these to electromagnetic multipoles~(\ref{eq:mp-f1})-(\ref{eq:mp-f4}).
From a review of some of the more frequently quoted works in this field,
we have found that the same symbol for a polarization asymmetry has been used by different
authors to refer to different experimental quantities; the magnitudes remain
the same across published works, but their signs vary (section~\ref{sec:review}).
For example, the definitions of the six observables 
$H$, $C_{x'}$, $C_{z'}$, $O_{x'}$, $O_{z'}$ and $L_{x'}$ in the MAID and SAID on-line PWA codes
is the negative of that used by BoGa and the present work.
This has already lead to confusion in the analysis of recent double-polarization data 
(figure~\ref{fig:cfits}).

We have used the assembled machinery to carry out a multipole analysis of 
the $\gamma p \to K^+ \Lambda$ reaction, free of model assumptions, and examined the impact 
of recently published measurements on 8 different observables. 
We have used a combined Monte Carlo sampling of the amplitude space, with gradient minimization, 
and have found a shallow $\chi^2$ valley pitted with a very large number of local minima.
This results in broad bands of multipole solutions, which are experimentally 
indistinguishable (figures~\ref{fig:mul-re} and~\ref{fig:mul-im}). 
Comparing to models that have recently reported a new $N^\ast(\sim 1900)$, we can
exclude PWA that incorporate a new $D_{13}$ since their amplitudes lie outside the 
model-independent solution bands in the associated multipoles. 
(These PWA were carried out before most of the data used in our analysis were available.)
Recent BoGa analyses have modeled the $N^\ast(\sim 1900)$ as a $P_{13}$ resonance. 
While their solution lies within our experimental multipole bands, we cannot yet validate it due 
to the significant width of the bands.

From our studies with published measurements, 
as well as simulations with mock data, we have seen that clusters of local 
minima in $\chi^2$ are often present. 
With the current collection of results on 8 observables, these minima are completely degenerate 
and experimentally indistinguishable. 
In studies with mock data we have seen that this degeneracy can be removed with high precision 
data on a large number of observables (section~\ref{sec:mock}). 
As determined in the present analysis, a greater number of different observables tend to be more effective 
in creating a global minimum than higher precision. 
We conclude that, while a general solution to the problem of determining an amplitude free of 
ambiguities may require 8 observables, as has been discussed by CT~\cite{chiang}, 
such requirements assume data of arbitrarily high precision. 
Experiments with realistically achievable uncertainties will require a significantly larger number.
Simulations using mock data with statistics comparable to what is anticipated from the new generation of
CLAS experiments reconstruct narrow bands that are quite well defined for almost all multipoles 
(figures~\ref{fig:mock4} and~\ref{fig:mock5}).
We expect such results to create a watershed in our understanding of the nucleon spectrum.

\ack
This work was supported
by the U.S. Department of Energy, Office of Nuclear Physics Division,
under Contract No. DE-AC05-060R23177
under which Jefferson Science Associates operates Jefferson Laboratory,
and also by the U.S. Department of Energy,
Office of Nuclear Physics Division, under
Contract No. DE-AC02-06CH11357 and
Contract No. DE-FG02-97ER41025.

\appendix

\section{\label{apx_eqs}General expression for the differential cross section 
with fixed polarizations}

We summarize here the derivation of an analytic expression for the differential cross section 
in pseudoscalar meson photoproduction with general values of the beam, target 
and recoil polarization. 
Following the formalism of the spin density matrices described by 
FTS~\cite{fasano}, one can write the general cross section~(\ref{eq:general-dcs}) as,
\begin{eqnarray}
d\sigma^{{\rm B,T,R}}(\vec P^\gamma,\vec P^T,\vec P^R) &=&
\rho_0(\rho^R)_{kn} (F_\mu)_{nm} (\rho^T)_{ml} (F^\dag_{\lambda})_{lk}(\rho^\gamma)_{\mu\lambda}.
\label{dcs-fts}
\end{eqnarray}
(Throughout this appendix the same indices in equations imply taking summation.)
Here $\rho_0=k/q$; 
$(F_\lambda)_{m_{s_\Lambda}m_{s_N}} = \bra{m_{s_\Lambda}}F_{{\rm CGLN}}\ket{m_{s_N}}$, in which 
the spin states of the initial and final baryons are quantized in the $z$-direction and 
the (unit) photon polarization vector is taken to be circularly polarized with 
the helicity $\lambda$.

The $2\times 2$ spin density matrix $\rho^X$ for $X=\gamma,T,R$ is given by
\begin{eqnarray}
\rho^\gamma &=& \frac{1}{2}[\mathbf{1}+\vec {\cal P}^\gamma \cdot \vec\sigma],
\label{rhog}
\\
\rho^T &=& \frac{1}{2}[\mathbf{1}+\vec P^T \cdot \vec\sigma],
\label{rhot}
\\
\rho^R &=& \frac{1}{2}[\mathbf{1}+\vec P^R \cdot \vec\sigma],
\label{rhor}
\end{eqnarray}
where $\vec\sigma$ is the Pauli spin vector, as in (\ref{eq:pauli}), and 
$\vec {\cal P}^\gamma$ is the so-called Stokes vector for the photon polarizations~\cite{fasano}.
Note that in the $x$-$y$-$z$ coordinate (see figure~\ref{fig:coord}),
$\vec {\cal P}^\gamma = (-P^\gamma_L \cos 2\phi_\gamma, -P^\gamma_L\sin 2\phi_\gamma, P^\gamma_c)$.

Substituting (\ref{rhog})-(\ref{rhor}) into (\ref{dcs-fts}), we have
\begin{eqnarray}
d\sigma^{{\rm B,T,R}}(\vec P^\gamma,\vec P^T,\vec P^R) 
&=&
\rho_0 \frac{1}{2}(\mathbf{1}+\vec P^R\cdot\vec\sigma)_{kn} 
(F_\mu)_{nm} \frac{1}{2}(\mathbf{1}+\vec P^T\cdot\vec\sigma)_{ml}(F^\dag_{\lambda})_{lk}
\nonumber\\
&&
\times
\frac{1}{2}(\mathbf{1}+\vec {\cal P}^\gamma\cdot\vec\sigma)_{\mu\lambda}
\nonumber\\
&=&
\frac{\rho_0}{8}
(\mathbf{1}+\vec P^R\cdot\vec\sigma)_{kn} 
\left[
(F_\lambda)_{nm}(F^\dag_{\lambda})_{mk}
\right.
\nonumber\\
&&
+
(F_\mu)_{nm} (F^\dag_{\lambda})_{mk}\vec {\cal P}^\gamma\cdot\vec\sigma_{\mu\lambda}
+
(F_\lambda)_{nm} \vec P^T\cdot\vec\sigma_{ml} (F^\dag_{\lambda})_{lk}
\nonumber\\
&&
+
\left.
(F_\mu)_{nm} \vec P^T \cdot\vec\sigma_{ml} 
(F^\dag_{\lambda})_{lk}\vec {\cal P}^\gamma\cdot\vec\sigma_{\mu\lambda}
\right].
\end{eqnarray}
Noting that $d\sigma_0 = (\rho_0 /4) {\cal N}$ where 
${\cal N}= (F_\lambda)_{nm}(F^\dag_{\lambda})_{mn}$, we can further expand the above equation as
\begin{eqnarray}
d\sigma^{{\rm B,T,R}}(\vec P^\gamma,\vec P^T,\vec P^R) &=&
\frac{d\sigma_0}{2}
\left\{
1 
+
(\vec {\cal P}^\gamma)^a 
\frac{(F_\mu)_{kn}(F^\dag_{\lambda})_{nk} \sigma^a_{\mu\lambda}}{{\cal N}}
\right.
\nonumber\\
&&
\quad\qquad
+
(\vec P^T)^a 
\frac{(F_\lambda)_{kn} \sigma^a_{nm} (F^\dag_{\lambda})_{mk}}{{\cal N}}
\nonumber\\
&&
\quad\qquad
+
(\vec P^R)^{a'} 
\frac{\sigma^{a'}_{kn}(F_\lambda)_{nm} (F^\dag_{\lambda})_{mk}}{{\cal N}}
\nonumber\\
&&
\quad\qquad
+ 
(\vec P^T)^a (\vec {\cal P}^\gamma)^b 
\frac{(F_\mu)_{km} \sigma^a_{ml}(F^\dag_{\lambda})_{lk}\sigma^b_{\mu\lambda}}{{\cal N}}
\nonumber\\
&&
\quad\qquad
+
(\vec P^R)^{a'} (\vec {\cal P}^\gamma)^b
\frac{\sigma^{a'}_{kn}(F_\mu)_{nl} (F^\dag_{\lambda})_{lk}\sigma^b_{\mu\lambda}}{{\cal N}}
\nonumber\\
&&
\quad\qquad
+
(\vec P^R)^{a'} (\vec P^T)^a
\frac{\sigma^{a'}_{kn}(F_\lambda)_{nm} \sigma^a_{ml}(F^\dag_{\lambda})_{lk}}{{\cal N}}
\nonumber\\
&&
\quad\qquad
+
\left.
(\vec P^R)^{a'} (\vec P^T)^a (\vec {\cal P}^\gamma)^b
\frac{\sigma^{a'}_{kn}(F_\mu)_{nm} \sigma^a_{ml}(F^\dag_{\lambda})_{lk}\sigma^b_{\mu\lambda}}{{\cal N}}
\right\}
\nonumber\\
&=&
\frac{d\sigma_0}{2}
\left\{
1 
+
(\vec {\cal P}^\gamma)^a {\bSigma}^a
+
(\vec P^T)^a \bi{T}^a
+ 
(\vec P^R)^{a'} \bi{P}^{a'}
\right.
\nonumber\\
&&
+ (\vec P^T)^a    (\vec {\cal P}^\gamma)^b C^{{\rm BT}}_{ab}
+ (\vec P^R)^{a'} (\vec {\cal P}^\gamma)^a C^{{\rm BR}}_{ab}
\nonumber\\
&&
\left.
+ (\vec P^R)^{a'} (\vec P^T)^a C^{{\rm TR}}_{a'b}
+ (\vec P^R)^{a'} (\vec P^T)^a (\vec {\cal P}^\gamma)^b C^{{\rm BTR}}_{a'ab}
\right\}.
\label{eq:dcs2}
\end{eqnarray}
In the last step we have introduced
\begin{eqnarray}
{\bSigma}^a &=&
\frac{(F_\mu)_{mn}(F^\dag_{\lambda})_{nm} \sigma^a_{\mu\lambda}}{{\cal N}},
\label{eq:s}
\\
\bi{T}^a &=&
\frac{(F_\lambda)_{kn} \sigma^a_{nm} (F^\dag_{\lambda})_{mk}}{{\cal N}},
\label{eq:t}
\\
\bi{P}^{a'} &=&
\frac{\sigma^{a'}_{kn}(F_\lambda)_{nm} (F^\dag_{\lambda})_{mk}}{{\cal N}},
\label{eq:p}
\\
C^{{\rm BT}}_{ab}&=&
\frac{ (F_\mu)_{kn} \sigma^a_{nm} (F^\dag_{\lambda})_{mk}\sigma^b_{\mu\lambda}}{{\cal N}},
\label{eq:cbt}
\\
C^{{\rm BR}}_{a'b}&=&
\frac{ \sigma^{a'}_{kn} (F_\mu)_{nm} (F^\dag_{\lambda})_{mk} \sigma^a_{\mu\lambda}}{{\cal N}},
\label{eq:cbr}
\\
C^{{\rm TR}}_{a'b}&=&
\frac{ \sigma^{a'}_{kn} (F_\lambda)_{nm} \sigma^a_{ml} (F^\dag_{\lambda})_{lk}}{{\cal N}},
\label{eq:ctr}
\\
C^{{\rm BTR}}_{a'ab}&=&
\frac{ \sigma^{a'}_{kn} (F_\mu)_{nm} \sigma^a_{ml}(F^\dag_{\lambda})_{lk}\sigma^b_{\mu\lambda}}{{\cal N}}.
\label{eq:cbtr}
\end{eqnarray}
In (\ref{eq:dcs2})-(\ref{eq:cbtr})
the Pauli matrices that are combined in products with beam and target
polarizations are defined in reference to the unprimed $x,y,z$ coordinates of figure~\ref{fig:coord}
with the beam along $+\hat z$, so that
\begin{eqnarray*}
\vec {\cal P}^{\gamma} \cdot \vec\sigma 
&=&(\vec {\cal P}^{\gamma})^a  \sigma^a 
\equiv(\vec {\cal P}^{\gamma})^x \sigma^x
+(\vec {\cal P}^{\gamma})^y \sigma^y
+(\vec {\cal P}^{\gamma})^z \sigma^z,
\end{eqnarray*}
\begin{eqnarray*}
\vec P^T \cdot \vec\sigma 
&=&(\vec P^T)^a  \sigma^a 
\equiv(P^T_x) \sigma^x
+(P^T_y) \sigma^y
+(P^T_z) \sigma^z .
\end{eqnarray*}
However, the Pauli matrices appearing in products with the recoil polarization vector are defined
in reference to the primed $x',y',z'$ coordinates of figure~\ref{fig:coord}
with the meson momentum along $+\hat z'$, so that
\begin{eqnarray*}
\vec P^R \cdot \vec\sigma 
&=&(\vec P^R)^{a'}  \sigma^{a'} 
\equiv(P^R_{x'}) \sigma^{x'}
+(P^R_{y'}) \sigma^{y'}
+(P^R_{z'}) \sigma^{z'}.
\end{eqnarray*}
[If the unprimed $x,y,z$ coordinates were also used to expand
$\vec P^R \cdot \vec\sigma$, then one would obtain a corresponding set of unprimed
observables that are related via equation~(\ref{eq:br-1}).]

We note that ${\bSigma}^a$, $\bi{T}^a$, $\bi{P}^a$, $C^{{\rm BT}}_{ab}$, $C^{{\rm BR}}_{a'b}$, 
and $C^{{\rm TR}}_{a'b}$ are exactly the same as those defined 
in \cite{fasano}.
The $C^{{\rm BTR}}_{a'ab}$ term was not included in \cite{fasano}, 
which did not consider the triple polarization case.
Each component in (\ref{eq:s})-(\ref{eq:cbtr}) can be related with
16 observables defined in tables~\ref{tab:1pol}-\ref{tab:trpol} of \ref{apx_tab}:

\begin{equation}
{\bSigma}^{x_B} = \Sigma,\qquad \bi{T}^{y_T} = T,\qquad \bi{P}^{y'_R} = P,
\end{equation}
\begin{equation}
\eqalign{
C^{{\rm BT}}_{z_Tz_B}= -E,\qquad C^{{\rm BT}}_{z_Ty_B}= -G,\qquad C^{{\rm BT}}_{x_Tz_B}=  F,\cr
C^{{\rm BT}}_{x_Ty_B}= -H,\qquad C^{{\rm BT}}_{y_Tx_B}=  P,
}
\end{equation}
\begin{equation}
\eqalign{
C^{{\rm BR}}_{z'_Rz_B}=  C_{z'},\qquad C^{{\rm BR}}_{z'_Ry_B}= -O_{z'},\qquad C^{{\rm BR}}_{x'_Rz_B}=  C_{x'},\cr
C^{{\rm BR}}_{x'_Ry_B}= -O_{x'},\qquad C^{{\rm BR}}_{y'_Rx_B}=  T,
}
\end{equation}
\begin{equation}
\eqalign{
C^{{\rm TR}}_{z'_Rz_T}=  L_{z'},\qquad C^{{\rm TR}}_{z'_Rx_T}=  T_{z'},\qquad C^{{\rm TR}}_{x'_Rz_T}=  L_{x'},\cr 
C^{{\rm TR}}_{x'_Rx_T}=  T_{x'},\qquad C^{{\rm TR}}_{y'_Ry_T}=  T,
}
\end{equation}
\begin{equation}
\eqalign{
C^{{\rm BTR}}_{y'_Rx_Ty_B}= -E,\qquad C^{{\rm BTR}}_{y'_Rx_Tz_B}=  G,\qquad C^{{\rm BTR}}_{y'_Rz_Ty_B}= -F,\cr
C^{{\rm BTR}}_{y'_Rz_Tz_B}= -H,\qquad C^{{\rm BTR}}_{x'_Ry_Ty_B}= -C_{z'},\qquad C^{{\rm BTR}}_{x'_Ry_Tz_B}= -O_{z'},\cr
C^{{\rm BTR}}_{z'_Ry_Ty_B}=  C_{x'},\qquad C^{{\rm BTR}}_{z'_Ry_Tz_B}=  O_{x'},\qquad C^{{\rm BTR}}_{x'_Rx_Tx_B}=  L_{z'},\cr
C^{{\rm BTR}}_{x'_Rz_Tx_B}= -T_{z'},\qquad C^{{\rm BTR}}_{z'_Rx_Tx_B}= -L_{x'},\qquad C^{{\rm BTR}}_{z'_Rz_Tx_B}=  T_{x'},\cr
C^{{\rm BTR}}_{y'_Ry_Tx_B}=  1.
}
\end{equation}
Here, all other components not explicitly shown are identically zero, due to
symmetry constraints.

Finally, we also note that the spin density matrices~(\ref{rhog})-(\ref{rhor})
can be expressed as
\begin{eqnarray}
\rho^\gamma &=& \sum_{\hat P = \hat P^\gamma_1,\hat P^\gamma_2}\mathfrak{p}^\gamma_{\hat P}
\frac{1}{2}[\mathbf{1} + \hat{\cal P}_{\hat P}\cdot \vec \sigma] , 
\label{rhog2}\\
\rho^T &=& \sum_{\hat Q = \pm \hat P^T}\mathfrak{p}^T_{\hat Q}
\frac{1}{2}[\mathbf{1} + \hat Q\cdot \vec \sigma] ,
\label{rhot2}\\
\rho^R &=& \sum_{\hat R = \pm \hat P^R}\mathfrak{p}^R_{\hat R}
\frac{1}{2}[\mathbf{1} + \hat R\cdot \vec \sigma] .
\label{rhor2}
\end{eqnarray}
Here $\mathfrak{p}^X_{\hat P}$ is the probability of observing particle $X$ polarized in 
the $\hat P$ direction; 
$\hat{\cal P}_{\hat P}$ is the Stokes vector specified by the unit photon polarization
vector $\hat P$; $\hat P^\gamma_2$ is a unit photon polarization vector perpendicular 
to $\hat P^\gamma_1 \equiv \hat P^\gamma$ for linearly polarized photons, while
$\hat P^\gamma_1$ and $\hat P^\gamma_2$ express two different helicity states
for circularly polarized photons.
The non-unit polarization vectors can be expressed with the unit polarization vectors as 
$\vec P^\gamma =(\mathfrak{p}^\gamma_{\hat P^\gamma_1}-\mathfrak{p}^\gamma_{\hat P^\gamma_2})\hat P^\gamma$,
$\vec P^T = (\mathfrak{p}^T_{+\hat P^T}-\mathfrak{p}^T_{-\hat P^T} ) \hat P^T$, and
$\vec P^R = (\mathfrak{p}^R_{+\hat P^R}-\mathfrak{p}^R_{-\hat P^R} ) \hat P^R$.
Substituting (\ref{rhog2})-(\ref{rhor2}) into (\ref{dcs-fts}),
one obtain the relation between the general cross sections with unit and non-unit
polarization vectors~(\ref{eq:general-dcs}).

\section{\label{apx_tab} Constructing observables from measurements}

We tabulate here the pairs of measurements needed to construct each of the
16 transverse photoproduction observables in terms of the polarization orientation 
angles of figure~\ref{fig:coord}. 
The photon beam is characterized either by its helicity, $h_\gamma$ for circular polarization,
or by $\phi_\gamma^L$ for linear polarization. 
Assuming 100\% polarizations, each observable $\hat A = Ad\sigma_0 $
is determined by a pair of
measurements, each denoted as $\sigma(B,T,R)$; 
``\upa'' indicates the
need to average over the initial spin states of the target and/or beam, and
to sum over the final spin states of the recoil baryon.
For observables involving only beam and/or target polarizations,
$d\sigma_0=(1/2)(\sigma_1+\sigma_2)$ and
$\hat{A} = (1/2)(\sigma_{1} - \sigma_{2})$.
For observables involving the final state recoil polarization,
$d\sigma_0=(\sigma_1+\sigma_2)$ and
$\hat{A} = (\sigma_{1} - \sigma_{2})$.

\begin{table}[hb]
\caption{\label{tab:1pol} The cross section and the observables
involving only one polarization in their leading terms in equation~(\ref{eq:gcs});
$d\sigma_0=\beta(\sigma_1+\sigma_2)$ and $\hat A=\beta(\sigma_1-\sigma_2)$,
where $\beta = 1$ ($\beta=1/2$) if recoil polarization is (is not) observed.}
\begin{indented}
\item[] \begin{tabular}{@{}llcccccc}
\br
\multicolumn{2}{l}{$d\sigma_0$, $\Sigma$, $T$, $P$} & \multicolumn{2}{c}{Beam} &
\multicolumn{2}{c}{Target} & \multicolumn{2}{c}{Recoil} \\ 
Observable & $(\sigma_{1}-\sigma_{2})$ & $h_{\gamma}$ & $\phi_{\gamma}^{L}$ &
$\theta_{p}$ & $\phi_{p}$ & $\theta_{p'}$ & $\phi_{p'}$ \\ 
\mr
$d\sigma_{0}$ & & \upa & \upa & \upa & \upa & \upa & \upa \\
&&&&&&&\\
$2\hat{\Sigma}$ & $\sigma_{1} = \sigma(\perp,0,0)$ & - & $\pi/2$ & \upa & \upa & \upa & \upa \\
 & $\sigma_{2} = \sigma(\parallel,0,0)$ & - & 0 & \upa & \upa & \upa & \upa \\
&&&&&&&\\
$2\hat{T}$ & $\sigma_{1} = \sigma(0,+y,0)$ & \upa & \upa & $\pi/2$ & $\pi/2$ & \upa & \upa \\
 & $\sigma_{2} = \sigma(0,-y,0)$ & \upa & \upa & $\pi/2$ & $3\pi/2$ & \upa & \upa \\
&&&&&&&\\
$\hat{P}$ & $\sigma_{1} = \sigma(0,0,+y')$ & \upa & \upa & \upa & \upa & $\pi/2$ & $\pi/2$ \\
& $\sigma_{2} = \sigma(0,0,-y')$ & \upa & \upa & \upa & \upa & $\pi/2$ & $3\pi/2$ \\
\br
\end{tabular}
\end{indented}
\end{table}

\begin{table}[tb]
\caption{\label{tab:btpol} Observables involving both beam and target polarizations
in their leading terms in equation~(\ref{eq:gcs});
$d\sigma_0=(1/2)(\sigma_1+\sigma_2)$ and $\hat A=(1/2)(\sigma_1-\sigma_2)$.}
\begin{indented}
\item[] \begin{tabular}{@{}llcccccc} 
\br
\multicolumn{2}{l}{$B$-$T$} & \multicolumn{2}{c}{Beam} &
\multicolumn{2}{c}{Target} & \multicolumn{2}{c}{Recoil} \\
Observable & $(\sigma_{1}-\sigma_{2})$ & $h_{\gamma}$ & $\phi_{\gamma}^{L}$ &
$\theta_{p}$ & $\phi_{p}$ & $\theta_{p'}$ & $\phi_{p'}$ \\ 
\mr
$2\hat{E}$ & $\sigma_{1} = \sigma(+1,-z,0)$ & $+1$ & - & $\pi$ & 0 & \upa & \upa \\
 & $\sigma_{2} = \sigma(+1,+z,0)$ & $+1$ & - & 0 & 0 & \upa & \upa \\ 
$2\hat{E}$ & $\sigma_{1} = \sigma(+1,-z,0)$ & $+1$ & - & $\pi$ & 0 & \upa & \upa \\ 
 & $\sigma_{2} = \sigma(-1,-z,0)$ & $-1$ & - & $\pi$ & 0 & \upa & \upa \\
&&&&&&&\\
$2\hat{G}$ & $\sigma_{1} = \sigma(+\pi/4,+z,0)$ & - & $\pi/4$ & 0 & 0 & \upa & \upa \\ 
 & $\sigma_{2} = \sigma(+\pi/4,-z,0)$ & - & $\pi/4$ & $\pi$ & 0 & \upa & \upa \\
$2\hat{G}$ & $\sigma_{1} = \sigma(+\pi/4,+z,0)$ & - & $\pi/4$ & 0 & 0 & \upa & \upa \\ 
 & $\sigma_{2} = \sigma(-\pi/4,+z,0)$ & - & $3\pi/4$ & 0 & 0 & \upa & \upa \\ 
&&&&&&&\\
$2\hat{F}$ & $\sigma_{1} = \sigma(+1,+x,0)$ & $+1$ & - & $\pi/2$ & 0 & \upa & \upa \\ 
 & $\sigma_{2} = \sigma(-1,+x,0)$ & $-1$ & - & $\pi/2$ & 0 & \upa & \upa \\
$2\hat{F}$ & $\sigma_{1} = \sigma(+1,+x,0)$ & $+1$ & - & $\pi/2$ & 0 & \upa & \upa \\ 
 & $\sigma_{2} = \sigma(+1,-x,0)$ & $+1$ & - & $\pi/2$ & $\pi$ & \upa & \upa \\ 
&&&&&&&\\
$2\hat{H}$ & $\sigma_{1} = \sigma(+\pi/4,+x,0)$ & - & $\pi/4$ & $\pi/2$ & 0 & \upa & \upa \\
 & $\sigma_{2} = \sigma(-\pi/4,+x,0)$ & - & $3\pi/4$ & $\pi/2$ & 0 & \upa & \upa \\ 
$2\hat{H}$ & $\sigma_{1} = \sigma(+\pi/4,+x,0)$ & - & $\pi/4$ & $\pi/2$ & 0 & \upa & \upa \\ 
 & $\sigma_{2} = \sigma(+\pi/4,-x,0)$ & - & $\pi/4$ & $\pi/2$ & $\pi$ & \upa & \upa \\
\br
\end{tabular}
\end{indented}
\end{table}

\begin{table}[tb]
\caption{\label{tab:brpol} Observables involving both beam and recoil polarizations
in their leading terms in equation~(\ref{eq:gcs});
$d\sigma_0=(\sigma_1+\sigma_2)$ and $\hat A=(\sigma_1-\sigma_2)$.}
\begin{indented}
\item[] \begin{tabular}{@{}llcccccc}
\br
\multicolumn{2}{l}{$B$-$R$} & \multicolumn{2}{c}{Beam} &
\multicolumn{2}{c}{Target} & \multicolumn{2}{c}{Recoil} \\
Observable & $(\sigma_{1}-\sigma_{2})$ & $h_{\gamma}$ & $\phi_{\gamma}^{L}$ &
$\theta_{p}$ & $\phi_{p}$ & $\theta_{p'}$ & $\phi_{p'}$ \\ 
\mr
$\hat{C}_{x'}$ & $\sigma_{1} = \sigma(+1,0,+x')$ & $+1$ & - & \upa & \upa & $\pi/2 + \theta_{K}$ & 0 \\
 & $\sigma_{2} = \sigma(-1,0,+x')$ & $-1$ & - & \upa & \upa & $\pi/2 + \theta_{K}$ & 0 \\
$\hat{C}_{x'}$ & $\sigma_{1} = \sigma(+1,0,+x')$ & $+1$ & - & \upa & \upa & $\pi/2 + \theta_{K}$ & 0 \\
 & $\sigma_{2} = \sigma(+1,0,-x')$ & $+1$ & - & \upa & \upa & $3\pi/2 + \theta_{K}$ & 0 \\
&&&&&&&\\
$\hat{C}_{z'}$ & $\sigma_{1} = \sigma(+1,0,+z')$ & $+1$ & - & \upa & \upa & $\theta_{K}$ & 0 \\
 & $\sigma_{2} = \sigma(-1,0,+z')$ & $-1$ & - & \upa & \upa & $\theta_{K}$ & 0 \\
$\hat{C}_{z'}$ & $\sigma_{1} = \sigma(+1,0,+z')$ & $+1$ & - & \upa & \upa & $\theta_{K}$ & 0 \\
 & $\sigma_{2} = \sigma(+1,0,-z')$ & $+1$ & - & \upa & \upa & $\pi + \theta_{K}$ & 0 \\
&&&&&&&\\
$\hat{O}_{x'}$ & $\sigma_{1} = \sigma(+\pi/4,0,+x')$ & - & $\pi/4$ & \upa & \upa & $\pi/2 + \theta_{K}$ & 0 \\
 & $\sigma_{2} = \sigma(-\pi/4,0,+x')$ & - & $3\pi/4$ & \upa & \upa & $\pi/2 + \theta_{K}$ & 0 \\
$\hat{O}_{x'}$ & $\sigma_{1} = \sigma(+\pi/4,0,+x')$ & - & $\pi/4$ & \upa & \upa & $\pi/2 + \theta_{K}$ & 0 \\
 & $\sigma_{2} = \sigma(+\pi/4,0,-x')$ & - & $\pi/4$ & \upa & \upa & $3\pi/2 + \theta_{K}$ & 0 \\
&&&&&&&\\
$\hat{O}_{z'}$ & $\sigma_{1} = \sigma(+\pi/4,0,+z')$ & - & $\pi/4$ & \upa & \upa & $\theta_{K}$ & 0 \\
 & $\sigma_{2} = \sigma(-\pi/4,0,+z')$ & - & $3\pi/4$ & \upa & \upa & $\theta_{K}$ & 0 \\
$\hat{O}_{z'}$ & $\sigma_{1} = \sigma(+\pi/4,0,+z')$ & - & $\pi/4$ & \upa & \upa & $\theta_{K}$ & 0 \\
 & $\sigma_{2} = \sigma(+\pi/4,0,-z')$ & - & $\pi/4$ & \upa & \upa & $\pi + \theta_{K}$ & 0 \\
\br
\end{tabular}
\end{indented}
\end{table}

\begin{table}[tb]
\caption{\label{tab:trpol} Observables involving both target and recoil polarizations
in their leading terms in equation~(\ref{eq:gcs});
$d\sigma_0=(\sigma_1+\sigma_2)$ and $\hat A=(\sigma_1-\sigma_2)$.}
\begin{indented}
\item[] \begin{tabular}{@{}llcccccc}
\br
\multicolumn{2}{l}{$T$-$R$} & \multicolumn{2}{c}{Beam} &
\multicolumn{2}{c}{Target} & \multicolumn{2}{c}{Recoil} \\
Observable & $(\sigma_{1}-\sigma_{2})$ & $h_{\gamma}$ & $\phi_{\gamma}^{L}$ &
$\theta_{p}$ & $\phi_{p}$ & $\theta_{p'}$ & $\phi_{p'}$ \\ 
\mr
$\hat{L}_{x'}$ & $\sigma_{1} = \sigma(0,+z,+x')$ & \upa & \upa & 0 & 0 & $\pi/2 + \theta_{K}$ & 0 \\
 & $\sigma_{2} = \sigma(0,-z,+x')$ & \upa & \upa & $\pi$ & 0 & $\pi/2 + \theta_{K}$ & 0 \\
$\hat{L}_{x'}$ & $\sigma_{1} = \sigma(0,+z,+x')$ & \upa & \upa & 0 & 0 & $\pi/2 + \theta_{K}$ & 0 \\ 
 & $\sigma_{2} = \sigma(0,+z,-x')$ & \upa & \upa & 0 & 0 & $3\pi/2 + \theta_{K}$ & 0 \\
&&&&&&&\\
$\hat{L}_{z'}$ & $\sigma_{1} = \sigma(0,+z,+z')$ & \upa & \upa & 0 & 0 & $\theta_{K}$ & 0 \\
 & $\sigma_{2} = \sigma(0,-z,+z')$ & \upa & \upa & $\pi$ & 0 & $\theta_{K}$ & 0 \\
$\hat{L}_{z'}$ & $\sigma_{1} = \sigma(0,+z,+z')$ & \upa & \upa & 0 & 0 & $\theta_{K}$ & 0 \\
 & $\sigma_{2} = \sigma(0,+z,-z')$ & \upa & \upa & 0 & 0 & $\pi + \theta_{K}$ & 0 \\
&&&&&&&\\
$\hat{T}_{x'}$ & $\sigma_{1} = \sigma(0,+x,+x')$ & \upa & \upa & $\pi/2$ & 0 & $\pi/2 + \theta_{K}$ & 0 \\
 & $\sigma_{2} = \sigma(0,-x,+x')$ & \upa & \upa & $\pi/2$ & $\pi$ & $\pi/2 + \theta_{K}$ & 0 \\
$\hat{T}_{x'}$ & $\sigma_{1} = \sigma(0,+x,+x')$ & \upa & \upa & $\pi/2$ & 0 & $\pi/2 + \theta_{K}$ & 0 \\
 & $\sigma_{2} = \sigma(0,+x,-x')$ & \upa & \upa & $\pi/2$ & 0 & $3\pi/2 + \theta_{K}$ & 0 \\
&&&&&&&\\
$\hat{T}_{z'}$ & $\sigma_{1} = \sigma(0,+x,+z')$ & \upa & \upa & $\pi/2$ & 0 & $\theta_{K}$ & 0 \\
 & $\sigma_{2} = \sigma(0,-x,+z')$ & \upa & \upa & $\pi/2$ & $\pi$ & $\theta_{K}$ & 0 \\
$\hat{T}_{z'}$ & $\sigma_{1} = \sigma(0,+x,+z')$ & \upa & \upa & $\pi/2$ & 0 & $\theta_{K}$ & 0 \\
 & $\sigma_{2} = \sigma(0,+x,-z')$ & \upa & \upa & $\pi/2$ & 0 & $\pi + \theta_{K}$ & 0 \\
\br
\end{tabular}
\end{indented}
\end{table}

\clearpage

\section{\label{apx_fier} The Fierz identities}

We list here the Fierz identities relating \textit{asymmetries}, with signs
consistent with the definition of observables in~\ref{apx_tab} and with the form of the general
cross sections in equation~(\ref{eq:gcs}).
The equation numbering sequence in \ref{apx_fier1}-\ref{apx_fier3}
is that of Chiang and Tabakin~\cite{chiang}.
Compared to the results given in~\cite{chiang}, our equation~(\ref{eq:L.0})
differs by a factor $4/3$ and the remaining expressions have different signs in all but 
(\ref{eq:L.1}), (\ref{eq:L.4})-(\ref{eq:L.6}),
(\ref{eq:Q.r}), (\ref{eq:Q.bt.3}), (\ref{eq:Q.tr.1}), (\ref{eq:Q.tr.2}); 
needless to say, the six \textit{Squared} relations are the same.
Sign changes in eight of the equations can be attributed to the
different definition for the $E$ asymmetry used by Fasano, Tabakin
and Saghai~\cite{fasano}, to which Chiang and Tabakin refer.

\subsection{\label{apx_fier1}Linear-quadratic relations}
\renewcommand{\theequation}{L.0}
\begin{eqnarray}
1 &=& \{ \Sigma^{2} + T^{2} + P^{2} + E^{2} + G^{2} + F^{2} + H^{2} \nonumber\\
  & &+ O_{x'}^{2} + O_{z'}^{2} + C_{x'}^{2} + C_{z'}^{2} 
     + L_{x'}^{2} + L_{z'}^{2} + T_{x'}^{2} + T_{z'}^{2}\}/3 .
\label{eq:L.0}
\end{eqnarray}
\renewcommand{\theequation}{L.TR}
\begin{equation}
\Sigma = + TP + T_{x'}L_{z'} - T_{z'}L_{x'} .
\label{eq:L.TR}
\end{equation}
\renewcommand{\theequation}{L.BR}
\begin{equation}
T = + \Sigma P - C_{x'}O_{z'} + C_{z'}O_{x'} .
\label{eq:L.BR}
\end{equation}
\renewcommand{\theequation}{L.BT}
\begin{equation}
P = + \Sigma T + GF + EH .
\label{eq:L.BT}
\end{equation}
\setcounter{equation}{0}
\renewcommand{\theequation}{L.\arabic{equation}}
\begin{equation}
G = + PF + O_{x'}L_{x'} + O_{z'}L_{z'} .
\label{eq:L.1}
\end{equation}
\begin{equation}
H = + PE + O_{x'}T_{x'} + O_{z'}T_{z'} .
\label{eq:L.2}
\end{equation}
\begin{equation}
E = + PH - C_{x'}L_{x'} - C_{z'}L_{z'} .
\label{eq:L.3}
\end{equation}
\begin{equation}
F = + PG + C_{x'}T_{x'} + C_{z'}T_{z'} .
\label{eq:L.4}
\end{equation}
\begin{equation}
O_{x'} = + TC_{z'} + GL_{x'} + HT_{x'} .
\label{eq:L.5}
\end{equation}
\begin{equation}
O_{z'} = - TC_{x'} + GL_{z'} + HT_{z'} .
\label{eq:L.6}
\end{equation}
\begin{equation}
C_{x'} = - TO_{z'} - EL_{x'} + FT_{x'} .
\label{eq:L.7}
\end{equation}
\begin{equation}
C_{z'} = + TO_{x'} - EL_{z'} + FT_{z'} .
\label{eq:L.8}
\end{equation}
\begin{equation}
T_{x'} = + \Sigma L_{z'} + HO_{x'} + FC_{x'} .
\label{eq:L.9}
\end{equation}
\begin{equation}
T_{z'} = - \Sigma L_{x'} + HO_{z'} + FC_{z'} .
\label{eq:L.10}
\end{equation}
\begin{equation}
L_{x'} = - \Sigma T_{z'} + GO_{x'} - EC_{x'} .
\label{eq:L.11}
\end{equation}
\begin{equation}
L_{z'} = + \Sigma T_{x'} + GO_{z'} - EC_{z'} .
\label{eq:L.12}
\end{equation}

\subsection{\label{apx_fier2}Quadratic relations}
\renewcommand{\theequation}{Q.b}
\begin{equation}
C_{x'}O_{x'} + C_{z'}O_{z'} + EG - FH = 0 .
\label{eq:Q.b}
\end{equation}
\renewcommand{\theequation}{Q.t}
\begin{equation}
GH - EF - L_{x'}T_{x'} - L_{z'}T_{z'} = 0 .
\label{eq:Q.t}
\end{equation}
\renewcommand{\theequation}{Q.r}
\begin{equation}
C_{x'}C_{z'} + O_{x'}O_{z'} - L_{x'}L_{z'} - T_{x'}T_{z'} = 0 .
\label{eq:Q.r}
\end{equation}
\setcounter{equation}{0}
\renewcommand{\theequation}{Q.bt.\arabic{equation}}
\begin{equation}
\Sigma G - TF - O_{z'}T_{x'} + O_{x'}T_{z'} = 0 .
\label{eq:Q.bt.1}
\end{equation}
\begin{equation}
\Sigma H - TE + O_{z'}L_{x'} - O_{x'}L_{z'} = 0 .
\label{eq:Q.bt.2}
\end{equation}
\begin{equation}
\Sigma E - TH + C_{z'}T_{x'} - C_{x'}T_{z'} = 0 .
\label{eq:Q.bt.3}
\end{equation}
\begin{equation}
\Sigma F - TG + C_{z'}L_{x'} - C_{x'}L_{z'} = 0 .
\label{eq:Q.bt.4}
\end{equation}
\setcounter{equation}{0}
\renewcommand{\theequation}{Q.br.\arabic{equation}}
\begin{equation}
\Sigma O_{x'} - PC_{z'} + GT_{z'} - HL_{z'} = 0 .
\label{eq:Q.br.1}
\end{equation}
\begin{equation}
\Sigma O_{z'} + PC_{x'} - GT_{x'} + HL_{x'} = 0 .
\label{eq:Q.br.2}
\end{equation}
\begin{equation}
\Sigma C_{x'} + PO_{z'} - ET_{z'} - FL_{z'} = 0 .
\label{eq:Q.br.3}
\end{equation}
\begin{equation}
\Sigma C_{z'} - PO_{x'} + ET_{x'} + FL_{x'} = 0 .
\label{eq:Q.br.4}
\end{equation}
\setcounter{equation}{0}
\renewcommand{\theequation}{Q.tr.\arabic{equation}}
\begin{equation}
TT_{x'} - PL_{z'} - HC_{z'} + FO_{z'} = 0 .
\label{eq:Q.tr.1}
\end{equation}
\begin{equation}
TT_{z'} + PL_{x'} + HC_{x'} - FO_{x'} = 0 .
\label{eq:Q.tr.2}
\end{equation}
\begin{equation}
TL_{x'} + PT_{z'} - GC_{z'} - EO_{z'} = 0 .
\label{eq:Q.tr.3}
\end{equation}
\begin{equation}
TL_{z'} - PT_{x'} + GC_{x'} + EO_{x'} = 0 .
\label{eq:Q.tr.4}
\end{equation}

\subsection{\label{apx_fier3}Squared relations}
\renewcommand{\theequation}{S.bt}
\begin{equation}
G^{2} + H^{2} + E^{2} + F^{2} + \Sigma^{2} + T^{2} - P^{2} = 1 .
\label{eq:S.bt}
\end{equation}
\renewcommand{\theequation}{S.br}
\begin{equation}
O_{x'}^{2} + O_{z'}^{2} + C_{x'}^{2} + C_{z'}^{2} + \Sigma^{2} - T^{2} + P^{2} = 1 .
\label{eq:S.br}
\end{equation}
\renewcommand{\theequation}{S.tr}
\begin{equation}
T_{x'}^{2} + T_{z'}^{2} + L_{x'}^{2} + L_{z'}^{2} - \Sigma^{2} + T^{2} + P^{2} = 1 .
\label{eq:S.tr}
\end{equation}
\renewcommand{\theequation}{S.b}
\begin{equation}
G^{2} + H^{2} - E^{2} - F^{2} - O_{x'}^{2} - O_{z'}^{2} + C_{x'}^{2} + C_{z'}^{2} = 0 .
\label{eq:S.b}
\end{equation}
\renewcommand{\theequation}{S.t}
\begin{equation}
G^{2} - H^{2} + E^{2} - F^{2} + T_{x'}^{2} + T_{z'}^{2} - L_{x'}^{2} - L_{z'}^{2} = 0 .
\label{eq:S.t}
\end{equation}
\renewcommand{\theequation}{S.r}
\begin{equation}
O_{x'}^{2} - O_{z'}^{2} + C_{x'}^{2} - C_{z'}^{2} - T_{x'}^{2} + T_{z'}^{2} - L_{x'}^{2} + L_{z'}^{2} = 0 .
\label{eq:S.r}
\end{equation}

\subsection{\label{apx_fier4} ARS-squared relations}
Here we include a set of squared relations discussed in Artru, Richard and Soffer (ARS)~\cite{art09}. 
These can be derived from combinations of relations in the preceding sections. 
For example, the first, (\ref{eq:ARS.S.bt}), can be obtained by combining (\ref{eq:S.bt}) 
and~(\ref{eq:L.BT}). 
Our relations differ in sign from ARS in those terms involving $F$, $C_{x'}$ and $C_{z'}$, 
and as a result there are sign differences in (\ref{eq:ARS.S.bt}),~(\ref{eq:ARS.S.br}) 
and~(\ref{eq:ARS.btr1}).
\renewcommand{\theequation}{ARS.S.bt}
\begin{equation}
(1 \pm P)^2 = (T \pm \Sigma)^2 + (E \pm H)^2 + (G \pm F)^2 .
\label{eq:ARS.S.bt}
\end{equation}
\renewcommand{\theequation}{ARS.S.br}
\begin{equation}
(1 \pm T)^2 = (P \pm \Sigma)^2 + (C_{x'} \mp O_{z'})^2 + (C_{z'} \pm O_{x'})^2 .
\label{eq:ARS.S.br}
\end{equation}
\renewcommand{\theequation}{ARS.S.tr}
\begin{equation}
(1 \pm \Sigma)^2 = (P \pm T)^2 + (L_{x'} \mp T_{z'})^2 + (L_{z'} \pm T_{x'})^2 .
\label{eq:ARS.S.tr}
\end{equation}
\renewcommand{\theequation}{ARS.btr1}
\begin{equation}
(1 \pm L_{z'})^2 = (\Sigma \pm T_{x'})^2 + (E \mp C_{z'})^2 + (G \pm O_{z'})^2 .
\label{eq:ARS.btr1}
\end{equation}
\makeatletter
\renewcommand{\theequation}{ARS.btr2}
\@addtoreset{equation}{section}
\makeatother
\begin{equation}
(1 \pm T_{x'})^2 = (\Sigma \pm L_{z'})^2 + (F \pm C_{x'})^2 + (H \pm O_{x'})^2 .
\label{eq:ARS.btr2}
\end{equation}

\section{\label{apx_born} Born amplitudes for $\gamma N \to  K \Lambda$}
\setcounter{equation}{0}
\renewcommand{\theequation}{\Alph{section}.\arabic{equation}}

In this Appendix, we summarize the Born amplitudes 
for $\gamma(q)+ p(p)\rightarrow  K^+(k')+\Lambda (p')$
in the center of mass energy ($\vec p = -\vec q$, $\vec p'=-\vec k'$), 
which are used to fix high partial waves ($4\leq L\leq 8$)
in the multipole analyses presented in section~\ref{sec:anal}.
We consider the following Born terms for $I^\mu\epsilon_\mu$ 
[see the paragraph including (\ref{eq:dsdo}) for 
the description of $I^\mu\epsilon_\mu$]:
\begin{eqnarray}
I^\mu\epsilon_\mu = I_a+I_b+I_c+I_d+I_e+I_f ,
\end{eqnarray}
where
\begin{eqnarray}
I_a &=& i\frac{f_{K N \Lambda}}{m_K}
           \sla{k'}\gamma_5 \frac{1}{\sla{p'}+\sla{k'}-m_N }\Gamma_N(q^2) 
F(|\vec k'|,\Lambda_{KN\Lambda}) , 
\\
I_b &=& i\frac{f_{KN\Lambda}}{m_K}
         \Gamma_\Lambda(q^2) \frac{1}{\sla{p}-\sla{k'}-m_\Lambda }\sla{k'}\gamma_5 
F(|\vec k'|,\Lambda_{KN\Lambda}) , 
\\
I_c &=&  i\frac{f_{KN\Sigma}}{m_K}
          \Gamma_{\Lambda\Sigma}(q^2) \frac{1}{\sla{p}-\sla{k'}-m_\Sigma}\sla{k'}\gamma_5 
F(|\vec k'|,\Lambda_{KN\Sigma}) , 
\\
I_d &=& -ie\frac{f_{KN\Lambda}}{m_K}\sla{\epsilon_\gamma}\gamma_5 
F(|\vec k'|,\Lambda_{KN\Lambda}) , 
\\
I_e &=& ie\frac{f_{KN\Lambda}}{m_K}\frac{\sla{\tilde{k}}\gamma_5}
          {\tilde{k}^2-m^2_K}(\tilde{k}+k')\cdot\epsilon_\gamma 
F(|\vec {\tilde k}|,\Lambda_{KN\Lambda}) , 
\\
I_f &=& -e\frac{g_{K^*N\Lambda}g_{K^\ast K^+\gamma}}{m_K}
 [\gamma^\delta+\frac{\kappa_{K^*N\Lambda}}{2(m_N+m_\Lambda)}
 (\gamma^\delta\sla{\tilde{k}}-\sla{\tilde{k}} \gamma^\delta)] 
\nonumber\\
&&
\times
  \epsilon_{\alpha\beta\eta\delta} {\tilde{k}}^\eta q^\alpha \epsilon^\beta_\gamma 
  \frac{1}{\tilde{k}^2-m^2_{K^*}} 
F(|\vec {\tilde k}|,\Lambda_{K^\ast N\Lambda}) , 
\end{eqnarray}
with $\tilde{k}=p-p'$ and 
\begin{eqnarray}
 \Gamma_N &=&e\{\sla{\epsilon_\gamma}-\frac{\kappa_N}{4m_N}
[\sla{\epsilon_\gamma}\sla{q}-\sla{q}\sla{\epsilon_\gamma}]\} , \\
 \Gamma_\Lambda &=&-e\frac{\kappa_\Lambda}{4m_N}
[\sla{\epsilon_\gamma}\sla{q}-\sla{q}\sla{\epsilon_\gamma}] , \\
 \Gamma_{\Lambda\Sigma} &=&-e\frac{\kappa_{\Lambda\Sigma}}{4m_N}
[\sla{\epsilon_\gamma}\sla{q}-\sla{q}\sla{\epsilon_\gamma}] .
\end{eqnarray}
Also, we have introduced the dipole form factors $F(|\vec k|,\Lambda)$ for the hadronic vertex
defined as
\begin{equation}
F(|\vec k|,\Lambda) = \left(\frac{\Lambda^2}{|\vec k|^2+\Lambda^2}\right)^2 .
\end{equation}

We make use of the SU(3) relation for the coupling constants,
\begin{eqnarray}
\frac{f_{KN\Lambda}}{m_K} &=& \frac{f_{\pi NN}}{m_\pi}\frac{-3+2d}{\sqrt{3}}, \\
\frac{f_{KN\Sigma}}{m_K}  &=& \frac{f_{\pi NN}}{m_\pi}\frac{3-4d}{\sqrt{3}},  \\
g_{K^*N\Lambda} &=& g_{\rho NN}\frac{-3+2d}{\sqrt{3}},  \\
\frac{\kappa_{K^*N\Lambda}}{m_N+m_\Lambda} &=& \frac{\kappa_\rho}{2m_N} ,
\end{eqnarray}
and take parameters as
$f_{\pi NN}=\sqrt{0.08\times 4\pi}$~\cite{jlms07},
$\kappa_{p}=\mu_p -1=1.79$~\cite{pdg10}, $d=0.635$~\cite{onl06}, 
$g_{\rho NN}=8.72$~\cite{jlms07}, $\kappa_\rho=2.65$~\cite{jlms07},
$g_{\gamma K^* K^+}/m_K=  0.254$GeV$^{-1}$~\cite{onl06},
$\kappa_{\Lambda}= -0.61$~\cite{onl06}, and
$\kappa_{\Lambda\Sigma}= -1.61$~\cite{onl06}.
As for the cutoff factors, we take
$\Lambda_{KN\Lambda} =\Lambda_{KN\Sigma} =\Lambda_{K^\ast N\Lambda}= 500$ MeV.


\section*{References}
\begin{thebibliography}{99}
\expandafter\ifx\csname url\endcsname\relax
  \def\url#1{\texttt{#1}}\fi
\expandafter\ifx\csname urlprefix\endcsname\relax\def\urlprefix{URL }\fi

\bibitem{chew}
Chew~G~R, Goldberger~M~L, Low~F~E and Nambu~Y 1957
{\it Phys. Rev.} {\bf 105} 1345

\bibitem{chiang}
Chiang W T and Tabakin F 1997
{\it Phys. Rev.} C {\bf 55} 2054

\bibitem{Frost} 
The~g9-FROST~series~of~experiments,
\urlprefix\url{http://clasweb.jlab.org/shift/g9/} .

\bibitem{HDice}
The~g14-HDice~experiment,
\urlprefix\url{http://www.jlab.org/exp\_prog/proposals/06/PR-06-101.pdf}

\bibitem{gr07}
Lleres~A et~al. 2007
{\it Eur. Phys. J.} A {\bf 31} 79

\bibitem{gr09}
Lleres~A et~al. (GRAAL Collaboration) 2009
{\it Eur. Phys. J.} A {\bf 39} 149

\bibitem{barker}
Barker~I~S, Donnachie~A and Storrow~J~K 1975
{\it Nucl. Phys.} {\bf B95} 347

\bibitem{donn66}
Donnachie~A and Shaw~G 1966
{\it Appl. Phys.} {\bf 37} 333

\bibitem{donn72}
Donnachie~A (edited by Bishop~E~H~S) 1972
{\it Pure and Applied Physics}
vol~25-V (Academic Press, New York) p~1

\bibitem{adel}
Adelseck~R~A and Saghai~B 1990
{\it Phys. Rev.} C {\bf 42} 108

\bibitem{fasano}
Fasano~C~G, Tabakin~F and Saghai~B 1992
{\it Phys. Rev.} C {\bf 46} 2430

\bibitem{drech92}
Drechsel~D and Tiator~L 1992
{\it J. Phys.} G {\bf 18} 449

\bibitem{knoch}
Kn\"{o}chlein~G, Drechsel~D and Tiator~L 1995
{\it Z. Phys.} A {\bf 352} 327

\bibitem{jackson}
Jackson~J~D 1975
{\it Classical Electrodynamics} (John Wiley \& Sons, New York)

\bibitem{bjdr}
Bjorken~J~D and Drell~S~D 1964
{\it Relativistic Quantum Mechanics} (McGraw-Hill, New York)

\bibitem{jw59}
Jacob~M and Wick~G~C 1959
{\it Ann. Phys.} {\bf 7} 404; 
Jacob~M and Wick~G~C 2000
{\it Ann. Phys.} {\bf 281} 774

\bibitem{bs68}
Brink~D~M and Satchler~G~R 1968
\textit{Angular Momentum} (Oxford University Press, Oxford)

\bibitem{worden}
Worden~R 1972
{\it Nucl. Phys.} {\bf B37} 253

\bibitem{legs09}
Hoblit~S et~al. (LEGS-Spin Collaboration) 2009
{\it Phys. Rev. Lett.} {\bf 102} 172002, and references therein.

\bibitem{g1c07}
Bradford~R~K et~al. (CLAS Collaboration) 2007
{\it Phys. Rev.} C {\bf 75} 035205

\bibitem{ly57}
Lee~T~D and Yang~C~N 1957
{\it Phys. Rev.} {\bf 108} 1645

\bibitem{pdg10}
K.~Nakamura et~al. (Particle Data Group) 2010
{\it J. Phys.} G {\bf 37} 075021

\bibitem{MAID}
MAID and Kaon-MAID isobar models of meson production,
\urlprefix\url{http://wwwkph.kph.uni-mainz.de/MAID/}.

\bibitem{drech07}
Drechsel~D, Kamalov~S~S and Tiator~L 2007
{\it Eur. Phys. J.} A {\bf 34} 69

\bibitem{mart}
Mart~T and Bennhold~C 1999
{\it Phys. Rev.} C {\bf 61} 012201

\bibitem{SAID}
SAID partial wave analysis facility,
\urlprefix\url{http://gwdac.phys.gwu.edu/}.

\bibitem{arndt}
Arndt~R~A, Briscoe~W~J, Strakovsky~I~I and Workman~R~L 2002
{\it Phys. Rev.} C {\bf 66} 055213

\bibitem{boga10}
Anisovich~A~V, Klempt~E, Nikonov~V~A, Matveev~M~A, Sarantsev~A~V and Thoma~U 2010
{\it Eur. Phys. J.} A {\bf 44} 203;
Bonn-Gatchina Partial Wave Analysis,
\urlprefix\url{http://pwa.hiskp.uni-bonn.de/}

\bibitem{bruno}
Juli\'{a}-D\'{i}az~B, Saghai~B, Lee~T~S~H and Tabakin~F 2006
{\it Phys. Rev.} C {\bf 73} 055204

\bibitem{art07}
Artru~X, Richard~J~M and Soffer~J 2007
{\it Phys. Rev.} C {\bf 75} 024002

\bibitem{art09}
Artru~X, Elchikh~M, Richard~J~M, Soffer~J and Teryaev~O~V 2009
{\it Phys. Rep.} \textbf{470} 1

\bibitem{ire10}
Ireland~D (private communication)

\bibitem{saph04}
Glander~K~K et~al. (SAPHIR Collaboration) 2004
{\it Eur. Phys. J.} A {\bf 19} 2

\bibitem{g11a10}
McCracken~M~E et~al. (CLAS Collaboration) 2010
{\it Phys. Rev.} C {\bf 81} 025201

\bibitem{g1c06}
Bradford~R~K et~al. (CLAS Collaboration) 2006
{\it Phys. Rev.} C {\bf 73} 035202

\bibitem{leps03}
Zegers~R~G~T et~al. (LEPS Collaboration) 2003
{\it Phys. Rev. Lett.} {\bf 91} 92001

\bibitem{legs01}
Blanpied~G et~al. (LEGS Collaboration) 2001
{\it Phys. Rev.} C  {\bf 64} 25203

\bibitem{bb75}
Bowcock~J~E and Burkhardt~H 1975
{\it Rep. Prog. Phys.} {\bf 38} 1099

\bibitem{boga07}
Anisovich~A~V, Kleber~V, Klempt~E, Nikonov~V~A, Sarantsev~A~V, and Thoma~U 2007
{\it Eur.\ Phys.\ J.} A {\bf 34} 243

\bibitem{hkamano}
H.~Kamano, hkamano@jlab.org

\bibitem{jlms07}
Juli\'{a}-D\'{i}az~B, Lee~T~S~H, Matsuyama~A and Sato~T 2007
{\it Phys. Rev.} C {\bf 76} 065201

\bibitem{onl06}
Oh~Y, Nakayama~K and Lee~T~S~H 2006
{\it Phys. Rep.} {\bf 423} 49

\end{thebibliography}
\end{document}